%% file: superphot_plus.tex
\documentclass[twocolumn,twocolappendix]{aastex631}
\usepackage{amsmath}

\input{data_values}

\begin{document}

\title{Superphot+: Realtime Fitting and Classification of Supernova Light Curves}

\author[0000-0002-9886-2834]{Kaylee M. de Soto}
\affil{Center for Astrophysics \textbar{} Harvard \& Smithsonian, 60 Garden Street, Cambridge, MA 02138-1516, USA}

\author[0000-0002-5814-4061]{Ashley Villar}
\affil{Center for Astrophysics \textbar{} Harvard \& Smithsonian, 60 Garden Street, Cambridge, MA 02138-1516, USA}

\author[0000-0002-9392-9681]{Edo Berger}
\affil{Center for Astrophysics \textbar{} Harvard \& Smithsonian, 60 Garden Street, Cambridge, MA 02138-1516, USA}
\affiliation{The NSF AI Institute for Artificial Intelligence and Fundamental Interactions}

\author[0000-0001-6395-6702]{Sebastian Gomez}
\affil{Space Telescope Science Institute, 3700 San Martin Dr, Baltimore, MD 21218, USA}

\author[0000-0002-0832-2974]{Griffin Hosseinzadeh}
\affil{Steward Observatory, University of Arizona, 933 North Cherry Avenue, Tucson, AZ 85721-0065, USA}

\author[0009-0009-7822-7110]{Doug Branton}
\affiliation{DiRAC Institute and the Department of Astronomy, University of Washington, 3910 15th Ave NE, Seattle, WA 98195, USA}

\author[0009-0007-9870-9032]{Sandro Campos}
\affiliation{McWilliams Center for Cosmology, Department of Physics, Carnegie Mellon University, Pittsburgh, PA 15213, USA}

\author[0000-0002-1074-2900]{Melissa DeLucchi}
\affiliation{McWilliams Center for Cosmology, Department of Physics, Carnegie Mellon University, Pittsburgh, PA 15213, USA}

\author[0009-0009-2281-7031]{Jeremy Kubica}
\affiliation{McWilliams Center for Cosmology, Department of Physics, Carnegie Mellon University, Pittsburgh, PA 15213, USA}

\author[0000-0001-5028-146X]{Olivia Lynn}
\affiliation{McWilliams Center for Cosmology, Department of Physics, Carnegie Mellon University, Pittsburgh, PA 15213, USA}

\author[0000-0001-7179-7406]{Konstantin Malanchev}
\affiliation{McWilliams Center for Cosmology, Department of Physics, Carnegie Mellon University, Pittsburgh, PA 15213, USA}

\author[0000-0002-8676-1622]{Alex I. Malz}
\affiliation{McWilliams Center for Cosmology, Department of Physics, Carnegie Mellon University, Pittsburgh, PA 15213, USA}

\begin{abstract}
Photometric classifications of supernova (SN) light curves have become necessary to utilize the full potential of large samples of observations obtained from wide-field photometric surveys, such as the Zwicky Transient Facility (ZTF) and the Vera C. Rubin Observatory. Here, we present a photometric classifier for SN light curves that does not rely on redshift information and still maintains comparable accuracy to redshift-dependent classifiers. Our new package, Superphot+, uses a parametric model to extract meaningful features from multiband SN light curves. We train a gradient-boosted machine with fit parameters from \numSNSpec\ ZTF SNe that pass data quality cuts and are spectroscopically classified as one of five classes: SN Ia, SN II, SN Ib/c, SN IIn, and SLSN-I. Without redshift information, our classifier yields a class-averaged F$_1$-score of \FNoRedshift\ and a total accuracy of \accNoRedshift. Including redshift information improves these metrics to \FRedshift\ and \accRedshift, respectively. We assign new class probabilities to \numSNPhot\ ZTF transients that show SN-like characteristics (based on the ALeRCE Broker light curve and stamp classifiers) but lack spectroscopic classifications. Finally, we compare our predicted SN labels with those generated by the ALeRCE light curve classifier, finding that the two classifiers agree on photometric labels for \agreementSpecPercent\ of light curves with spectroscopic labels and \agreementPhotPercent\ of light curves without spectroscopic labels. Superphot+ is currently classifying ZTF SNe in real time via the ANTARES Broker, and is designed for simple adaptation to six-band Rubin light curves in the future.
\end{abstract}


\input{sec1_intro}

\begin{figure}[t]
    \plotone{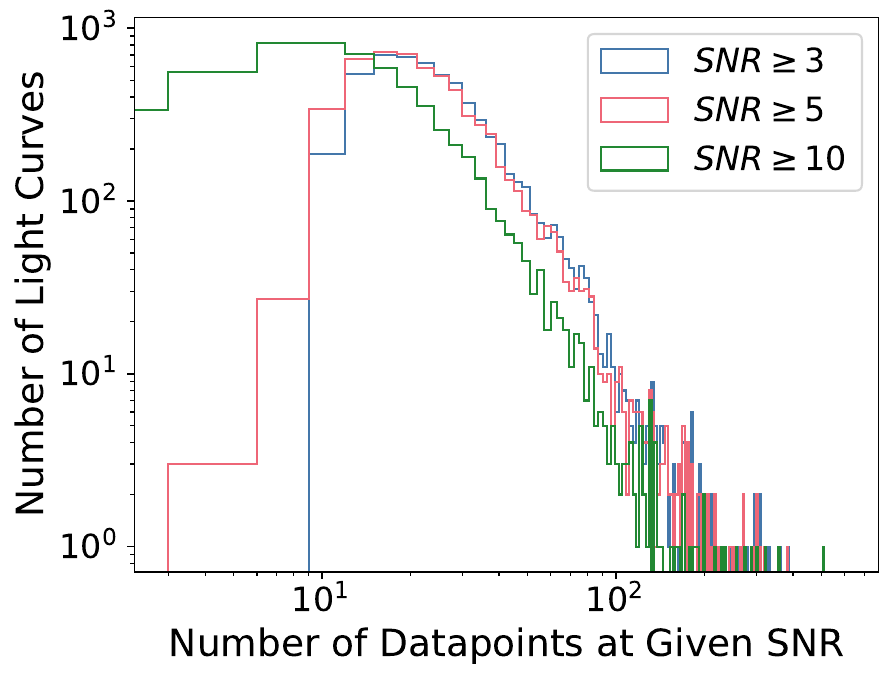} 
    \caption{The distribution of spectroscopically classified SNe in our dataset as a function of the number of data points above a given SNR. Any light curves with fewer than 10 $\mathrm{SNR} \geq 3$ data points are removed during our pruning process. Most light curves that pass our quality cuts (\fracLCtwentyPoints\%) have at least 20 $\mathrm{SNR} \geq 3$ data points, and most $\mathrm{SNR} \geq 3$ points (\fracPointsSNRFive\%) also have SNRs $\geq 5$. This highlights the high quality of our remaining light curves.}
    \label{fig:snr_hist}
\end{figure}

\begin{figure}[t]
    \plotone{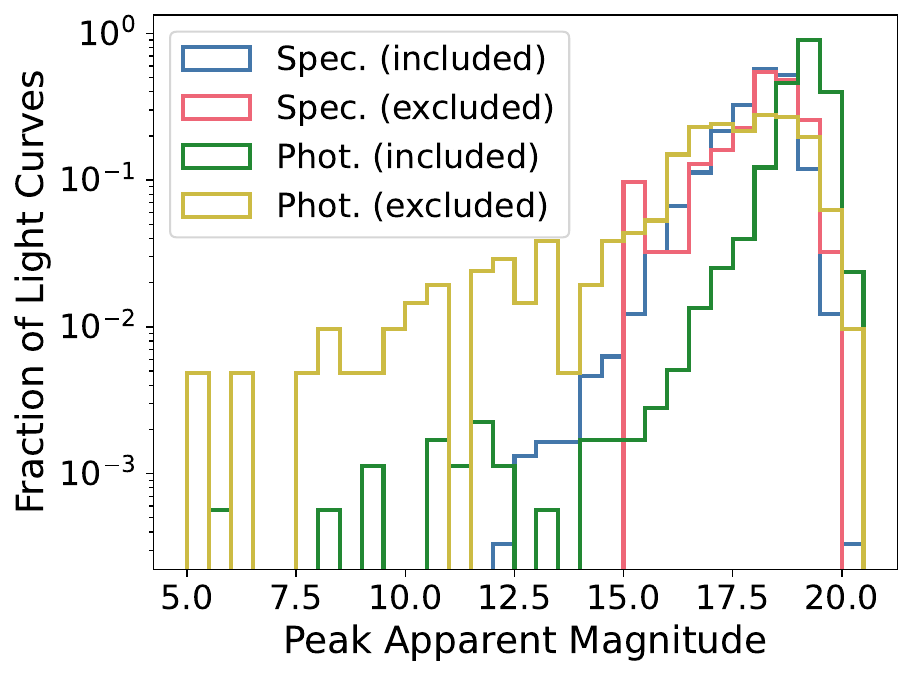} 
    \caption{The distribution of peak $r$-band apparent magnitude ($m_r$) of the spectroscopic (blue and pink) and photometric (green and yellow) datasets. The spectroscopic and photometric datasets are pre-processed identically. The pink and yellow histograms (\numRemovedChisqCut\ and \numPhotRemovedChisqCut\ objects) correspond to poorly fitted light curves that are cut from the final sample, whereas the blue and green light curves pass the fit quality cuts. Extremely bright sources within the photometric set ($m_r<10$) and other probable false detections are successfully excluded by the reduced chi-squared cut. The remaining photometric set has a similarly shaped apparent magnitude distribution to the spectroscopic set but shifted toward fainter magnitudes, which reflects ZTF BTS's (and therefore our spectroscopic sample's) brighter limiting magnitude compared to ZTF's (and thus the photometric sample's) limiting magnitude.}
    \label{fig:appm_hist_unc}
\end{figure}

\begin{deluxetable*}{cccccc}
 \tablecaption{Data Pruning Summary}
    \tablehead{\colhead{Object Type} & \colhead{Original Num.} & \colhead{\textit{N}-Obs Cut} & \colhead{Variability Cut} & \colhead{Num. Remaining} & \colhead{Percent Removed by Cuts}}
    
    \startdata
        \multicolumn{6}{c}{Used in Training}\\
        \hline \hline
        SLSN-I & 101 & 5 & 13 & 83 & 17.8\% \\
        SLSN-II & 55 & 3 & 4 & 48 & 12.7\% \\
        SN II & 1326 & 283 & 136 & 907 & 31.6\% \\
        SN IIL & 2 & 1 & 0 & 1 & 50.0\% \\
        SN IIP & 126 & 22 & 34 & 70 & 44.4\% \\
        SN IIn & 265 & 35 & 21 & 209 & 21.1\% \\
        SN Ia & 6128 & 1585 & 198 & 4345 & 29.1\% \\
        SN Ia-91T-like & 219 & 50 & 9 & 160 & 26.9\% \\
        SN Ia-91bg-like & 59 & 34 & 2 & 23 & 61.0\% \\
        SN Ia-CSM & 20 & 2 & 0 & 18 & 10.0\% \\
        SN Ib & 157 & 59 & 8 & 90 & 42.7\% \\
        SN Ib-Ca-rich & 6 & 5 & 1 & 0 & 100.0\% \\
        SN Ib/c & 40 & 19 & 4 & 17 & 57.5\% \\
        SN Ic & 170 & 59 & 9 & 102 & 40.0\% \\
        SN Ic-BL & 82 & 23 & 9 & 50 & 39.0\% \\
        \hline
        \multicolumn{6}{c}{Excluded from Training, Analyzed in Sec.~\ref{sec:results}}\\\hline
        \hline
        LBV & 10 & 3 & 0 & 7 & 30.0\% \\
        SN IIb & 121 & 53 & 8 & 60 & 50.4\% \\
        SN Iax & 18 & 5 & 1 & 12 & 33.3\% \\
        SN Ibn & 33 & 13 & 0 & 20 & 39.4\% \\
        TDE & 64 & 4 & 5 & 55 & 14.1\% \\
        \hline
        \multicolumn{6}{c}{Excluded from Training}\\
        \hline
        \hline
        AGN & 57 & 6 & 4 & 47 & 17.5\% \\
        CV & 215 & 106 & 3 & 106 & 50.7\% \\
        Galaxy & 19 & 17 & 0 & 2 & 89.5\% \\
        ILRT & 3 & 2 & 0 & 1 & 66.7\% \\
        LRN & 3 & 1 & 0 & 2 & 33.3\% \\
        M dwarf & 6 & 5 & 0 & 1 & 83.3\% \\
        Nova & 36 & 20 & 0 & 16 & 55.6\% \\
        QSO & 5 & 0 & 1 & 4 & 20.0\% \\
        SN & 28 & 14 & 1 & 13 & 53.6\% \\
        SN I & 29 & 12 & 2 & 15 & 48.3\% \\
        SN II-pec & 10 & 3 & 2 & 5 & 50.0\% \\
        SN IIn-pec & 2 & 0 & 1 & 1 & 50.0\% \\
        SN Ia-Ca-rich & 1 & 1 & 0 & 0 & 100.0\% \\
        SN Ia-SC & 5 & 1 & 0 & 4 & 20.0\% \\
        SN Ia-pec & 46 & 12 & 6 & 28 & 39.1\% \\
        SN Ib-pec & 5 & 1 & 1 & 3 & 40.0\% \\
        SN Ibn/Icn & 2 & 0 & 0 & 2 & 0.0\% \\
        SN Ic-Ca-rich & 1 & 0 & 0 & 1 & 0.0\% \\
        SN Ic-pec & 1 & 0 & 0 & 1 & 0.0\% \\
        SN Icn & 5 & 3 & 0 & 2 & 60.0\% \\
        Varstar & 19 & 11 & 0 & 8 & 57.9\% \\
        Other & 26 & 10 & 3 & 13 & 50.0\% 
    \enddata
    \tablecomments{Results of data quality cuts on each transient type from our original spectroscopic TNS sample. The first cut (``$N$-Obs Cut'') requires at least 5 datapoints of $SNR \geq 3$ in each band, while the second cut (``variability cut'') ensures the amplitude and flux variations in each band sufficiently exceed the average flux uncertainty. \label{table:data_cuts}}
\end{deluxetable*}

\input{sec2_data}

\begin{figure*}[t]
    \plotone{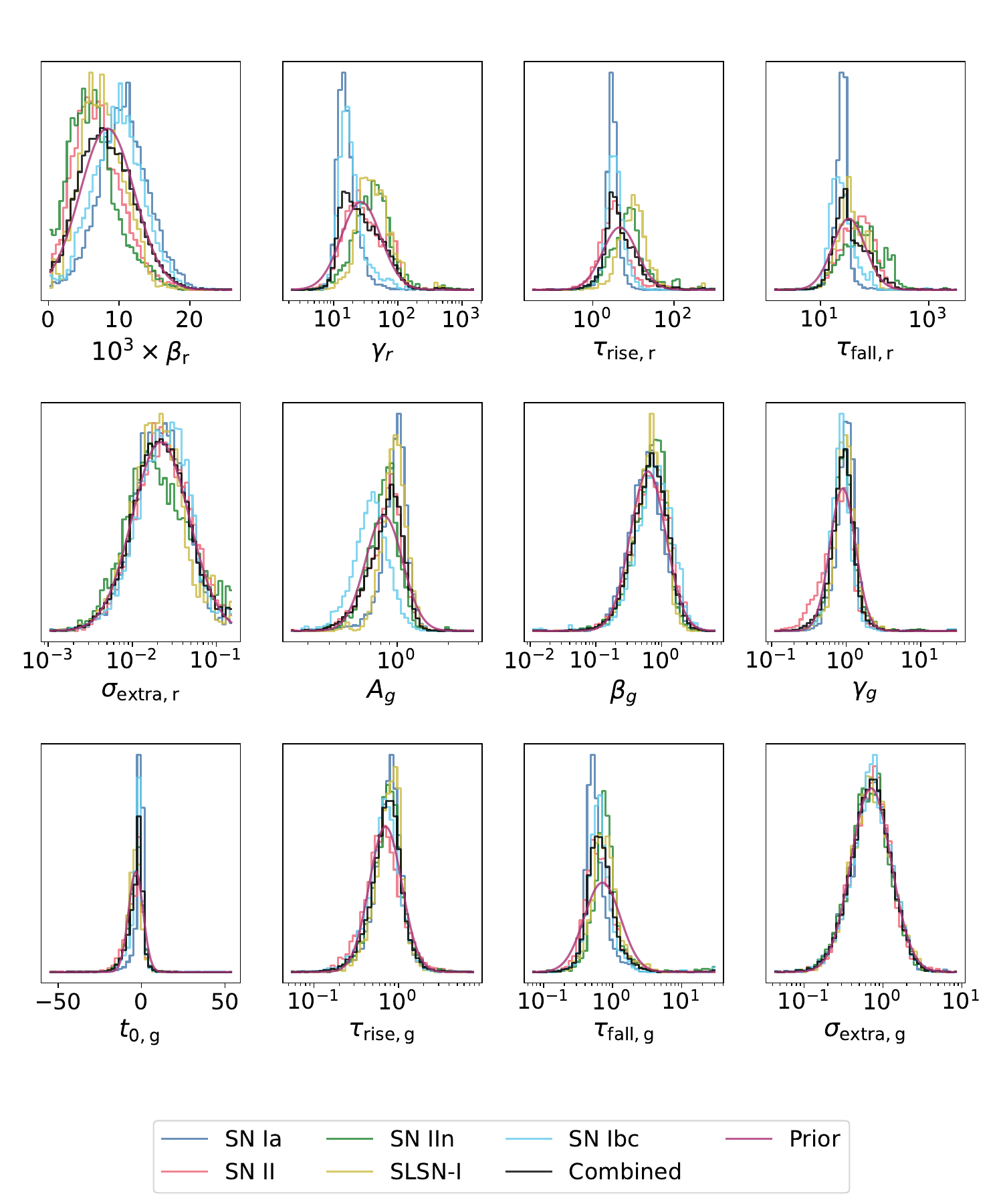} 
    \caption{Marginal distributions of each fit parameter in the oversampled (equal contribution from each class) dataset. We exclude the $r$-band $A$ and $t_0$ parameters as they are not used as inputs for our classifier without redshift information; the $g$-band versions of these parameters are log ratios between the two bands. Some parameters (e.g., $\beta_\mathrm{g}$, $\gamma_g$, $\tau_\mathrm{rise,g}$) are distributed similarly for each SN type, whereas others (e.g., $\gamma_r$, $\tau_\mathrm{fall}$) are clearly more distinguishing. Final priors, shown in purple, are set as the Gaussians or log-Gaussians that most closely match the combined population distributions.}
    \label{fig:1d_distributions}
\end{figure*}

\begin{figure*}[t] 
    \plottwo{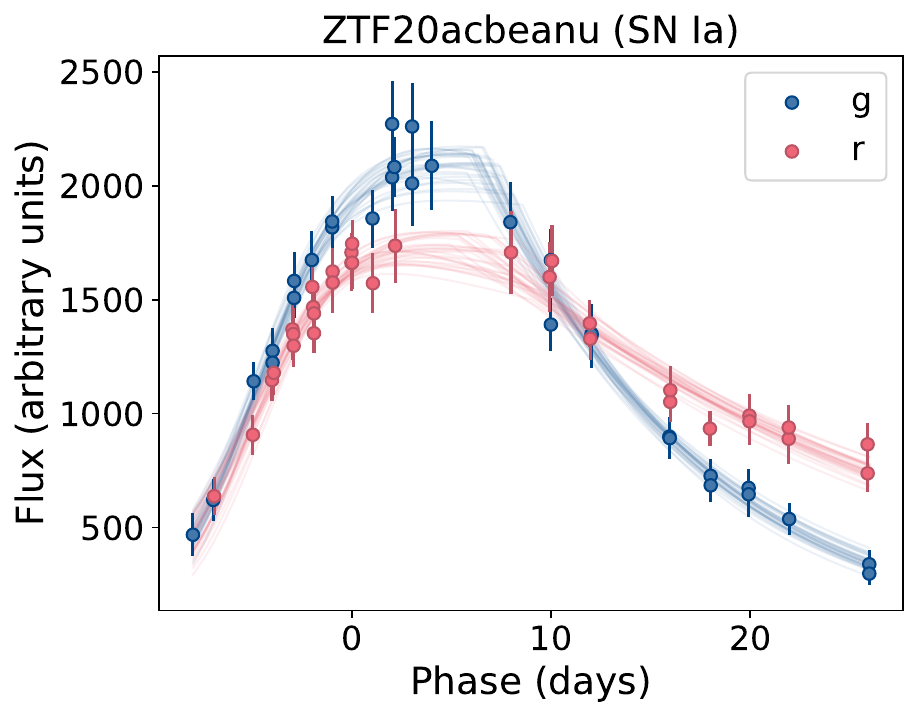}{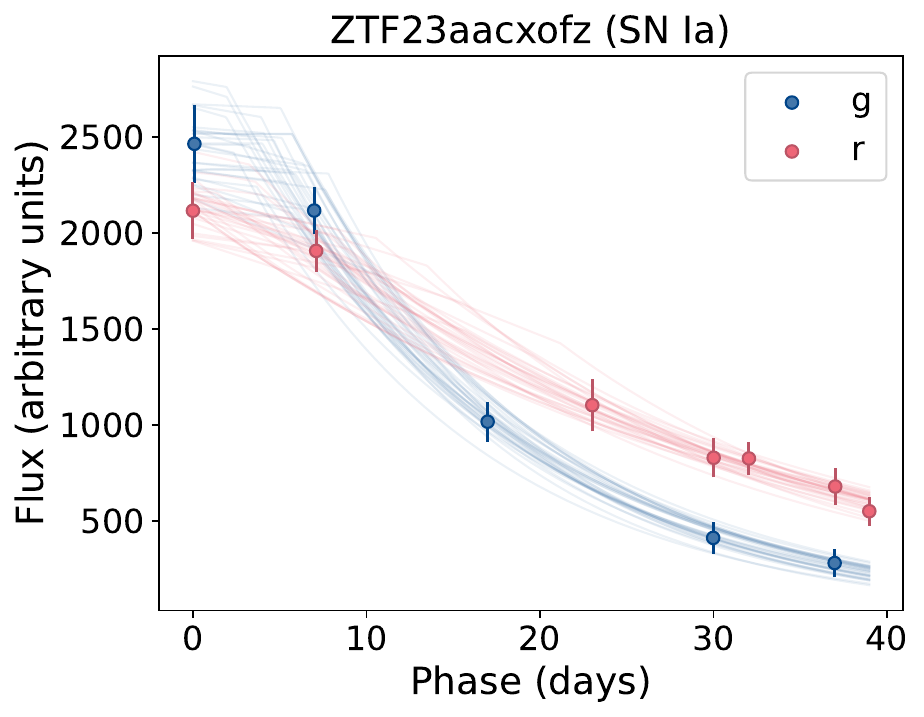}
    \plottwo{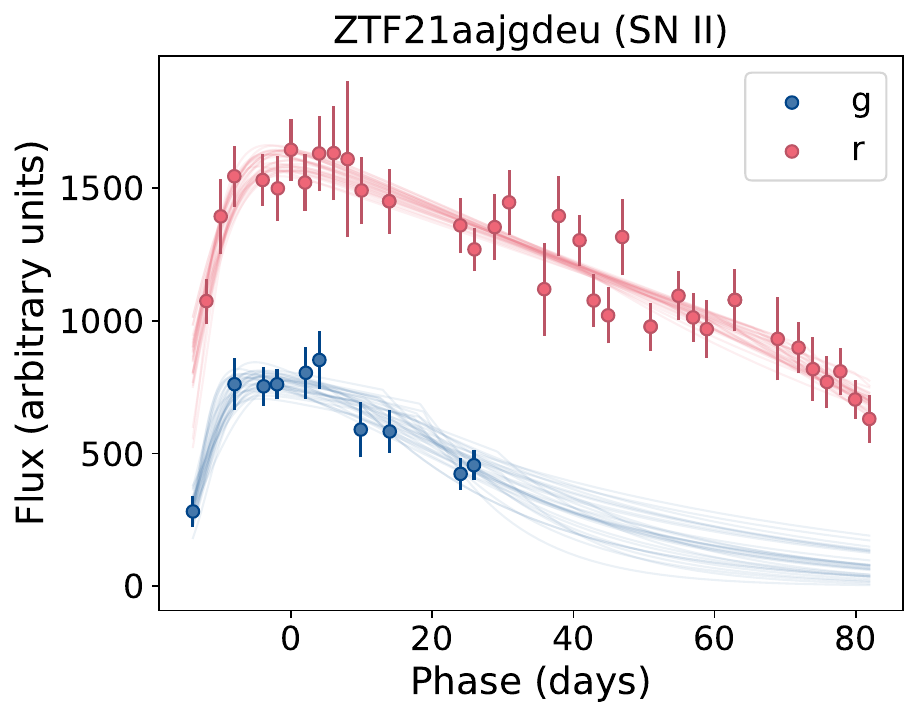}{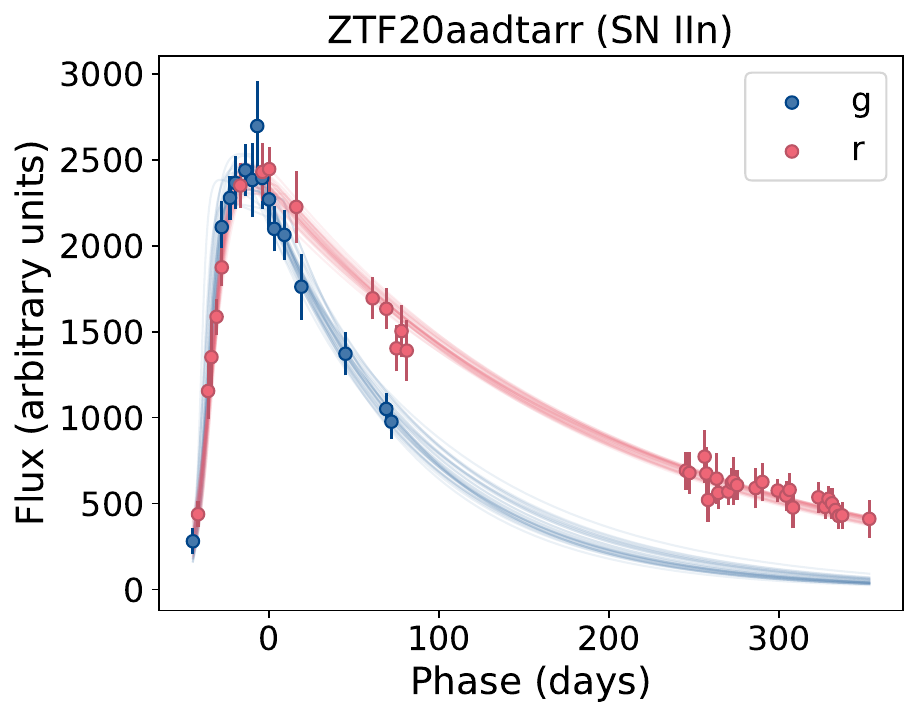}
    \plottwo{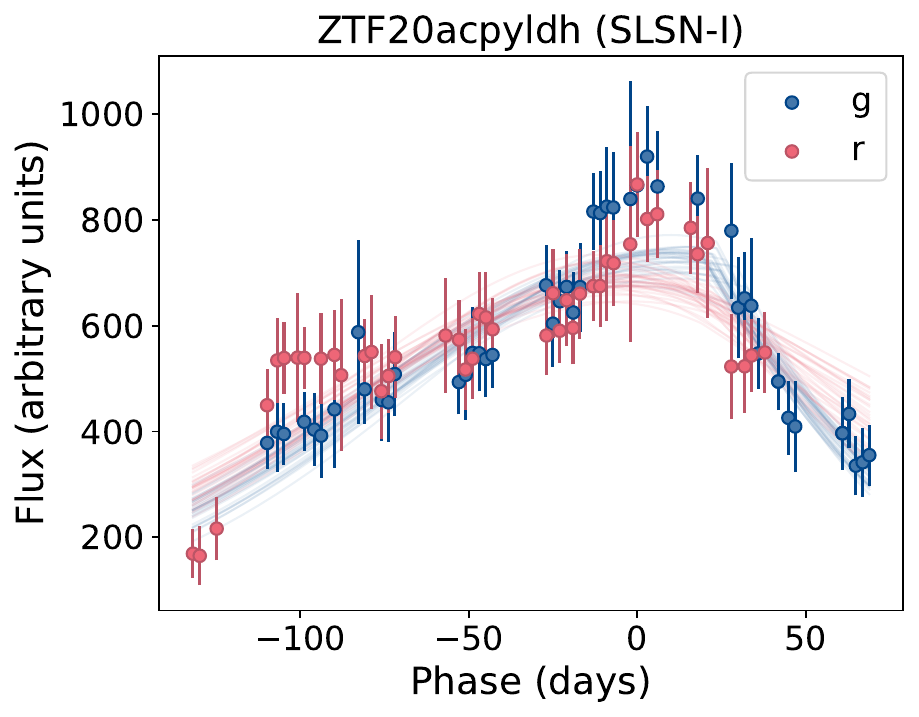}{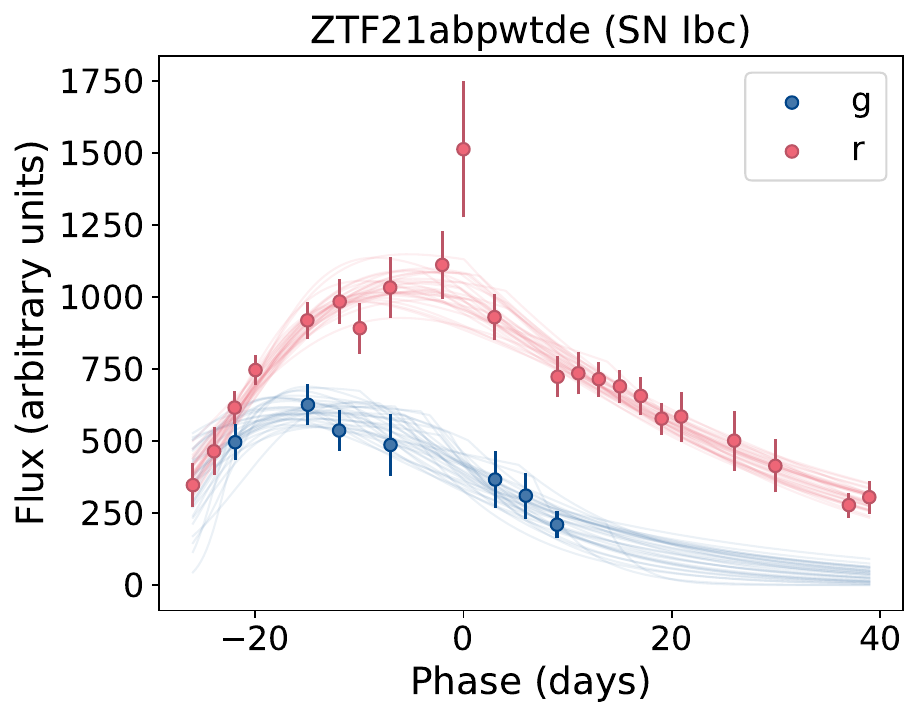}
    
    \caption{Representative model fits for six classified SN light curves. We show two SNe Ia (top row), a SN II (middle left), a SN IIn (middle right), a SLSN-I (bottom left), and a SN Ib/c (bottom right). Thirty posterior draws are shown for each light curve, capturing the uncertainties and scatter in our model fits.}
    \label{fig:ex_lc}
\end{figure*}

\begin{figure}[t] 
    \plotone{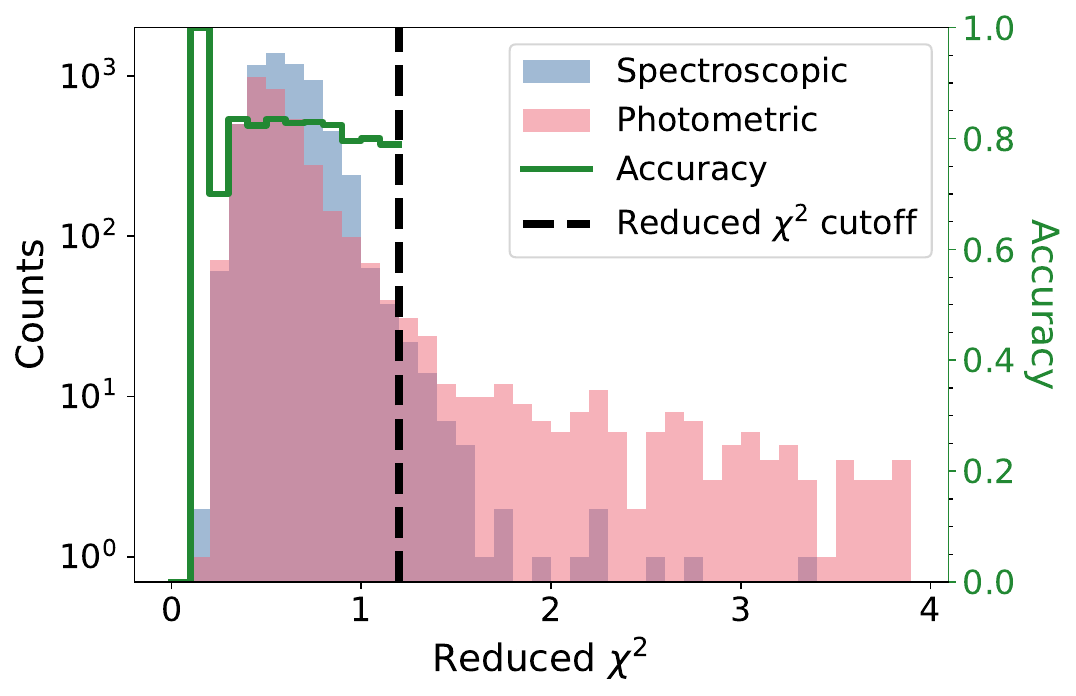}
    \caption{The distributions of reduced chi-squared values from the light curve model fits for the spectroscopically and photometrically classified samples. Because a very small fraction of the spectroscopic set has a reduced chi-squared above 1.2, we can reasonably assume that most light curves with values above 1.2 are either not SNe or SNe of much lower data quality. We therefore exclude them from our final datasets. The right axis shows classification accuracy as a function of reduced chi-squared for the pruned spectroscopic set, which further supports our choice of cutoff.}
    \label{fig:chisq_cutoff}
\end{figure}

\begin{deluxetable*}{cccccc}
\tablecaption{Model Fit Priors}
    \tablehead{\colhead{Parameter} & \colhead{Distribution} & \colhead{Mean} & \colhead{Standard Deviation} & \colhead{Truncated Min} & \colhead{Truncated Max}} 
    \startdata
        $A$ & Log-Gaussian & \priorLogA \\
        $\beta$ & Gaussian & \priorBeta \\
        $\gamma$ & Log-Gaussian & \priorGamma \\
        $t_0$ & Gaussian & \priorTnaught \\
        $\tau_\mathrm{rise}$ & Log-Gaussian & \priorTauRise \\
        $\tau_\mathrm{fall}$ & Log-Gaussian & \priorTauFall \\
        $\sigma_{\mathrm{extra}}$ & Log-Gaussian & \priorExtraSigma \\
        $A_g$ & Log-Gaussian & \priorAg \\
        $\beta_g$ & Log-Gaussian & \priorBetag \\
        $\gamma_g$ & Log-Gaussian & \priorGammag \\
        $t_{0,g} - 1$ & Log-Gaussian & \priorTnaughtg \\
        $\tau_{\mathrm{rise},g}$ & Log-Gaussian & \priorTauRiseg \\
        $\tau_{\mathrm{fall},g}$ & Log-Gaussian & \priorTauFallg \\
        $\sigma_{\mathrm{extra}, g}$ & Log-Gaussian & \priorExtraSigmag
    \enddata
    \tablecomments{The prior distributions for each fit parameter, which are sampled to explore the posterior probability space during nested sampling. For the log-Gaussian distributions, the provided mean, standard deviation, and truncated limits are of the underlying Gaussian distribution before exponentiation. \label{table:priors}}
\end{deluxetable*}

\input{sec3_fitting}

\input{sec4_classifier}

\begin{figure}[t] 
    \plotone{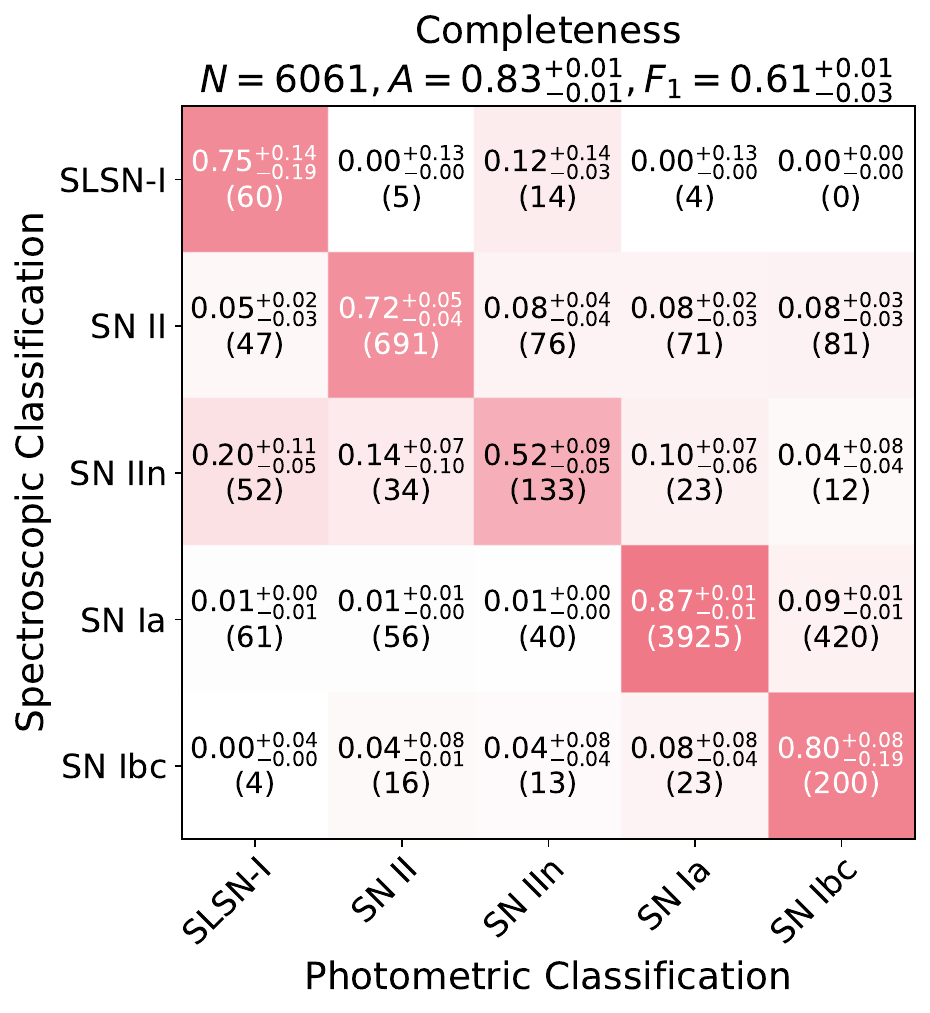}
    \plotone{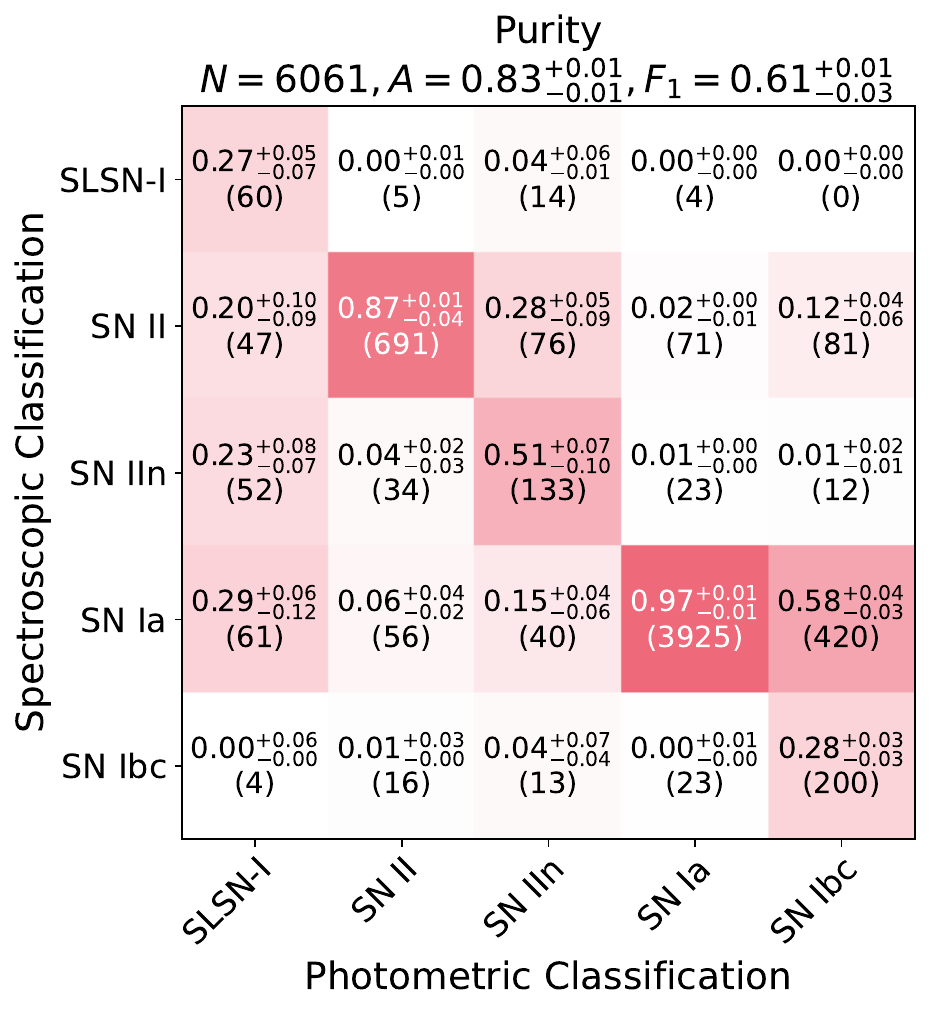}
    \caption{The confusion matrices of the trained classifier on the training set, using $K$-fold cross-validation. Here, we set $K=10$ folds. The completeness matrix is normalized to sum to one across each true class, while the purity matrix is normalized along each predicted class. Our classifier achieves high completeness across the five classes. Note that the most populous classes leak into the rarer predicted sample sets, reducing the associated purities.}
    \label{fig:no_redshift_cm}
\end{figure}

\begin{figure*}[t] 
    \plottwo{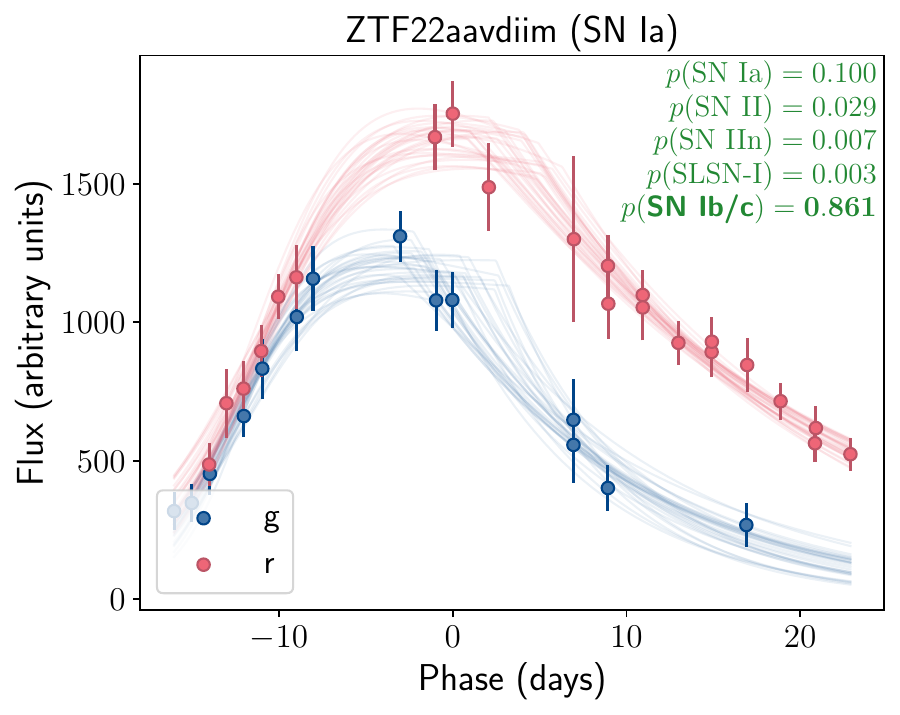}{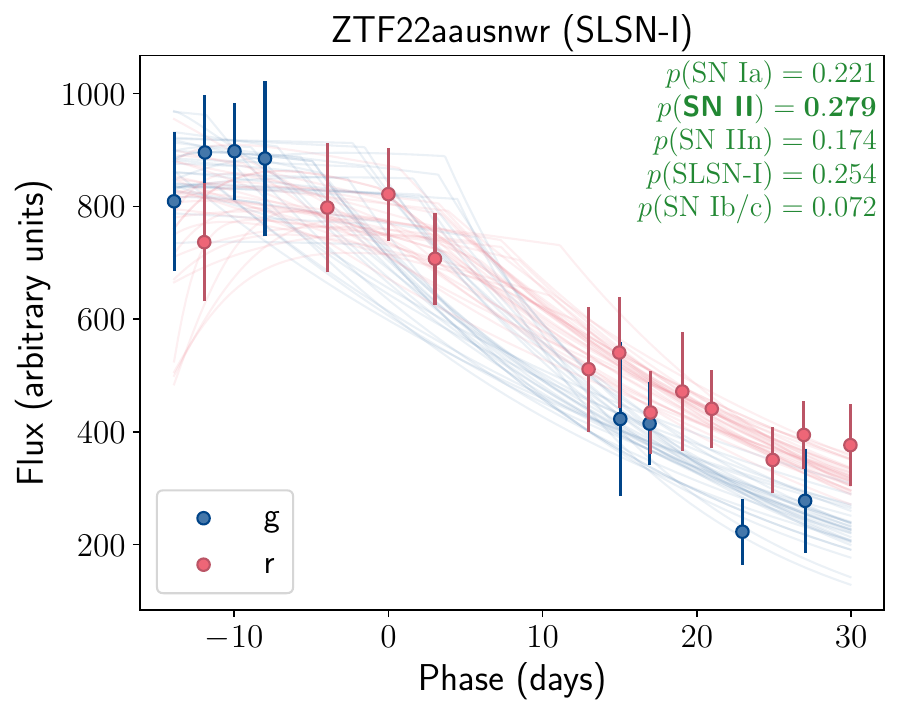}
    \plottwo{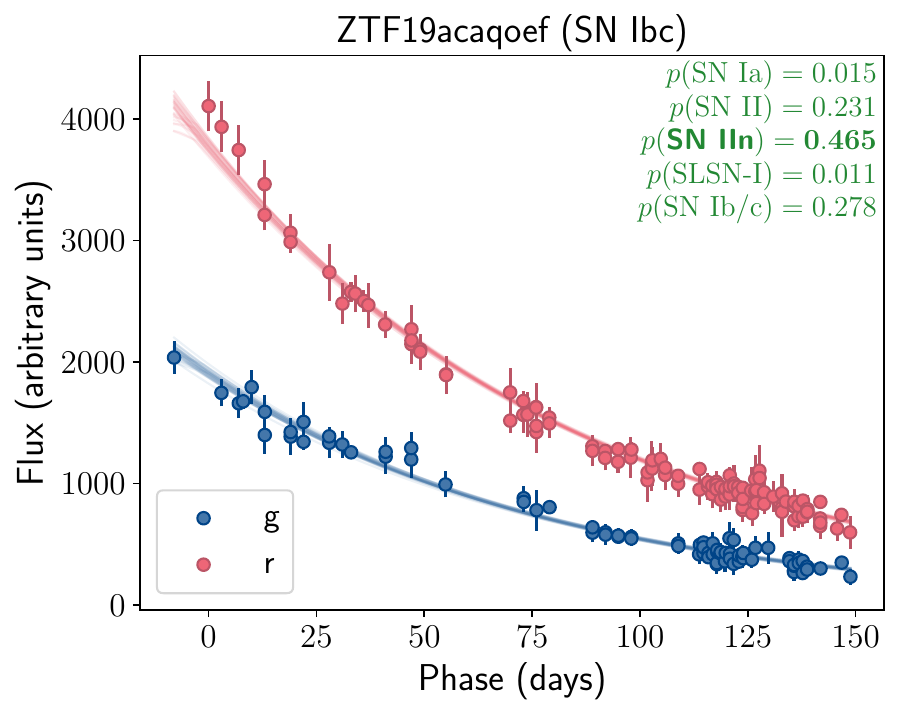}{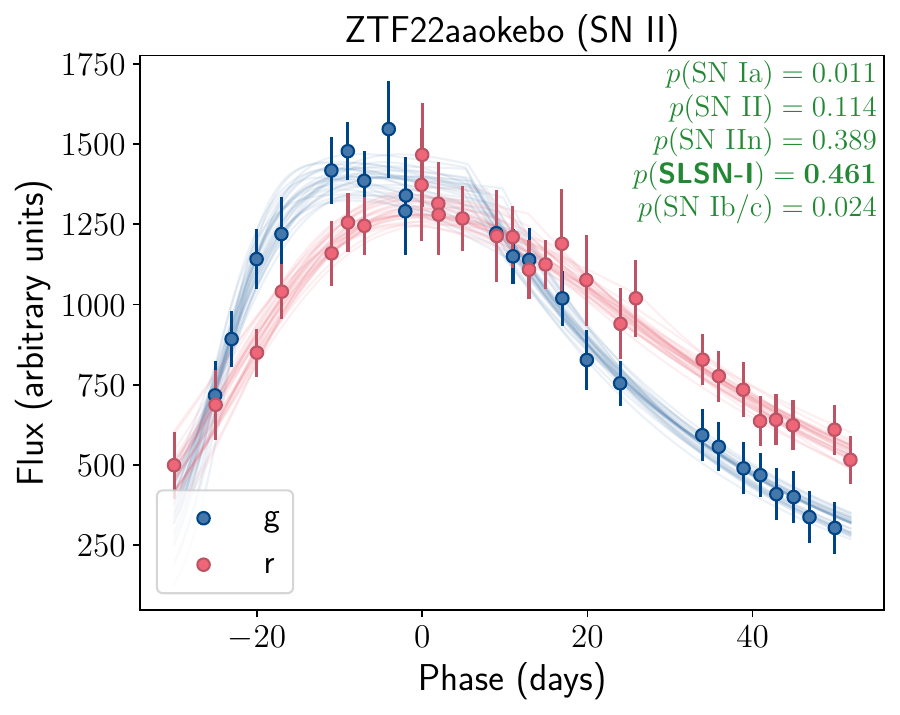}
    \caption{Examples of light curves that are misclassified by Superphot+, where the spectroscopic labels are specified in the titles and the probabilities output by Superphot+ are in green. Probable causes of misclassification include redder color (top left), unsampled regions (top right), and exceptionally long evolution timescales (bottom row).}
    \label{fig:misclassified_lc}
\end{figure*}

\begin{figure}[t] 
    \plotone{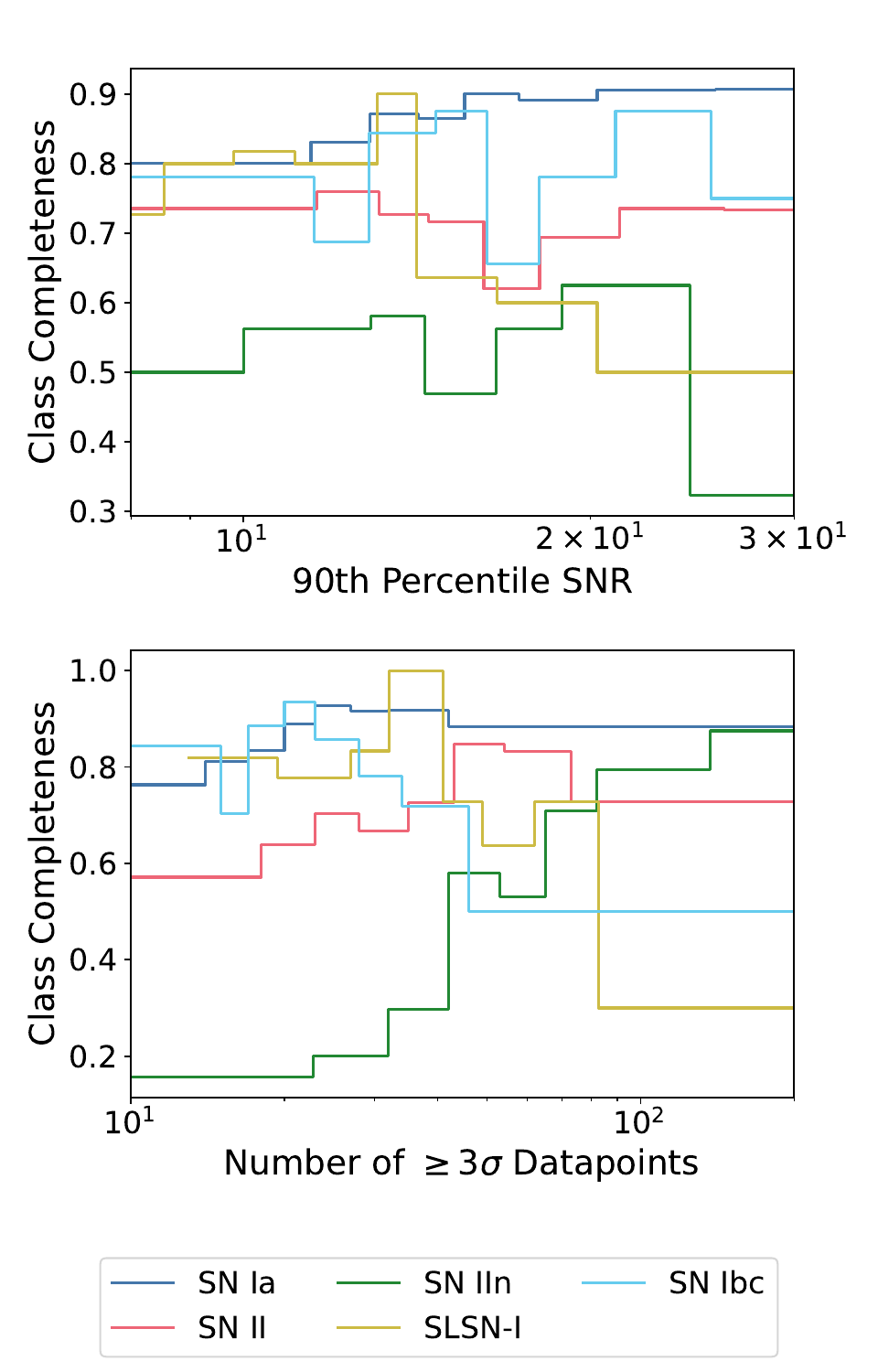}
    \caption{The spectroscopic set's binned classification accuracies as a function of 90th percentile signal-to-noise ratio (top) and number of datapoints with SNR $\geq$ 3 (bottom). The bin widths are chosen to contain an equal number of events per bin within each class. There is a weak increase in accuracy as a function of 90th-percentile SNR that plateaus around 20. The exception is SLSN-I, whose highest SNR light curves are longer-lived and therefore misclassified as SNe II or SNe IIn. For SNe II and SNe IIn, more datapoints correlate with higher classification accuracies, likely due to better sampled plateaus which are make them easier to classify. SLSNe-I and SNe Ib/c both show decreased classifier performance with excessive number of datapoints.}
    \label{fig:snr_vs_accuracy}
\end{figure}

\begin{figure}[t] 
    \plotone{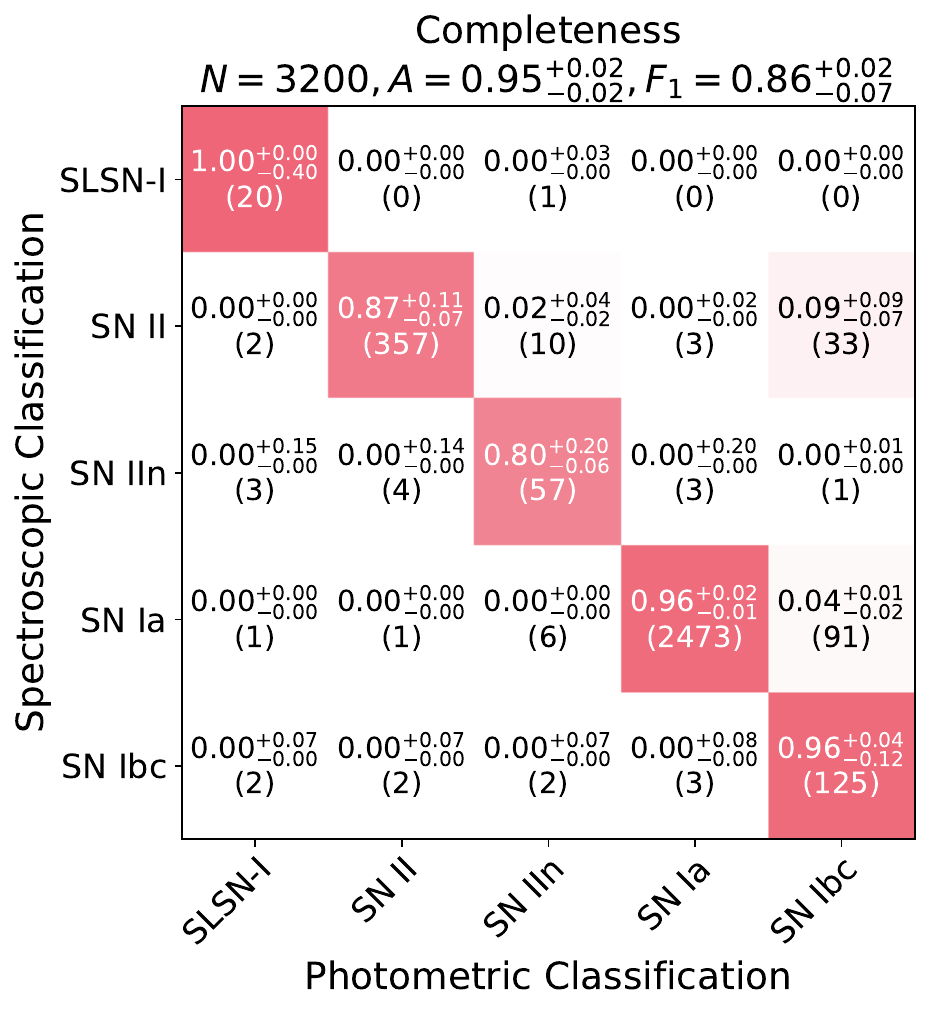}
    \plotone{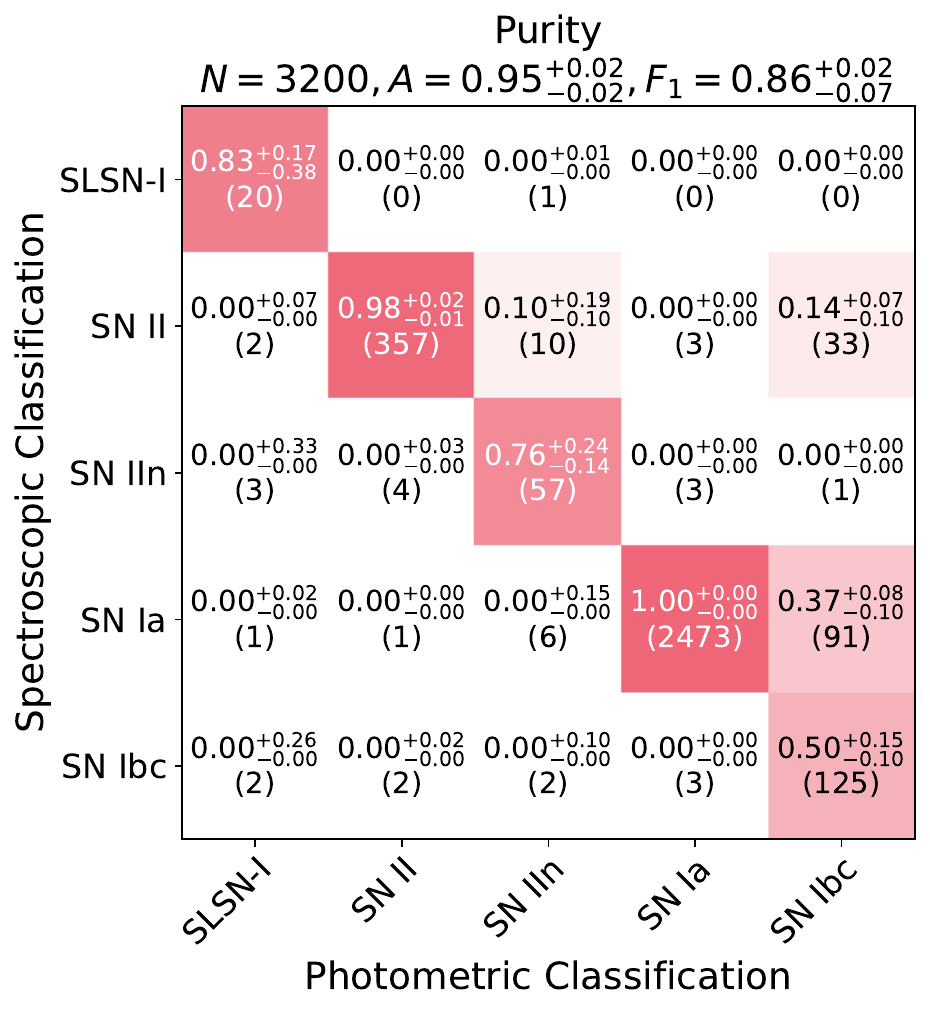}
    \caption{Same as Figure \ref{fig:no_redshift_cm}, but only including light curves classified with confidence greater than or equal to 0.7. This includes \numHCNoRedshift of the \numSNSpec total events, and both the completeness and purity confusion matrices show significant improvement. The F$_1$-score increases from \FNoRedshift\ to \FNoRedshiftHC, and accuracy increases from \accNoRedshift\ to \accNoRedshiftHC. These plots especially highlight the difficulty with discerning SNe IIn from SLSNe-I and SNe II without redshift information (much fewer events included), and in maintaining a high-purity set of Type Ib/c SNe with our imbalanced dataset.}
    \label{fig:no_redshift_p07}
\end{figure}

\begin{figure}[t] 
    \plotone{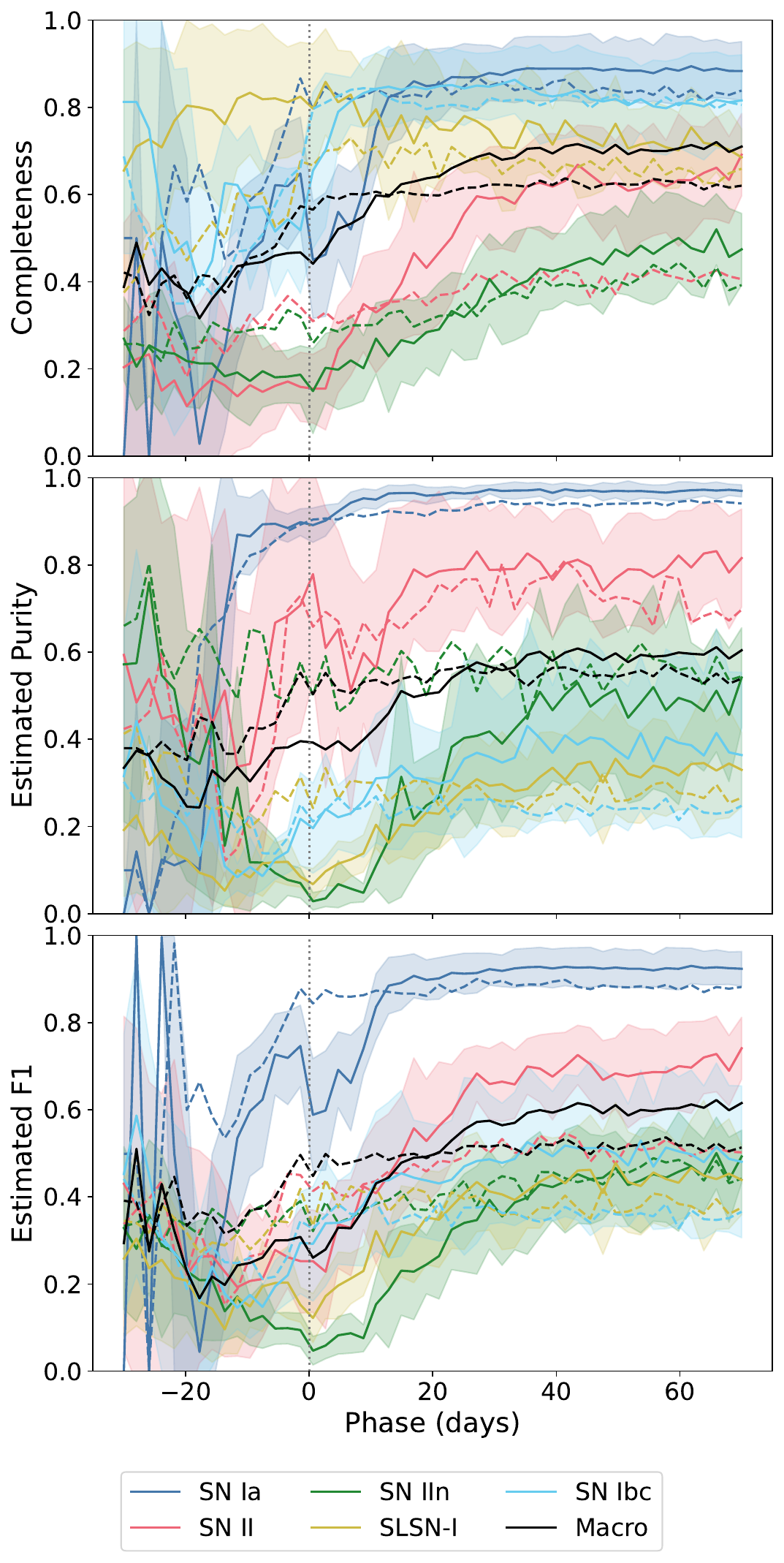}
    \caption{Classification completeness, purity, and F$_1$-score as a function of light curve phase for each SN type. Our early-phase classifier is shown by the dashed lines, whereas the full-phase classifier is shown by the solid lines and associated $1\sigma$ $K$-fold uncertainty margins. Here, phase is relative to peak magnitude in $r$-band. The purities and F$_1$-scores are estimated from class prevalences and completeness values. Twenty light curves of each type per fold are randomly sampled to generate this data, so final accuracies differ slightly from those of the entire spectroscopic set. While SNe Ib/c classification accuracy improves significantly a few days before peak magnitude, light curves are only consistently labeled as SNe II or SNe IIn $\simeq$15 days after peak, when the post-peak behavior can be adequately measured. The early-phase classifier outperforms the full-phase classifier before phase $\sim$20 days.}
    \label{fig:phase_vs_acc}
\end{figure}

\begin{figure}[t] 
    \plotone{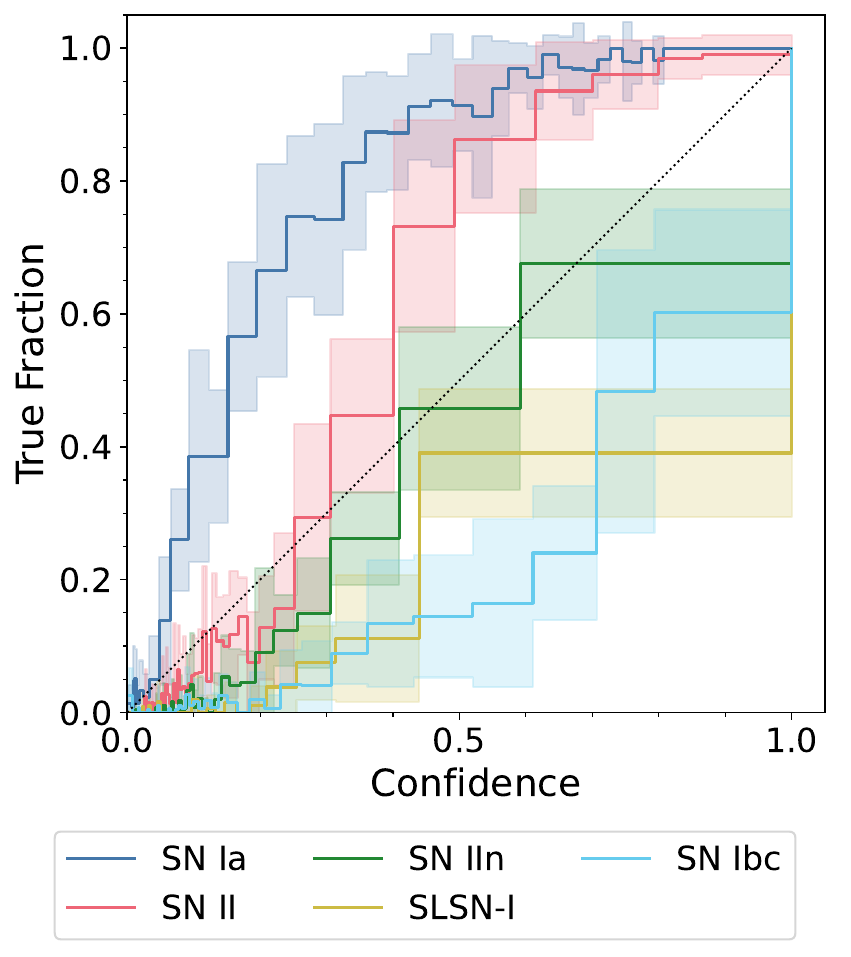}
    \caption{The calibration curve for each SN class, treating Superphot+ as a binary classifier. A calibration curve plots the fraction of events correctly classified at different confidence levels. A well-calibrated classifier would follow the $y=x$ diagonal for each class. We see that Superphot+ is overconfident about SN Ib/c probabilities and underconfident about SN Ia probabilities, while fairly well-calibrated for all other classes.}
    \label{fig:calibration}
\end{figure}

\begin{figure}[t] 
    \plotone{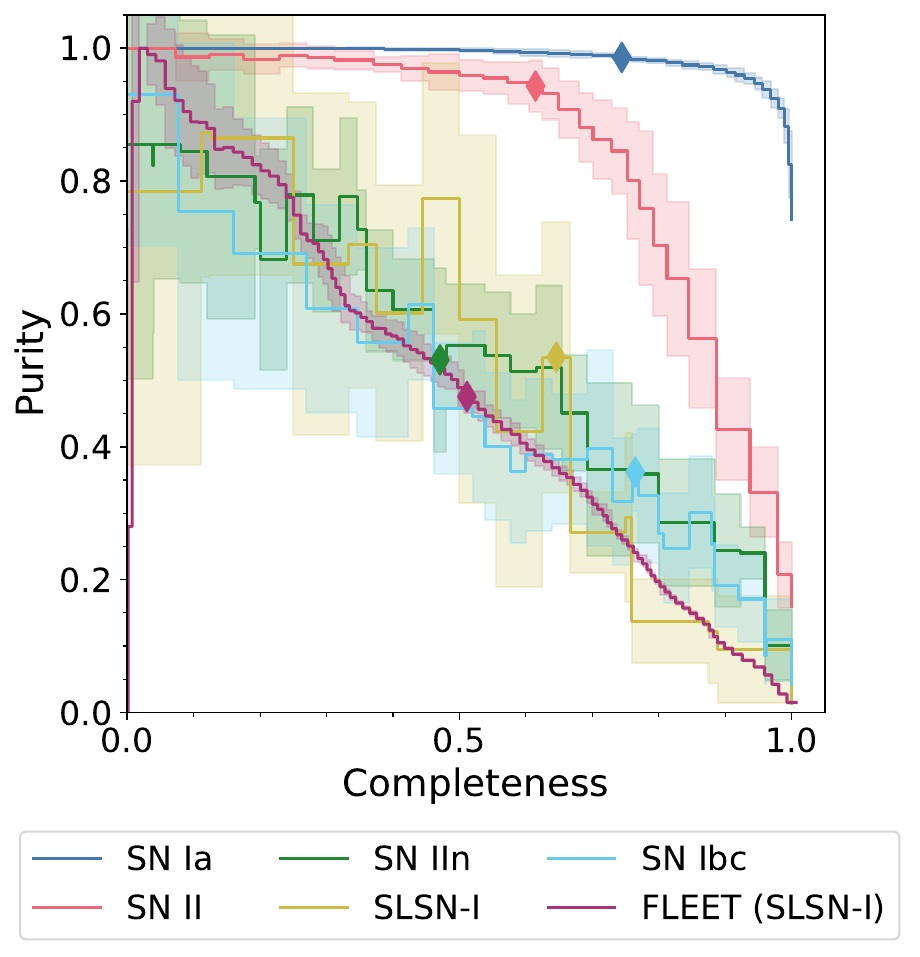}
    \caption{The purity-completeness curve for each SN class, treating Superphot+ as a binary classifier. We calculate single-class metrics by compressing all probabilities from classes other than the target class into a single probability. Each point along the curve corresponds to a different confidence threshold for classification. A ``perfect" classifier (i.e., one with 100\% confidence and accuracy) would follow the top-right corner of the axes, and a completely random classifier follows a horizontal line scaled to the target class's prevalence.}
    \label{fig:pr_curve}
\end{figure}

\begin{figure}[t]
    \plotone{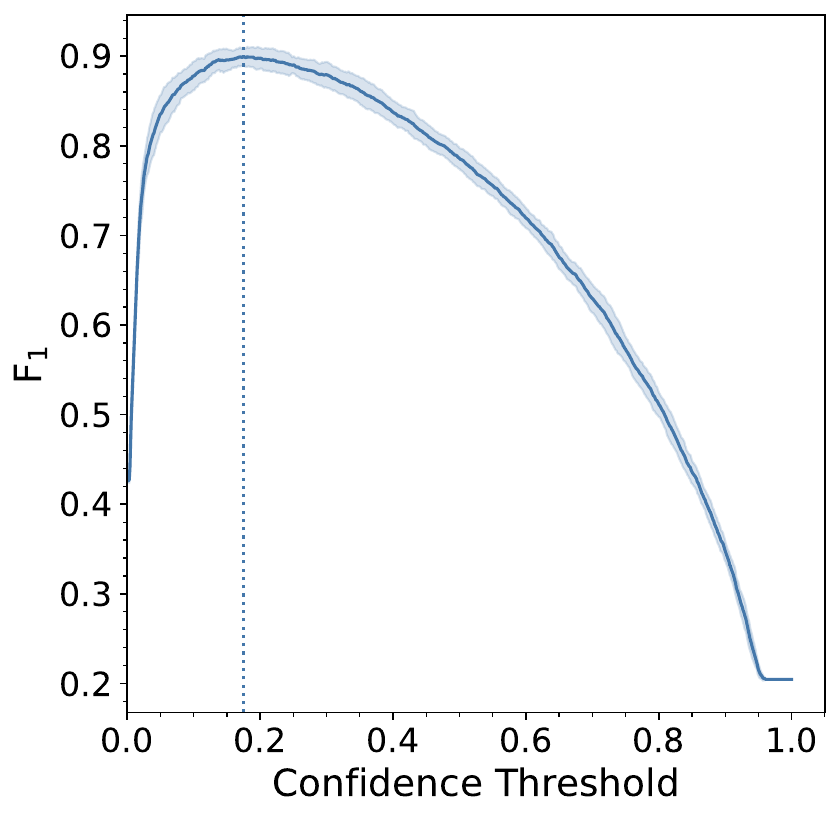} 
    \plotone{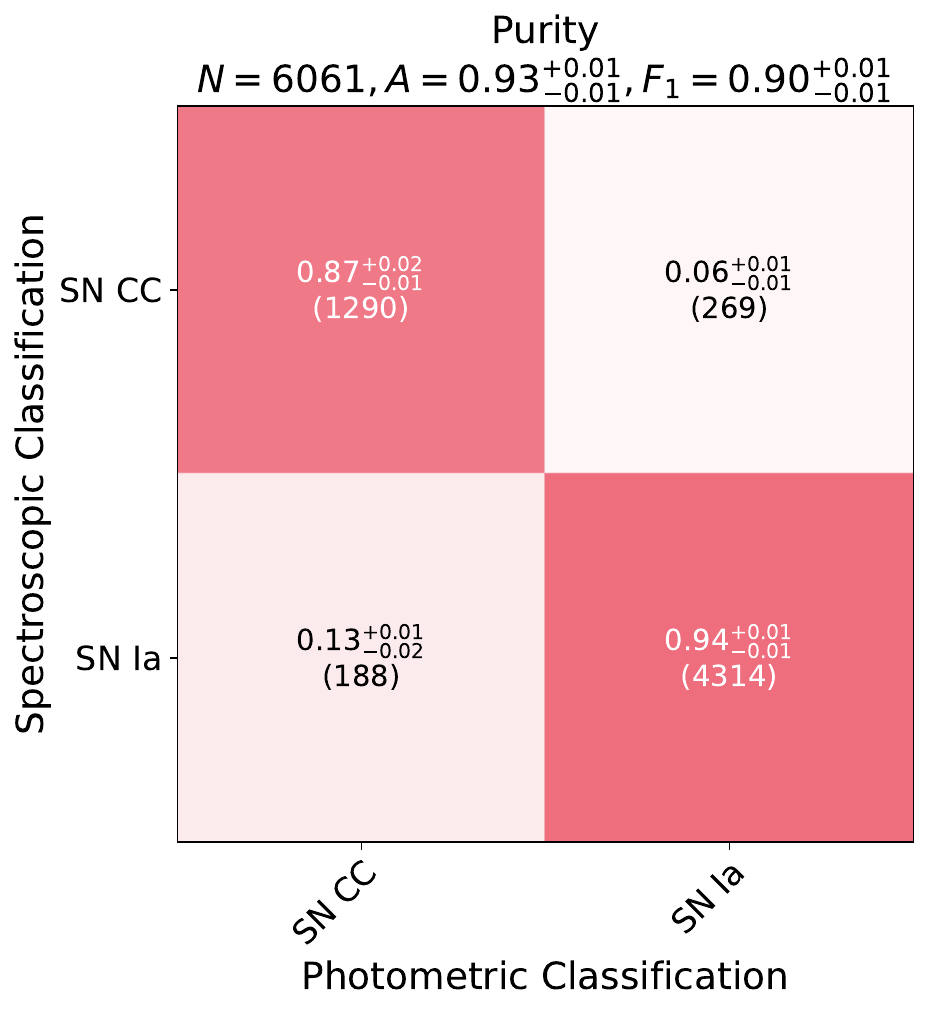}
    \caption{Top: The binary F$_1$-score of our SN Ia versus core-collapse SN classifier as a function of confidence threshold. Higher thresholds mean the classifier must return a higher pseudo-probability for SN Ia for us to assign the object a SN Ia photometric label. We find that F$_1$ is optimized for a SN Ia threshold of $p\geq0.142$. Bottom: The corresponding purity matrix when we use the optimal confidence threshold. We see very successful differentiation between SNe Ia and CC SNe, with an F$_1$-score of \FBinary\ and a SN Ia purity of \binaryPurityIa.}
    \label{fig:f1_vs_thresh}
\end{figure}

\begin{figure}[t]
    \plotone{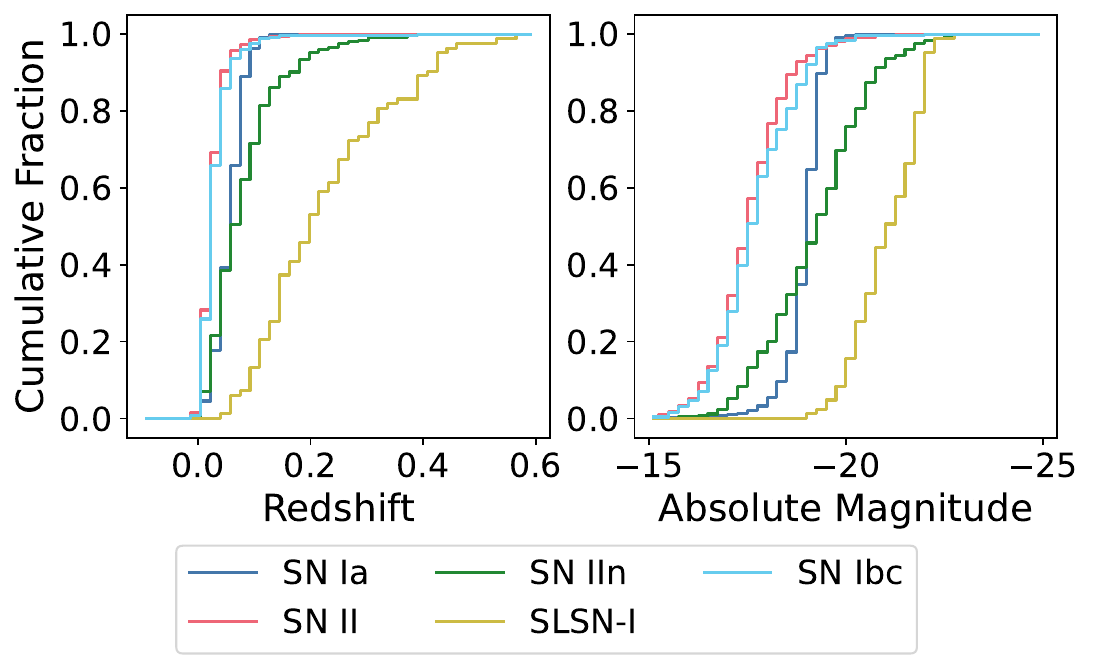} 
    \caption{Cumulative distributions of the redshifts and corresponding absolute peak $r$-band magnitudes of our spectroscopic subset with redshift information. As expected, some SLSNe-I have brighter absolute magnitudes and therefore can be detected at farther redshifts. The bulk of our dataset is low-redshift light curves ($z<0.2$) with absolute magnitudes between -16 and -20.}
    \label{fig:abs_mags}
\end{figure}

\begin{figure}[t]
    \plotone{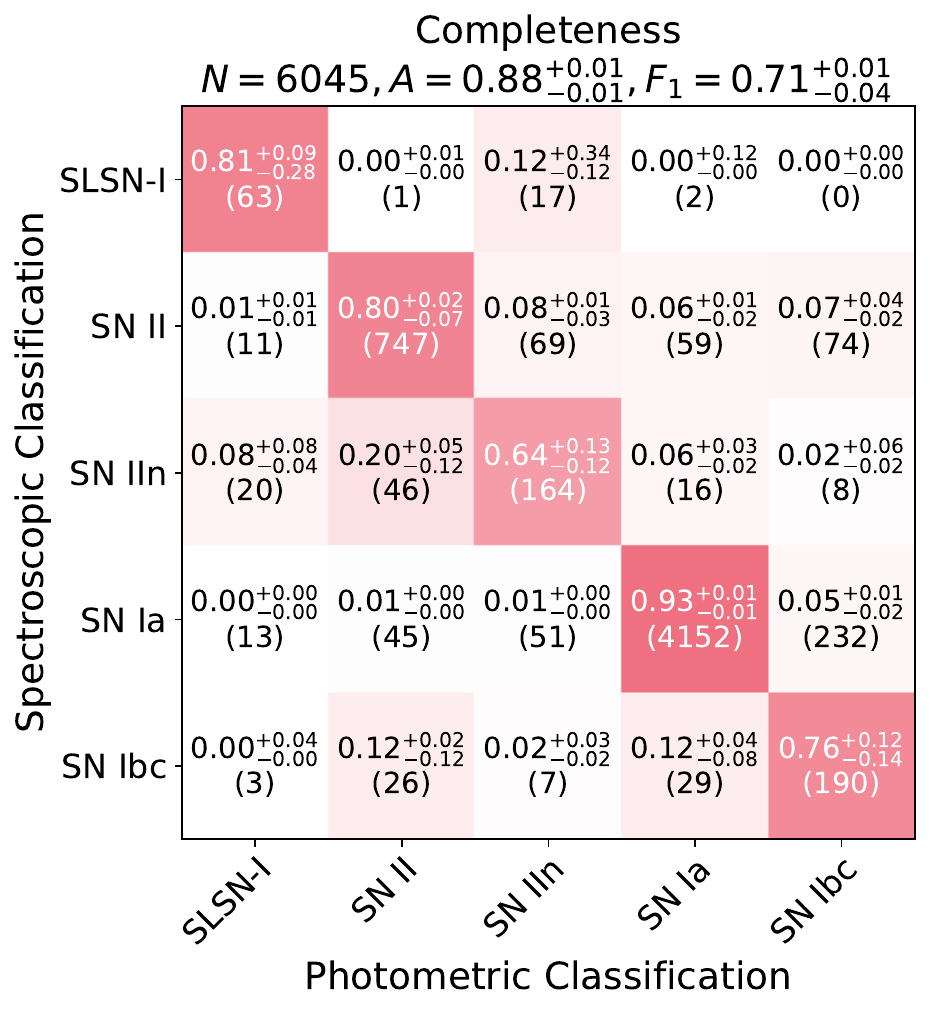} 
    \plotone{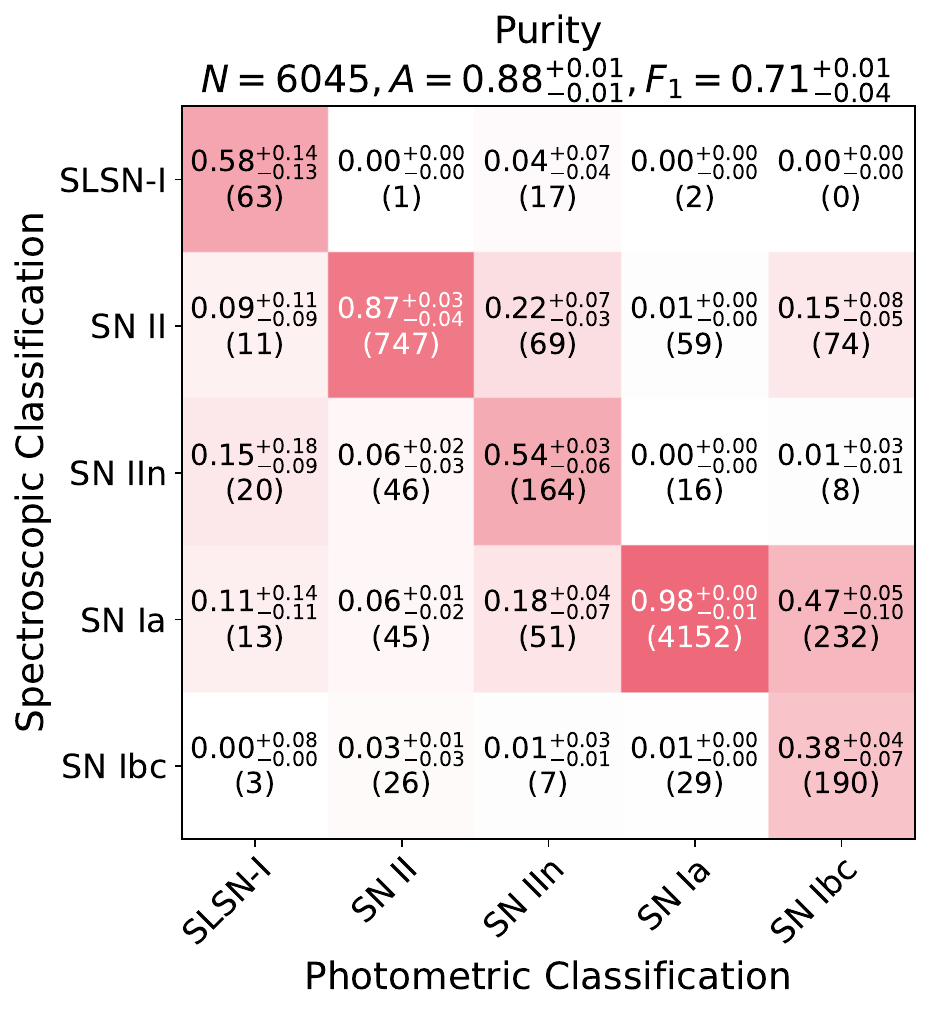}
    \caption{The completeness and purity confusion matrices for the redshift-inclusive classifier. Adding redshift information improves the F$_1$-score from \FNoRedshift\ to \FRedshift. There is significant improvement in SLSN-I, SN II, and SN IIn classification, which demonstrates the importance of luminosity in distinguishing between light curves of these classes
    }
    \label{fig:redshift_cm}
\end{figure}

\begin{figure}[t]
    \plotone{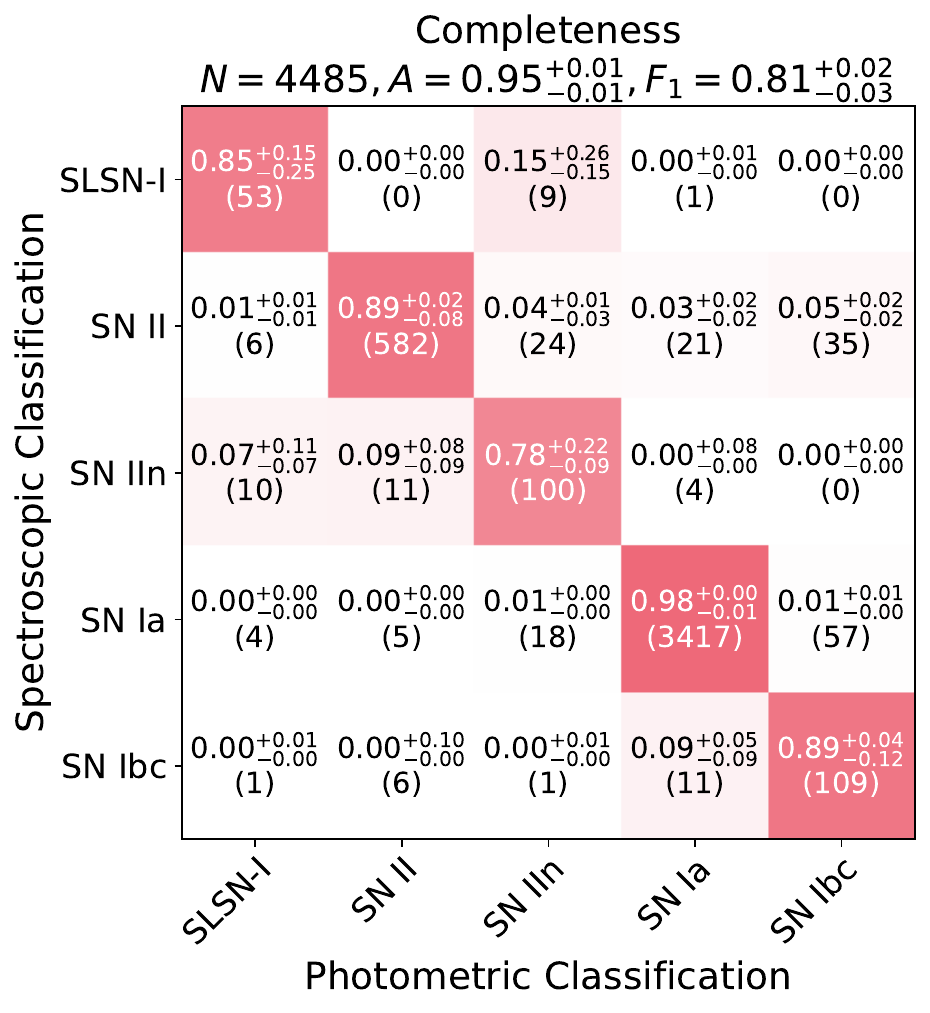} 
    \plotone{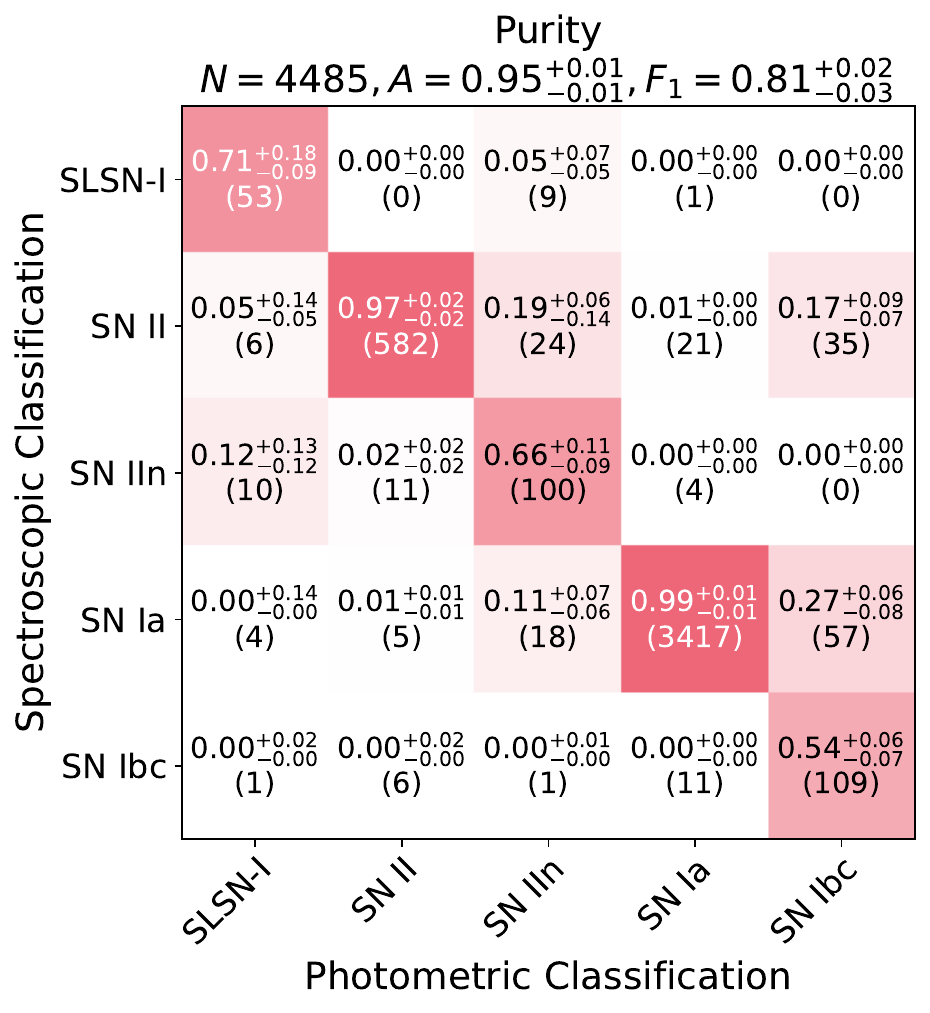}
    \caption{Same as Figure \ref{fig:redshift_cm}, but only including light curves classified with confidence greater than or equal to 0.7. While high-confidence performance is worse than when excluding redshift for the same cutoff, a much higher fraction (\numWRedshiftsHC\ out of \numWRedshifts) of light curves are classified with high confidence compared to the redshift-exclusive classifier. This highlights the potential for inaccurate but confident classifications from redshift biases.}
    \label{fig:redshift_p07}
\end{figure}

\input{sec5_results}

\begin{figure*}[t] 
    \plottwo{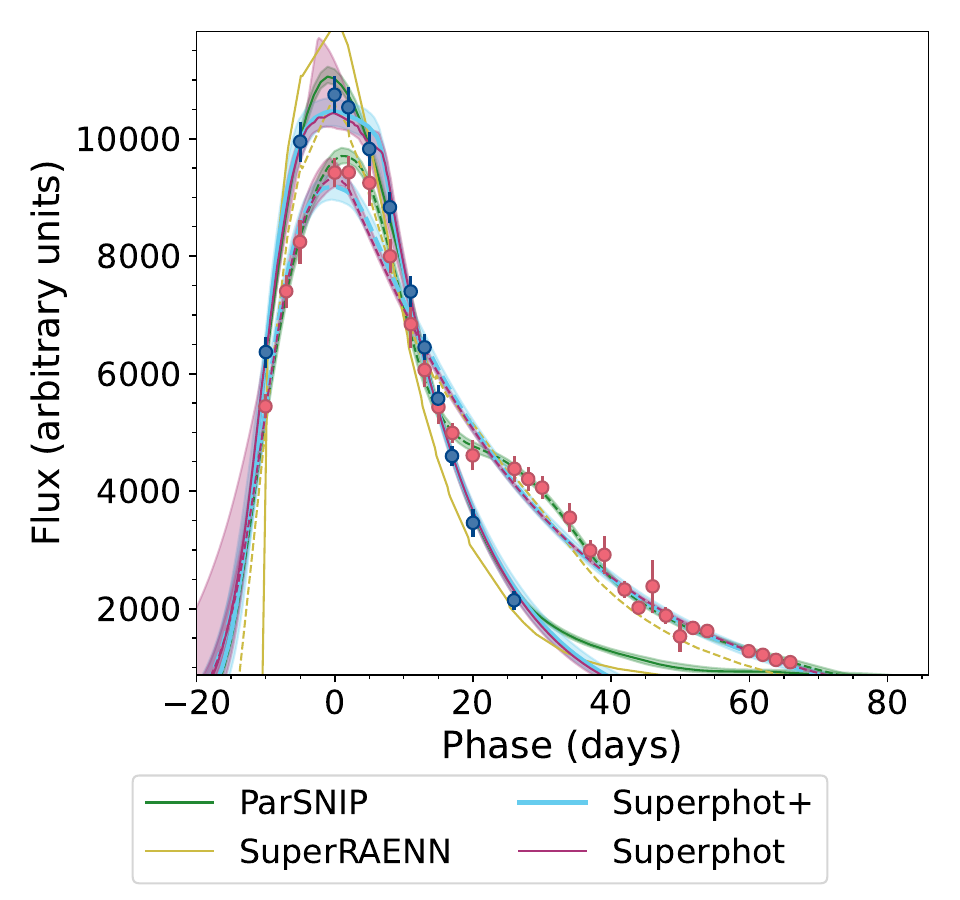}{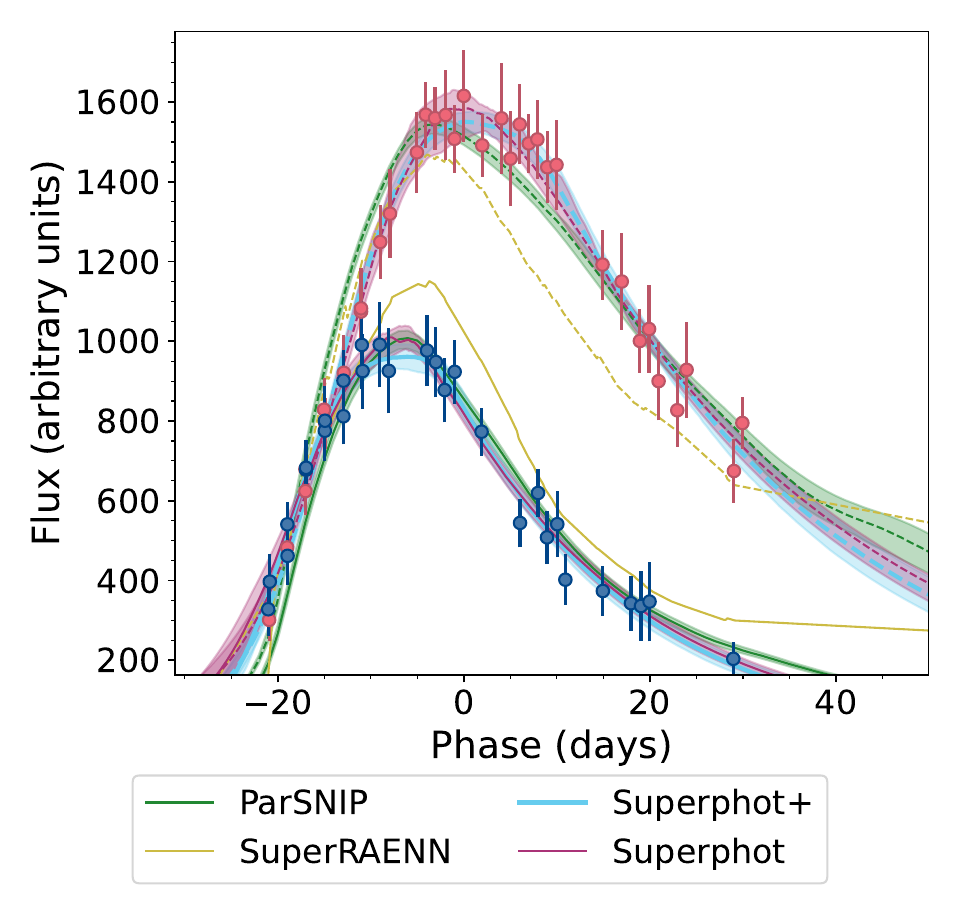}
    \plottwo{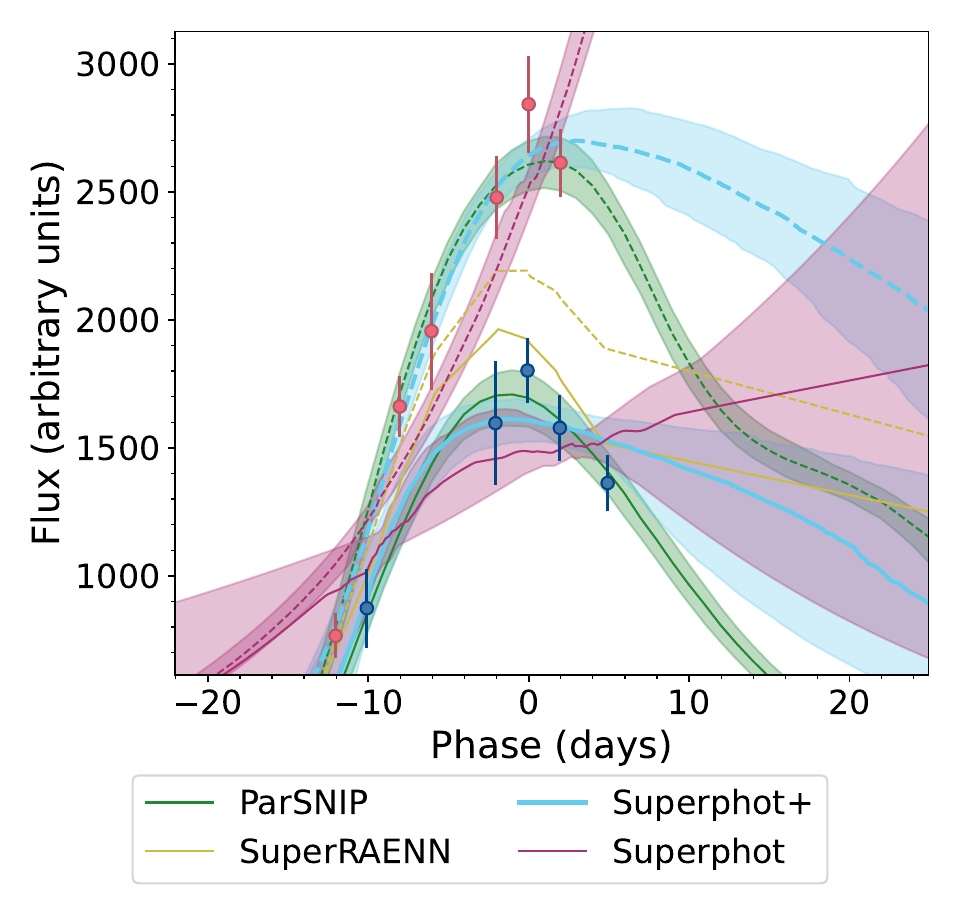}{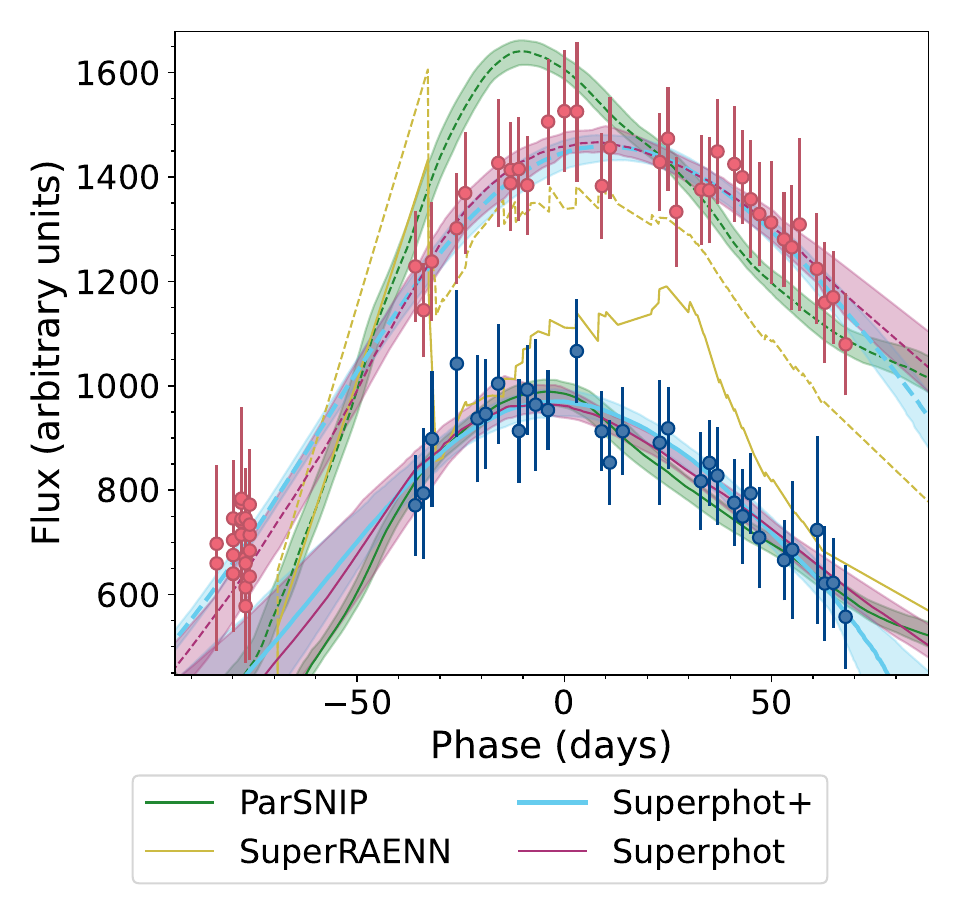}
    \caption{Three SN Ia and one SN IIn light curve, in that order, fitted with each of the four pipelines compared in Table~\ref{table:binary_comparison}. The top left light curve is fitted fairly well by Superphot+, SuperRAENN, and Superphot, but ParSNIP is the only pipeline that adequately captures the SN Ia's secondary $r$-band bump. All pipelines fit the top right light curve adequately. The bottom left light curve is fit reasonably by SuperRAENN, ParSNIP and Superphot+, but Superphot suffers from its too broad priors. The bottom right (SN IIn) light curve has an exceptionally slow rise timescale and suffers from uneven rise sampling; the parametric classifiers Superphot+ and Superphot recreate this light curve best.}
    \label{fig:classifier_compare_lc}
\end{figure*}

\begin{deluxetable*}{cccccc} 
    \tablecaption{Comparison to Other Classifiers}
    \tablehead{\colhead{Classifier} & \colhead{Multi-Class Accuracy} & \colhead{Multi-Class F$_1$} & \colhead{Binary Accuracy} & \colhead{Binary F$_1$} & \colhead{Citation}} 
    \startdata
        Superphot+ (no $z$) & \accNoRedshift & \FNoRedshift & \accBinary & \FBinary & This Work \\
        Superphot+ (with $z$) & \accRedshift & \FRedshift &  \accBinaryRedshift & \FBinaryRedshift & This Work \\
        Superphot & \accSuperphot & \FSuperphot & \accSuperphotBin & \FSuperphotBin & \cite{superphot} \\
        SuperRAENN & \accSuperraenn & \FSuperraenn & \accSuperraennBin & \FSuperraennBin & \cite{superraenn} \\
        SuperRAENN (only RAENN) & \accRaenn & \FRaenn & \accRaennBin & \FRaennBin & \cite{superraenn} \\
        ParSNIP & \accParsnip & \FParsnip & \accParsnipBin & \FParsnipBin & \cite{boone_2021} \\
        Only $z$, $M_{peak}$, and $A_g$ & \accAbsMag & \FabsMag & \accAbsMagBin & \FabsMagBin & This Work
    \enddata
    \tablecomments{Comparison of Superphot+'s multi-class and SN Ia binary classification metrics (with and without redshifts) with those from training identical LightGBMs on feature sets from other SN pipelines that use redshift information. All classifiers are retrained on identical ZTF datasets. The ALeRCE pipeline does not use redshift information and is therefore not included in this table. \label{table:binary_comparison} }
\end{deluxetable*}

\begin{figure*}[t] 
    \plottwo{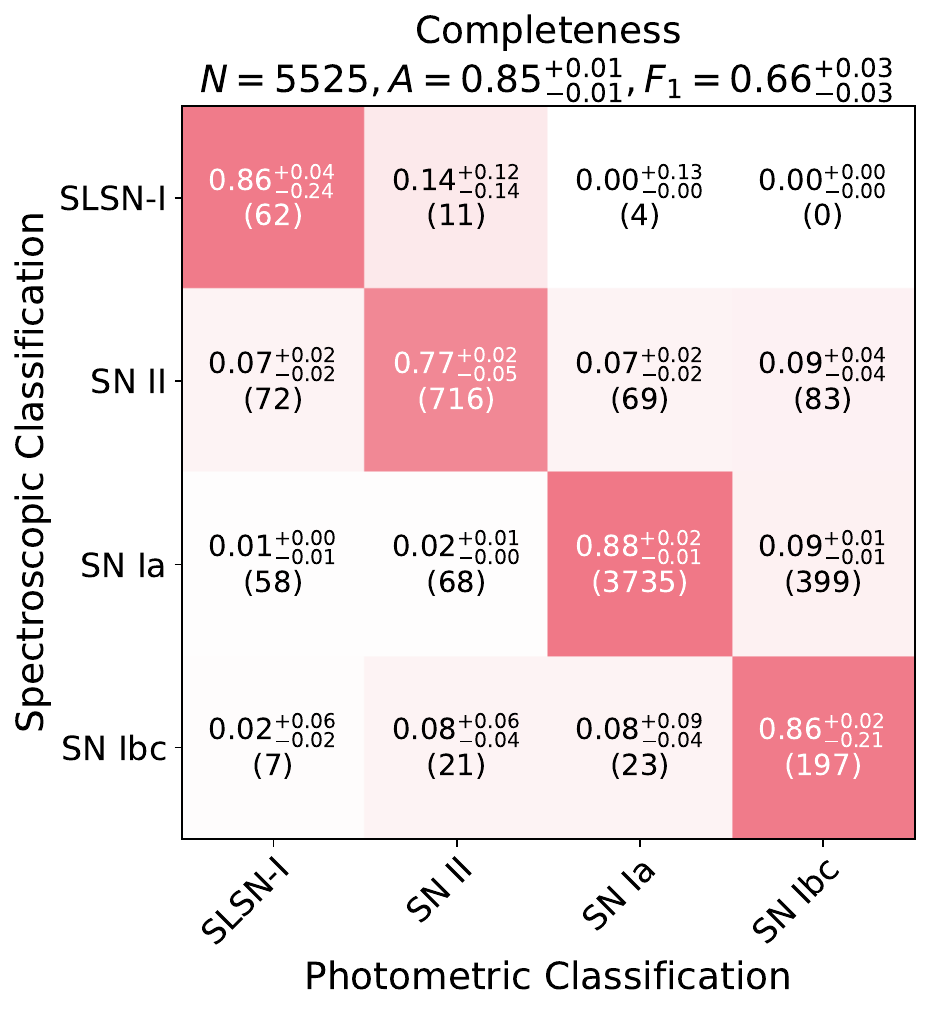}{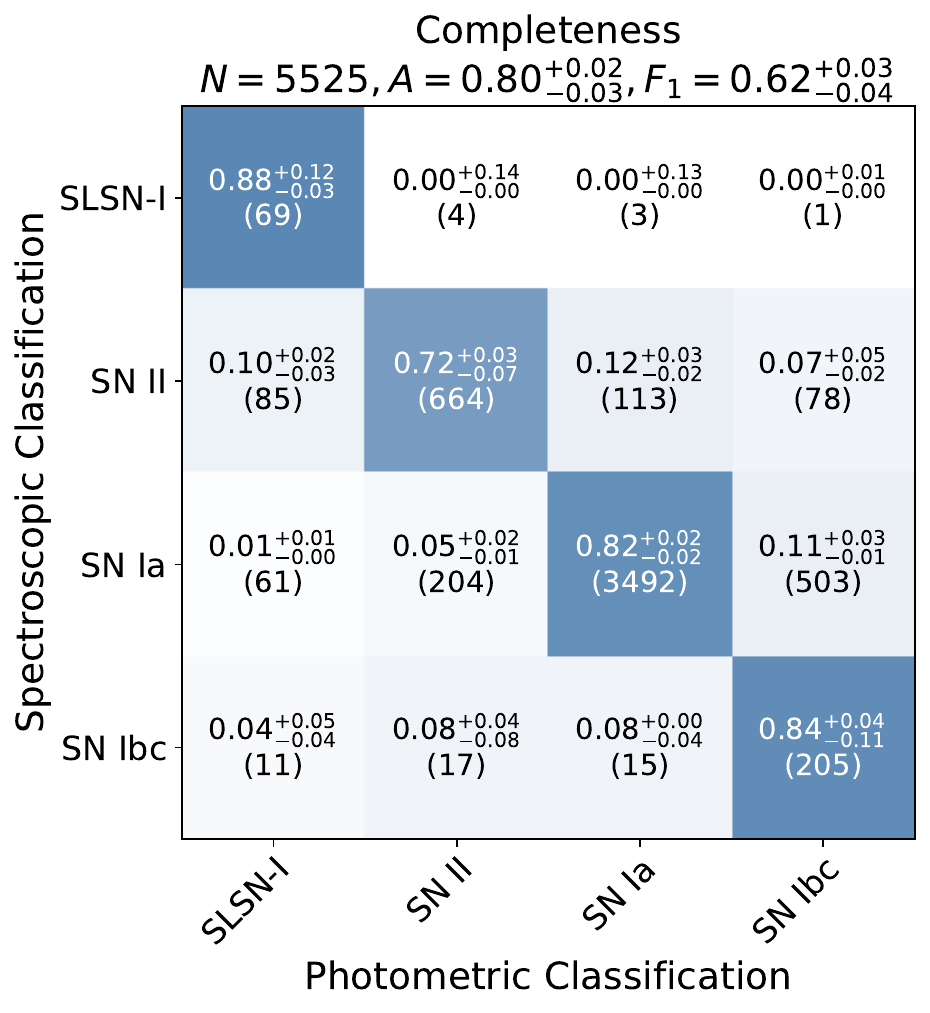} 
    \plottwo{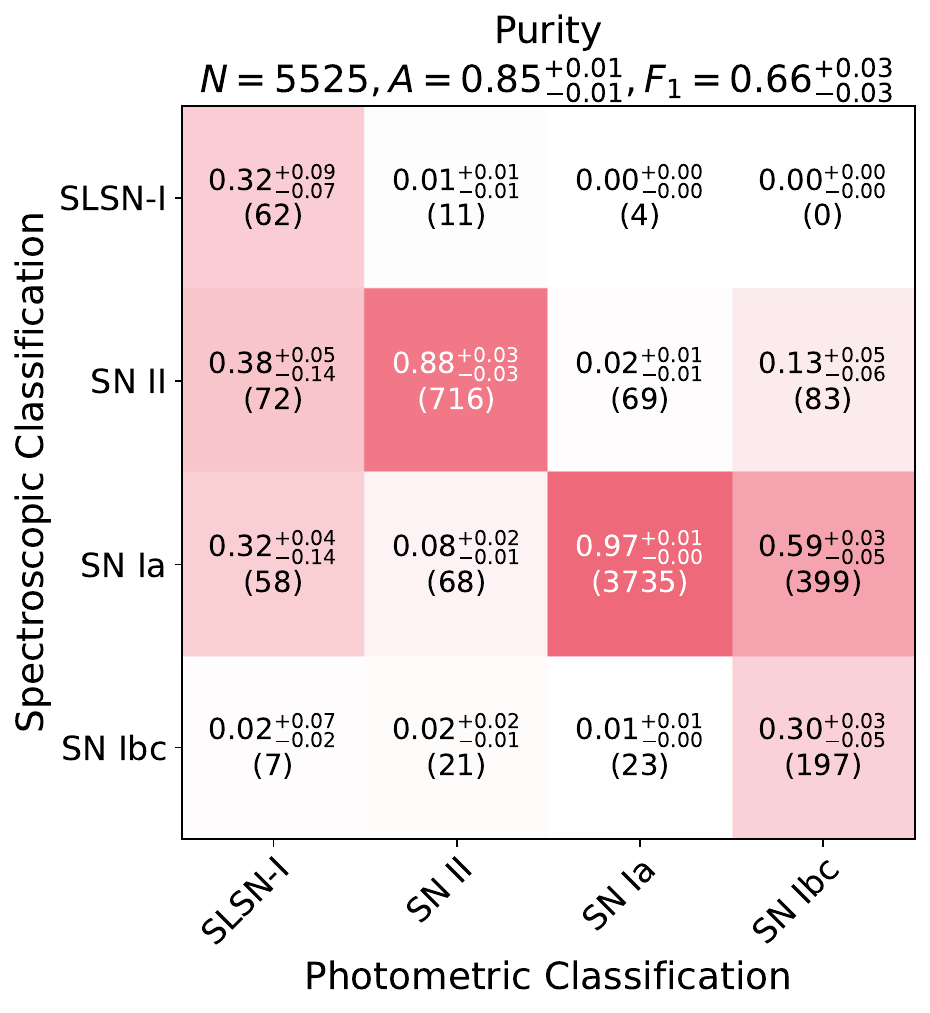}{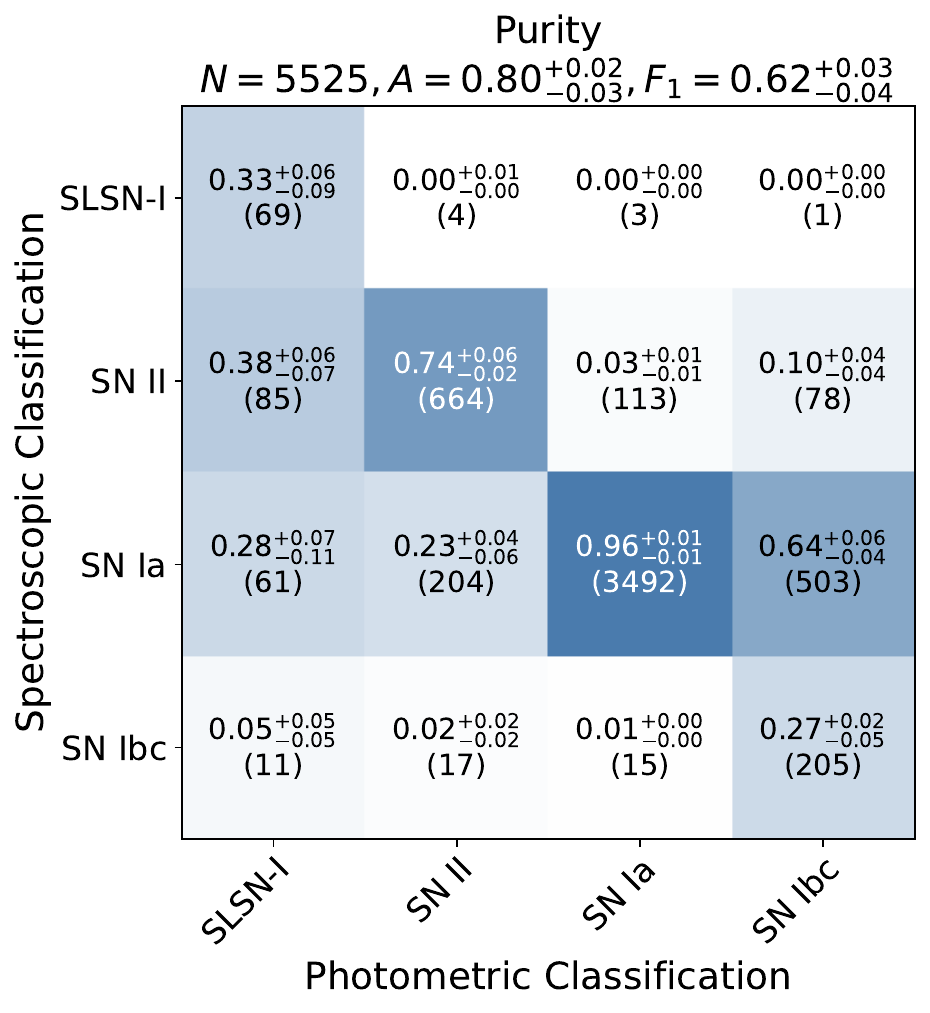}
    \caption{The four-class confusion matrices for Superphot+ (left) versus ALeRCE-SN (right). The Superphot+ confusion matrices are condensed into four classes by combining the second-highest probability label for objects predicted to be SNe IIn, and excluding samples with SN IIn true labels. This is done because ALeRCE-SN does not include a SN IIn label. Even with this regrouping, the Superphot+ F$_1$-score is higher than ALeRCE-SN's.}
    \label{fig:four_type}
\end{figure*}

\input{sec7_alerce}

\begin{deluxetable*}{ccccccccc} 
\tablecaption{Superphot+ Probabilities Assigned to the Photometric Dataset}
    \tablehead{
    \colhead{ZTF Name} &
    \colhead{IAU Name} &
    \colhead{Fit Reduced $\chi^2$} &
    \colhead{ALeRCE-SN Label} &
    \colhead{$p($SN Ia$)$} & \colhead{$p($SN II$)$} & \colhead{$p($SN IIn$)$} & \colhead{$p($SLSN-I$)$} & \colhead{$p($SN Ib/c$)$}
    }
    
    \startdata
        ZTF18aaanaev & 2022wkv & 0.607 & SN Ibc & 0.226 & 0.161 & 0.104 & 0.129 & 0.38 \\ 
        ZTF18aabdajx & 2018mac & 0.914 & SN Ia & 0.277 & 0.116 & 0.116 & 0.467 & 0.024 \\ 
        ZTF18aabeszt & & 0.382 & SN Ia & 0.933 & 0.026 & 0.014 & 0.011 & 0.016 \\ 
        ZTF18aacsudg & 2019pxz & 0.876 & SN II & 0.029 & 0.241 & 0.249 & 0.031 & 0.451 \\ 
        ZTF18aaczmob & 2023ecq & 0.549 & SN Ia & 0.434 & 0.071 & 0.041 & 0.012 & 0.441 \\ 
        ZTF18aadrhsi & 2018hzz & 0.869 & SN Ia & 0.301 & 0.158 & 0.089 & 0.358 & 0.094 \\ 
        ZTF18aaexyql & 2019tpu & 0.596 & SN Ibc & 0.121 & 0.082 & 0.029 & 0.021 & 0.747 \\ 
        ZTF18aagtwyh & & 0.606 & SN Ia & 0.445 & 0.257 & 0.132 & 0.029 & 0.136 \\ 
        ZTF18aahsuyl & 2021jqj & 0.569 & SN Ia & 0.934 & 0.022 & 0.014 & 0.005 & 0.025 \\ 
        ZTF18aahyute & & 0.706 & SLSN-I & 0.029 & 0.252 & 0.691 & 0.013 & 0.015
    \enddata
    \tablecomments{The first ten rows of our photometric classification results of the non-spectroscopically classified dataset, which are sorted in alphabetical order. Superphot+ and ALeRCE agree on 60\% of these classifications. A full table is available online. \label{table:new_classes}}
\end{deluxetable*}

\begin{figure*}[t]
    \plottwo{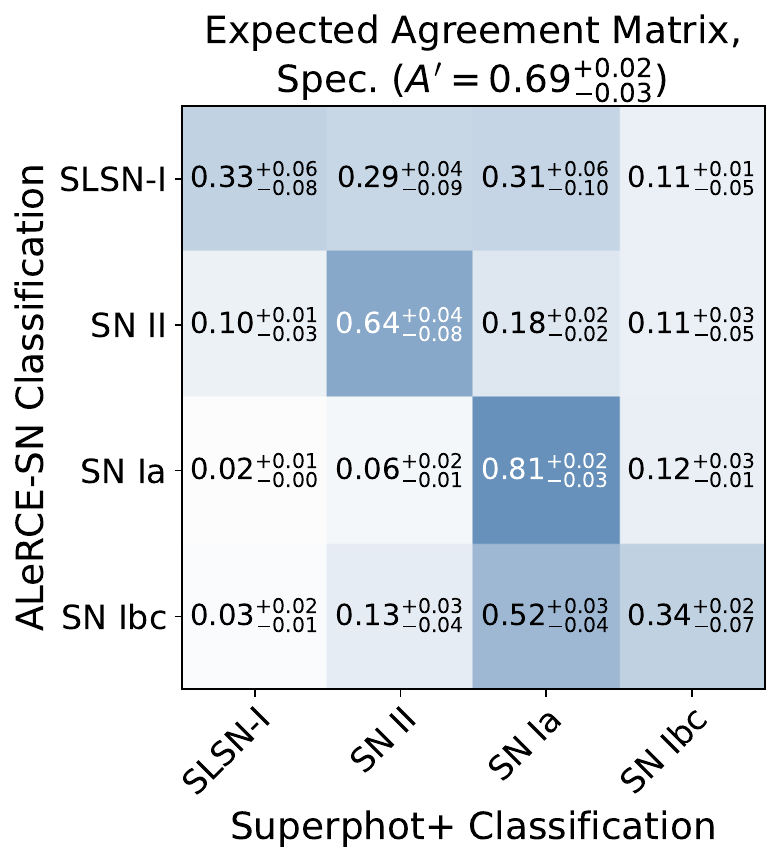}{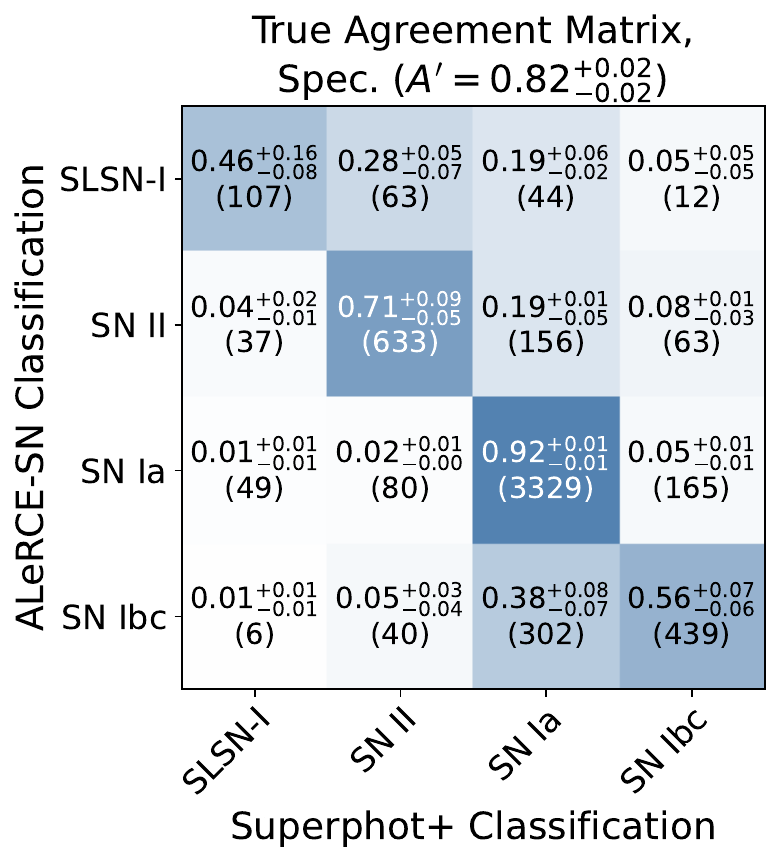}
    \epsscale{0.5} 
    \plotone{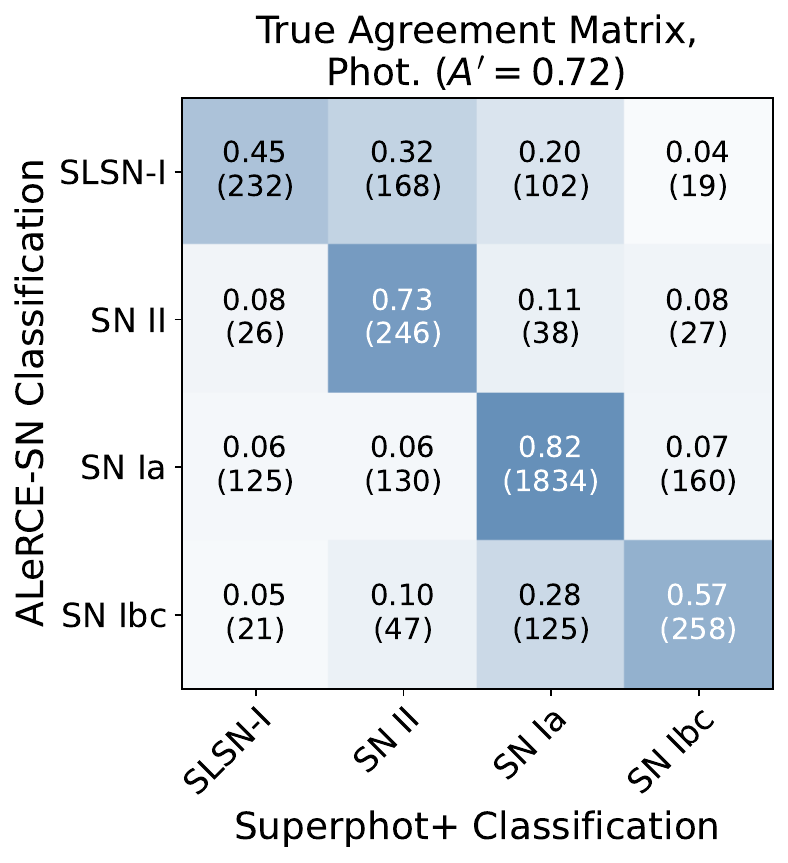}
    \caption{The expected (top left) and true (top right) agreement matrix for the spectroscopic dataset between Superphot+ and ALeRCE-SN, as well as the true agreement matrix for the photometric dataset (bottom). The classifiers agree more than expected when labeling objects as SNe II, but less than expected among ALeRCE's labeled SLSNe. The photometric set's agreement is worse than the spectroscopic set's, which may be due to contamination by non-transients or light curve differences compared to light curves with spectroscopic classifications. All three matrices suggest Superphot+ is more likely to classify unsure light curves as a common type (SN Ia or SN II), whereas ALeRCE-SN favors SLSN labels, validating Superphot+'s higher purity.}
    \label{fig:true_agreement_phot}
\end{figure*}

\begin{figure*}[t] 
    \plotone{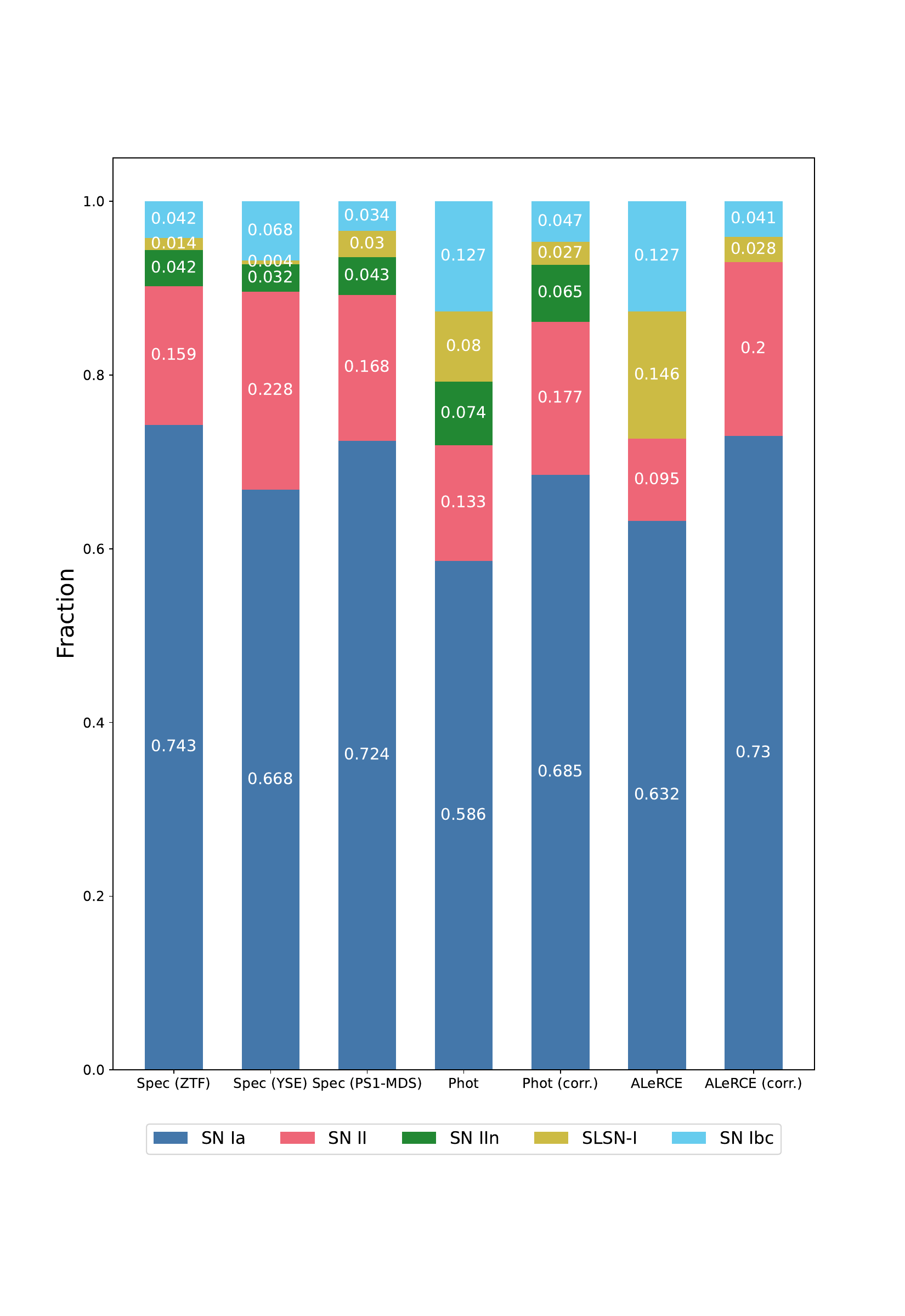}
    \caption{The class fractions of our spectroscopic dataset (``Spec ZTF") compared with those from the Young Supernova Data Release 1 dataset (``Spec YSE") and the PS1-MDS (``Spec PS1-MDS") set used in~\cite{superraenn}. We compare these fractions with those derived from Superphot+'s and ALeRCE-SN's photometric classifications (``Phot" and ``ALeRCE", respectively). Additionally, the corrected class fractions (using the confusion matrices from validation) are included for comparison.}
    \label{fig:class_fractions}
\end{figure*}

\input{sec8_newlabels}

\input{sec9_conclusions}

\section*{Acknowledgements}
KdS and VAV acknowledge support by the NSF through grant AST-2108676 and from LSSTC through grant 2023-SFF-LFI-02-Villar. This research received support through Schmidt Sciences. KdS and GH thank the LSSTC Data Science Fellowship Program, which is funded by LSSTC, NSF Cybertraining Grant \#1829740, the Brinson Foundation, and the Moore Foundation; their participation in the program has benefited this work.

\software{
    ANTARES \citep{antares, antares2018, antares2021, antares_docs},
    ALeRCE \citep{Forster_2021, alerce_docs},
    Astropy \citep{astropy:2013, astropy:2018, astropy:2022, astropy_docs},
    \texttt{dustmaps} \citep{dustmaps_docs, dustmaps_docs2},
    \texttt{dynesty} \citep{speagle_2020, dynesty_docs},
    \texttt{extinction}, \citep{extinction_docs},
    FLEET \citep{fleet, fleet_docs},
    \texttt{imbalanced-learn} \citep{imblearn, imblearn_docs},
    \texttt{iminuit} \citep{iminuit, iminuit_docs},
    JAX \citep{jax_paper, jax2018github},
    \texttt{light-curve} \citep{lc_python, lc_python_docs},
    LightGBM \citep{lightgbm, lightgbm_docs},
    Matplotlib \citep{matplotlib, matplotlib_software},
    NumPy \citep{numpy, numpy_docs},
    NumPyro \citep{phan2019composable, bingham2019pyro, numpyro_docs},
    \texttt{pandas} \citep{pandas, pandas_docs},
    ParSNIP \citep{boone_2021, parsnip_docs},
    \texttt{scikit-learn} \citep{sklearn, sklearn_docs}
    Superphot+ \citep{spp_zenodo},
    Superphot \citep{superphot, superphot_docs},
    SuperRAENN \citep{superraenn, superraenn_docs},
    Pytorch \citep{pytorch},
    Tensorflow Probability \citep{tfp},
}

\begin{figure}[t!]
    \plotone{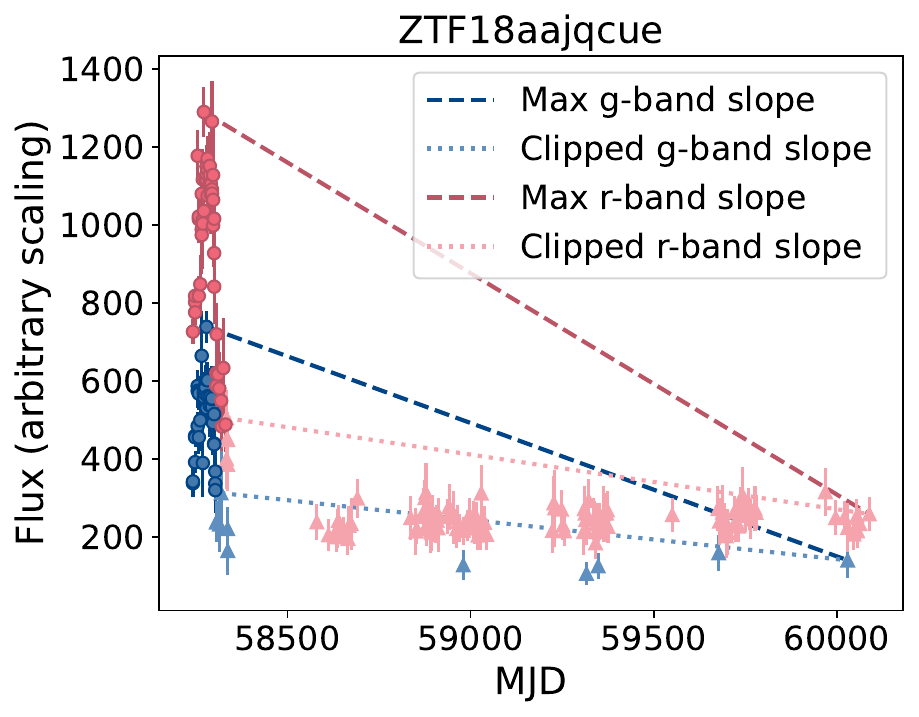} 
    \caption{Example ZTF light curve for a SLSN-I with an artificial extended ``tail" ($|\mathrm{slope}| \leq 0.2 |\mathrm{slope}_\mathrm{max}|$); this late-time excess flux is due to improper template subtraction. These tails are clipped to maintain fitting integrity and correct improper template subtraction. The maximum and tail slopes are indicated by dashed/dotted lines, respectively, and the clipped datapoints are represented by triangles.}
    \label{fig:lc_clip_demo}
\end{figure}

\begin{figure}[t!] 
    \plotone{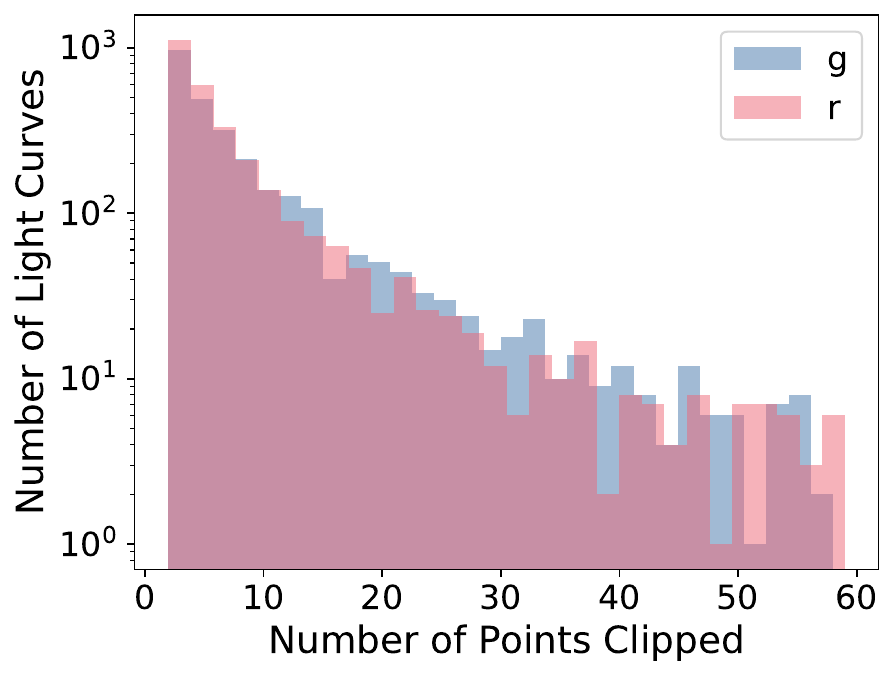}
    \caption{A histogram of the number of points clipped in each band from the light curves in our spectroscopic dataset before pruning. }
    \label{fig:clip_hist}
\end{figure}

\begin{figure}[t] 
    \plotone{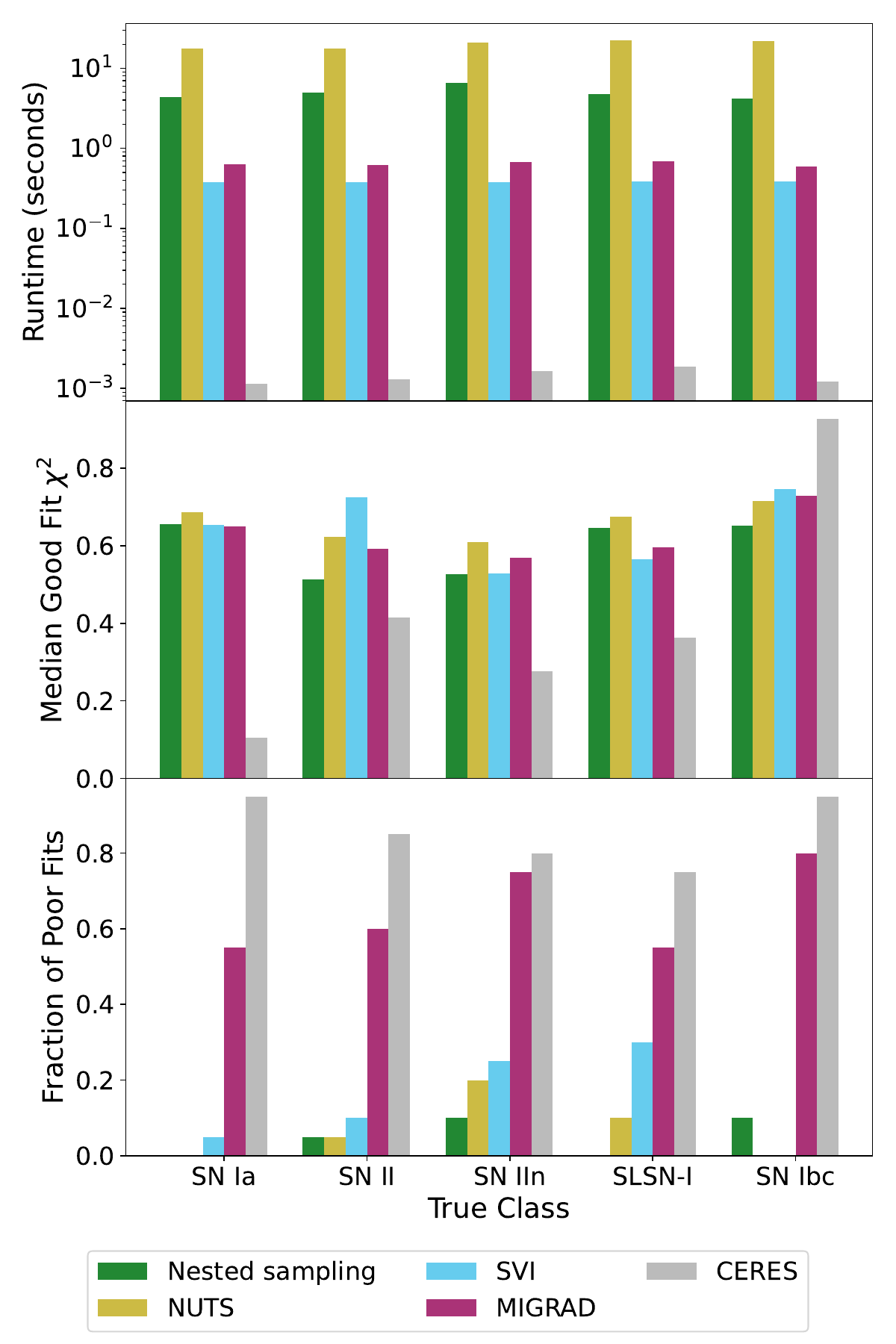}
    \caption{Comparison of the mean fitting runtime, median reduced $\chi^2$ value, and fraction of $\chi^2_{red} > 1.2$ fits from all available Superphot+ samplers, averaged across twenty light curves of each true spectroscopic class. Nested sampling and NUTS are the slowest but most accurate options, with almost no poor fits. SVI is much faster and still correctly fits most light curves. MIGRAD and CERES struggle with the enforced joint constraints inherent during fitting, so fail to fit $>50\%$ of the light curves.}
    \label{fig:runtime_compare}
\end{figure}

\begin{figure}[t] 
    \plotone{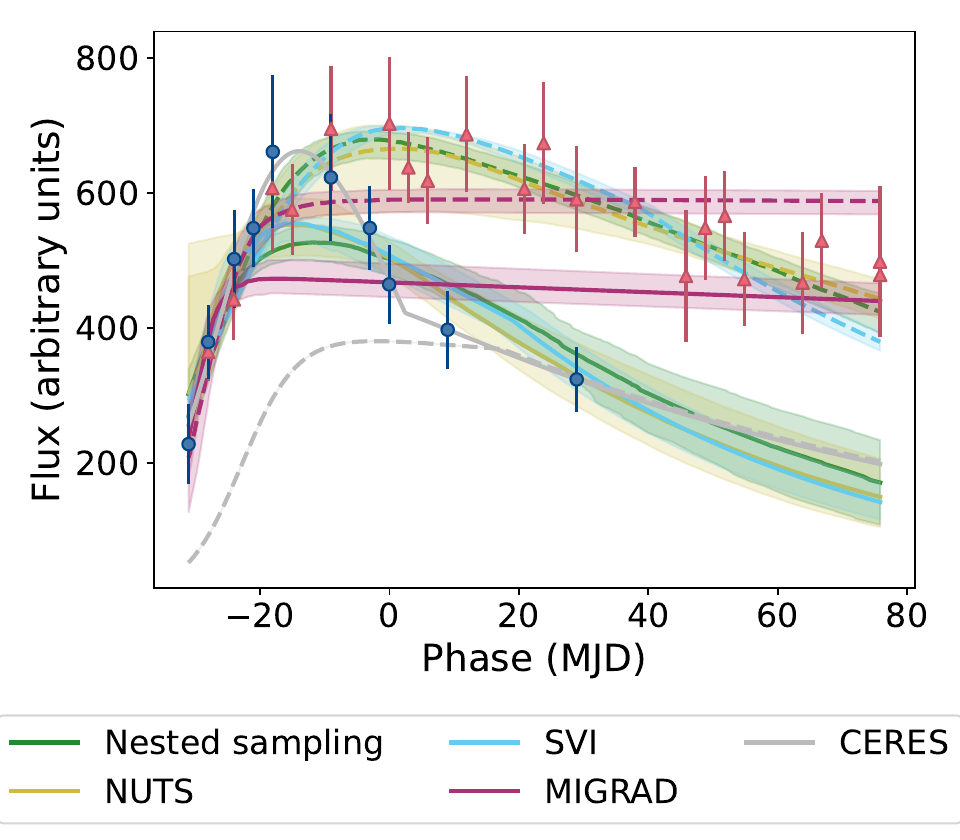}
    \plotone{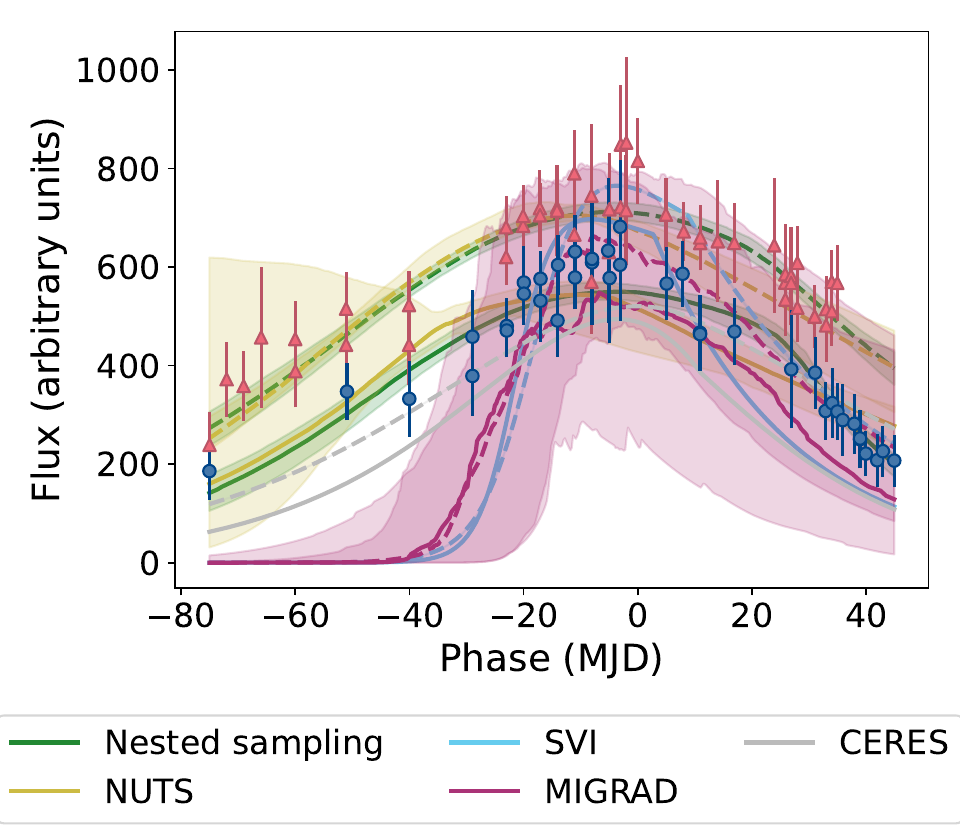}
    \caption{Median model fluxes (solid line) along with the 16th and 84th percentiles (shaded region) for each sampler in Superphot+, for two light curves in our data set. CERES and MIGRAD fail at fitting both light curves, whereas SVI only fits the top light curve correctly. Only NUTS and nested sampling fit both light curves well.}
    \label{fig:sampler_fit_compare}
\end{figure}

\begin{figure}[t] 
    \plotone{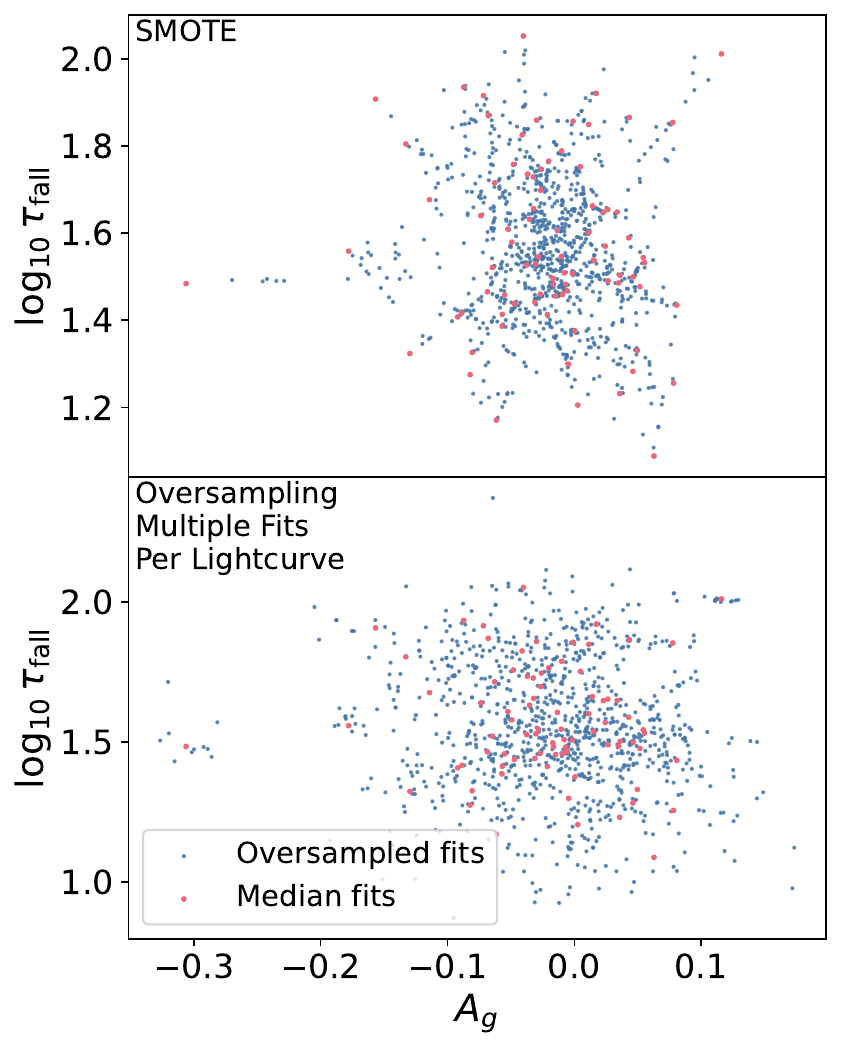}
    \caption{Comparison of oversampling techniques to balance our imbalanced spectroscopic dataset. Here we show only SLSNe-I, which is our smallest class in the sample, for visual clarity. SMOTE (top) draws samples from hyperplanes connecting pairs of observed SNe. This leads to a low diversity of samples when connecting outliers. In contrast, sampling directly from the set of equally-weighted fits (bottom) for each light curve creates smoother oversampling.}
    \label{fig:oversample_compare}
\end{figure}

\input{appendices}

\bibliography{superphot_plus}{}
\bibliographystyle{aasjournal}



\end{document}

%% file: data_values.tex
\newcommand{\agreementSpecPercent}{$82 \pm 2\%$} 
\newcommand{\agreementPhotPercent}{72\%} 

\newcommand{\numBeforeProcessing}{9,526} 

\newcommand{\numLCclipped}{4,324}
\newcommand{\fracLCclipped}{45\%}

\newcommand{\numPrunedNObs}{2,488} 
\newcommand{\fracPrunedNObs}{26\%} 
\newcommand{\numRemainingNObs}{7,038} 
\newcommand{\numPrunedVar}{486} 
\newcommand{\fracPrunedVar}{7\%} 
\newcommand{\numPrunedAllTypes}{6,552} 

\newcommand{\numSNother}{429} 
\newcommand{\numSNpeculiar}{38} 
\newcommand{\numSpecBeforeChisq}{6,123} 
\newcommand{\numSNrare}{154} 

\newcommand{\numLCAlerceTL}{21,781} 
\newcommand{\numPrunedPhotNObs}{5,805} 
\newcommand{\numPrunedPhotVar}{5,854} 
\newcommand{\numPhotBeforeStamp}{10,122} 
\newcommand{\numPhotBeforeChisq}{3,973} 

\newcommand{\numIa}{4,546} 
\newcommand{\numII}{978} 
\newcommand{\numIIn}{257} 
\newcommand{\numSLSN}{83} 
\newcommand{\numIbc}{259} 
\newcommand{\fracIa}{75.0} 
\newcommand{\fracII}{16.1} 
\newcommand{\fracIIn}{4.2} 
\newcommand{\fracSLSN}{1.4} 
\newcommand{\fracIbc}{4.3} 
\newcommand{\fracInBTS}{66\%} 
\newcommand{\fracBrightNotBTS}{91\%} 
\newcommand{\fracSLSNBTS}{0.75} 
\newcommand{\fracLCtwentyPoints}{69} 
\newcommand{\fracPointsSNRFive}{93} 
\newcommand{\numOnlyRise}{91} 
\newcommand{\fracOnlyRise}{1.5\%} 
\newcommand{\numOnlyFall}{503} 
\newcommand{\fracOnlyFall}{8.3\%} 
\newcommand{\numOnlyRisePhot}{135} 
\newcommand{\fracOnlyRisePhot}{3.4\%} 
\newcommand{\numOnlyFallPhot}{267} 
\newcommand{\fracOnlyFallPhot}{6.7\%} 

\newcommand{\numPartial}{594} 
\newcommand{\fracPartial}{9.8\%} 

\newcommand{\numRemovedChisqCut}{62} 
\newcommand{\IaCutChisq}{44} 
\newcommand{\IICutChisq}{12} 
\newcommand{\IbcCutChisq}{3} 
\newcommand{\IInCutChisq}{3} 
\newcommand{\numBadTemplates}{48}
\newcommand{\numSubparFits}{4}
\newcommand{\numGoodRemovedChisq}{15}
\newcommand{\numSNSpec}{6,061} 
\newcommand{\numSNPhot}{3,558} 
\newcommand{\fracRemainingPhotChisq}{89.6\%} 
\newcommand{\numPhotRemovedChisqCut}{415} 

\newcommand{\FNoRedshift}{0.61 $\pm$ 0.02} 
\newcommand{\accNoRedshift}{0.83 $\pm$ 0.01} 
\newcommand{\avgCompleteness}{0.73 $\pm$ 0.08} 
\newcommand{\IaCompleteness}{0.87 $\pm$ 0.01}  
\newcommand{\IInCompleteness}{0.52 $\pm$ 0.07} 
\newcommand{\avgPurity}{0.58 $\pm$ 0.04} 
\newcommand{\IIPurity}{0.84 $\pm$ 0.03} 
\newcommand{\IaPurity}{0.97 $\pm$ 0.01} 
\newcommand{\approxSLSNPurity}{$\simeq$0.25-0.30} 
\newcommand{\fracIaPredIbc}{9\%} 
\newcommand{\numIaPredIbc}{420} 
\newcommand{\numPredIbc}{713} 
\newcommand{\numHCNoRedshift}{3,200} 
\newcommand{\fracHCNoRedshift}{52.7\%} 
\newcommand{\FNoRedshiftHC}{0.86 $\pm$ 0.05} 
\newcommand{\accNoRedshiftHC}{0.95 $\pm$ 0.02} 
\newcommand{\avgCompletenessHC}{0.92 $\pm$ 0.10} 
\newcommand{\avgPurityHC}{0.81 $\pm$ 0.12} 
\newcommand{\SLSNcompleteness}{0.75 $\pm$ 0.17} 
\newcommand{\SLSNcompletenessHC}{1.00 $\pm$ 0.20} 
\newcommand{\IInCompletenessHC}{0.80 $\pm$ 0.13} 
\newcommand{\SLSNPurity}{0.27 $\pm$ 0.06} 
\newcommand{\SLSNPurityHC}{0.83 $\pm$ 0.28} 
\newcommand{\IInPurity}{0.51 $\pm$ 0.09} 
\newcommand{\IInPurityHC}{0.76 $\pm$ 0.19} 
\newcommand{\approxSLSNFracCutHC}{75\%} 
\newcommand{\IaFracCutHC}{43\%} 
\newcommand{\IbcPurityHC}{0.50 $\pm$ 0.13} 

\newcommand{\numLCsRealtime}{20} 
\newcommand{\IInEarlyPurityZero}{0.50 $\pm$ 0.25} 
\newcommand{\IICompleteness}{0.72 $\pm$ 0.05} 
\newcommand{\IInFullPurityZero}{0.03 $\pm$ 0.02} 
\newcommand{\IaEarlyCompletenessZero}{0.79 $\pm$ 0.09} 
\newcommand{\IaFullCompletenessZero}{0.45 $\pm$ 0.13} 
\newcommand{\IIEarlyCompletenessZero}{0.31 $\pm$ 0.11} 
\newcommand{\IIFullCompletenessZero}{0.15 $\pm$ 0.08} 
\newcommand{\IIEarlyCompletenessLate}{0.41 $\pm$ 0.13} 
\newcommand{\IbcPurity}{0.28 $\pm$ 0.03} 

\newcommand{\fleetAUPC}{0.49 $\pm$ 0.03} 
\newcommand{\AUPC}{0.52 $\pm$ 0.19} 
\newcommand{\fracBranchNormal}{96.7\%} 
\newcommand{\numIaBG}{23} 
\newcommand{\numIaT}{159} 
\newcommand{\numIaCSM}{18} 
\newcommand{\bestIaThresh}{0.175} 
\newcommand{\FBinary}{0.90 $\pm$ 0.01} 
\newcommand{\accBinary}{0.93 $\pm$ 0.01} 
\newcommand{\binaryPurityIa}{0.94 $\pm$ 0.01} 
\newcommand{\fracIaLabeledCC}{4 $\pm$ 1\%} 

\newcommand{\numRareCutChisq}{10} 
\newcommand{\numSNrareChisq}{144} 
\newcommand{\numRareLabeledIbc}{44} 
\newcommand{\fracPredIbcActuallyRare}{5.8\%} 

\newcommand{\accRedshift}{0.88 $\pm$ 0.01} 
\newcommand{\FRedshift}{0.71 $\pm$ 0.02} 
\newcommand{\SLSNPurityRedshift}{0.58 $\pm$ 0.14} 
\newcommand{\IbcPurityRedshift}{0.38 $\pm$ 0.06} 
\newcommand{\numWRedshifts}{6,045} 
\newcommand{\numWRedshiftsHC}{4,485} 
\newcommand{\fracRedshiftHC}{74\%} 
\newcommand{\FRedshiftHC}{0.81 $\pm$ 0.03} 
\newcommand{\bestIaThreshRedshift}{0.160} 
\newcommand{\FBinaryRedshift}{0.92 $\pm$ 0.01} 
\newcommand{\accBinaryRedshift}{0.94 $\pm$ 0.01} 
\newcommand{\binaryPurityIaRedshift}{0.95 $\pm$ 0.01} 

\newcommand{\FabsMag}{0.56 $\pm$ 0.02} 
\newcommand{\accAbsMag}{0.77 $\pm$ 0.01} 
\newcommand{\FSuperphot}{0.55 $\pm$ 0.02}
\newcommand{\accSuperphot}{0.80 $\pm$ 0.01}
\newcommand{\FSuperraenn}{0.71 $\pm$ 0.03} 
\newcommand{\accSuperraenn}{0.89 $\pm$ 0.01} 
\newcommand{\FRaenn}{0.62 $\pm$ 0.04} 
\newcommand{\accRaenn}{0.83 $\pm$ 0.02} 
\newcommand{\FParsnip}{0.71 $\pm$ 0.03} 
\newcommand{\accParsnip}{0.89 $\pm$ 0.02} 

\newcommand{\FabsMagBin}{0.86 $\pm$ 0.02} 
\newcommand{\accAbsMagBin}{0.89 $\pm$ 0.01} 
\newcommand{\FSuperphotBin}{0.88 $\pm$ 0.02}
\newcommand{\accSuperphotBin}{0.90 $\pm$ 0.01}
\newcommand{\FSuperraennBin}{0.93 $\pm$ 0.02} 
\newcommand{\FRaennBin}{0.89 $\pm$ 0.02} 
\newcommand{\accRaennBin}{0.92 $\pm$ 0.02} 
\newcommand{\accSuperraennBin}{0.94 $\pm$ 0.02} 
\newcommand{\FParsnipBin}{0.95 $\pm$ 0.01} 
\newcommand{\accParsnipBin}{0.96 $\pm$ 0.01} 

\newcommand{\numSNAlerce}{5525} 
\newcommand{\fracSNAlerce}{91.2\%} 
\newcommand{\numAlercePredIIn}{137} 
\newcommand{\numAlercePredIInRepredII}{74} 
\newcommand{\numAlercePredIInRepredSLSN}{45} 
\newcommand{\numAlercePredIInRepredIa}{10} 
\newcommand{\Ffourclass}{0.66 $\pm$ 0.03} 
\newcommand{\Falerce}{0.62 $\pm$ 0.04} 

\newcommand{\predFracIa}{58.6} 
\newcommand{\predFracII}{13.3} 
\newcommand{\predFracIIn}{7.4} 
\newcommand{\predFracSLSNI}{8.0} 
\newcommand{\predFracIbc}{12.7} 

\newcommand{\numAlercePhotPredIIn}{262} 
\newcommand{\alercePredSLSN}{14.6} 
\newcommand{\alercePredII}{9.5} 
\newcommand{\expectedAgreement}{0.69 $\pm$ 0.03} 
\newcommand{\agreementSpec}{0.82 $\pm$ 0.02} 
\newcommand{\agreementPhot}{0.72} 

\newcommand{\fracIInPredIIn}{51 $\pm$ 9\%} 
\newcommand{\fracIIPredIIn}{28 $\pm$ 7\%} 
\newcommand{\fracIaPredIIn}{15 $\pm$ 5\%} 
\newcommand{\corrPredFracIa}{68.5} 
\newcommand{\corrPredFracII}{17.7} 
\newcommand{\corrPredFracIIn}{6.5} 
\newcommand{\corrPredFracSLSNI}{2.7} 
\newcommand{\corrPredFracIbc}{4.7} 

\newcommand{\priorLogA}{0.096 & 0.058 & $-$0.3 & 0.5}
\newcommand{\priorBeta}{$8.3\times10^{-3}$ & $3.9\times10^{-3}$ & 0 & 0.03}
\newcommand{\priorGamma}{1.43 & 0.31 & 0 & 3.5 }
\newcommand{\priorTnaught}{$t(F_{max})-17.9$ & 9.9 & $t(F_{max})-100$ & $t(F_{max})+30$ }
\newcommand{\priorTauRise}{0.67 & 0.43 & $-$2 & 4 }
\newcommand{\priorTauFall}{1.53 & 0.30 & 0 & 4 }
\newcommand{\priorExtraSigma}{$-$1.66 & 0.34 & $-$3 & $-$0.8 }

\newcommand{\priorAg}{$-$0.08 & 0.11 & $-$1 & 1 }
\newcommand{\priorBetag}{$-$0.21 & 0.27 & $-$2 & 1 }
\newcommand{\priorGammag}{$-$0.05 & 0.17 & $-$1.5 & 1.5 }
\newcommand{\priorTnaughtg}{$-$3.4 & 4.4 & $-$50 & 30 }
\newcommand{\priorTauRiseg}{$-$0.15 & 0.19 & $-$1.5 & 1.5 }
\newcommand{\priorTauFallg}{$-$0.15 & 0.26 & $-$1.5 & 1.5 }
\newcommand{\priorExtraSigmag}{$-$0.15 & 0.25 & $-$1.5 & 1.0 }

\newcommand{\FMLP}{$0.59 \pm 0.03$}

%% file: sec1_intro.tex
\section{Introduction}\label{sec:intro}

Currently, $\sim$10,000 supernova-like (SN-like) transients are photometrically detected every year by the Zwicky Transient Facility (ZTF;~\citealt{ztf}), the Panoramic Survey Telescope and Rapid Response System (Pan-STARRS; \citealt{panstarrs}), and the Asteroid Terrestrial-impact Last Alert System (ATLAS; \citealt{atlas}), among other facilities. Global resources can spectroscopically follow up on $\sim$10\% of these transients. Wide-field surveys planned for this decade, including the Vera C. Rubin Observatory's Legacy Survey of Space and Time (LSST; \citealt{tyson_2002}) and the \textit{Nancy Grace Roman Space Telescope} High Latitude Time Domain Survey (HLTDS; \citealt{romanhighlat}), are expected to increase annual SN detections by a factor of $\sim100$. With spectroscopic resources not expected to increase exponentially in the same time frame, 99.9\% of new SNe light curves will lack traditional \textit{spectroscopic} classifications (see e.g., \citealt{filippenko} for review).

\par In response to this limitation, several works have implemented algorithms that classify SNe using only photometric information \citep{rapid,Villar_2019,superphot,superraenn,boone_2021,alerce}, or a combination of photometry and host galaxy information~\citep{fleet, fleet2, fleet_tde, gagliano2023, Kisley_2023}. Many of these classifiers show successful performance with simulated light curves (e.g., RAPID, \citealt{rapid}; ParSNIP, \citealt{boone_2021}). However, simulated light curves typically lack the observed population diversity in real datasets; it is therefore challenging to predict classifier performance on real data \citep{yse_dr1}. \cite{boone_2021} highlights a particular failure mode, in which a classifier is able to distinguish between SNe II simulated from discrete models. Among classifiers that do train on real data (e.g., Superphot, \citealt{superphot, Villar_2019} on Pan-STARRS data; SuperRAENN, \citealt{superraenn}, on Pan-STARRS data; FLEET, \citealt{fleet, fleet2, fleet_tde}, on ZTF and Open Supernova Catalog data; GHOST, \citealt{gagliano2023}, on SDSS-II, ESSENCE, and SNLS data; \citealt{Kisley_2023}, on THEx data), only the Automatic Learning for the Rapid Classification of Events (ALeRCE, \citealt{alerce}) and Fink (\citealt{Leoni_2022}) pipelines, both trained on ZTF data, are currently being run in realtime with publicly accessible predictions. Currently, the latter is also limited to binary classification between Type Ia and non-Ia supernovae.

\par Thus, we aim to design a publicly-available, multi-class classifier that is trained on real data. This pipeline should be designed for easy adaptation to Rubin light curves in the future. Of particular concern is the current requirement for spectroscopic redshift information for the vast majority of photometric classifiers (with the exception of FLEET and both of ALeRCE's classifiers). Even with new spectroscopic surveys (e.g., 4MOST; \citealt{de20194most}), which will obtain spectroscopic redshifts for millions of galaxies, only a small fraction of the $\sim$$10^{10}$ galaxies detected by LSST~\citep{lsst_sb_ch9} will have associated spectroscopic redshift information. Additionally, according to the LSST Science Requirements Document \citep{lsst_srd}, Rubin is not expected to meet its minimum target criteria for accurate photometric redshifts within its first three years of operation \citep{graham_2018, plasticc_redshift}. This lack of reliable redshifts for early Rubin observations, along with a preference for very dim host galaxies among exotic SNe (e.g., Type I SLSNe; \citealt{hsu2023slsn}), necessitates a SN classifier that does not use redshift information. Instead, we rely only on the light curve shape and color to differentiate between SN classes. 

\par Here, we present the SN classification pipeline Superphot+, which (1) empirically fits SN light curves to a parametric model and (2) trains a gradient-boosted machine (GBM) classifier on the best-fit model parameters. Superphot+ improves on the Superphot~\citep{Villar_2019,superphot} pipeline by accelerating fitting for real-time light curve processing, improving class re-weighting, and enabling classification without redshift information. In this work, we train and apply Superphot+ to ZTF light curves observed through October 2023, though we emphasize Superphot+'s adaptability for other photometric datasets. This paper is organized as follows. In Section~\ref{sec:data}, we describe the selection and pruning of our training and test datasets. We describe the details of the light curve fitting and choice of sampling algorithm in Section~\ref{sec:fitting}. In Section~\ref{sec:classifier}, we describe feature selection (including and excluding redshift information), optimization of the classifier architecture, and oversampling of the training dataset. In Section~\ref{sec:results}, we summarize multi-class and binary classifier performance with and without redshift information, emphasizing accuracy as a function of classification confidence. We also consider performance on partial light curves for real-time classification through the ANTARES Broker \citep{antares2018, antares2021}. In Section~\ref{sec:compare}, we compare Superphot+'s performance without redshift information to that of the Automatic Learning for the Rapid Classification of Events (ALeRCE; \citealt{alerce}) light curve classifier, one of the only comparable redshift-independent classifiers currently available in the literature. We also compare our training results using redshifts with those from previous pipelines that require redshift. In Section~\ref{sec:newlabels}, we use our trained classifier to assign photometric labels to \numSNPhot\ ZTF SN-like transients which lack spectroscopic classification but were labeled likely SNe by ALeRCE's light curve and stamp classifiers. We compare our photometric predictions with those from ALeRCE. Finally, we discuss conclusions and avenues for future work in Section~\ref{sec:conclusion}. Our code is publicly available on GitHub\footnote{\url{https://github.com/VTDA-Group/superphot-plus}} and the Python Package Index as \verb+superphot-plus+.

%% file: sec2_data.tex
\section{Dataset Generation}\label{sec:data}

\subsection{Photometric Data from the Zwicky Transient Facility}

\par This work uses light curve data from ZTF~\citep{ztf}, a wide-field survey conducted with a 48-inch telescope located at Palomar Observatory. ZTF photometrically identifies $\sim$5,000-10,000 new (likely) extragalactic transients every year, $\sim$20\% with spectroscopic classifications. Both ZTF public surveys (the Northern Sky Survey and the Galactic Plane Survey) image a combined $\approx$25,000 square-degrees of the northern sky at a high cadence of $\sim$2 days, in both the $g$- and $r$-bands.

\par To train our photometric classifier, we first collate SNe that (1) are spectroscopically labeled as one of Superphot+'s output classes and (2) have associated ZTF light curves of sufficient quality. We refer to this set as the training or ``spectroscopic'' dataset interchangeably throughout the paper. To create this training set, we first query the Transient Name Server (TNS; ~\citealt{tns}) for all spectroscopically classified transients with ZTF internal names and photometry. This yields \numBeforeProcessing\ events. While most of this set are SNe, there are also non-SN transients, like active galactic nuclei (AGNs) and tidal disruption events (TDEs), which are pruned as described in Sec.~\ref{subsec:sel}. We keep all TNS classes in our dataset for the time being to determine the classes (both SN and non-SN) that are not used directly for training but have sufficient ``high-quality'' light curves to merit further analysis in Section~\ref{sec:misctype}. We then download the $g$- and $r$-band light curves for these events through a Python API maintained by ALeRCE \citep{Forster_2021}'s LSST (currently ZTF) alert broker. 

\subsection{Data Pre-processing}
\par Superphot+'s parametric function captures light curves in flux (rather than magnitudes). We thus convert our photometry from magnitudes to fluxes using $f = 10^{-0.4(m-zp)}$ and a constant zeropoint of $zp=26.3$, which is halfway between the $r$-band and $g$-band median zeropoints as stated by the ZTF collaboration \citep{ztf_data, ztf_supplement}. We correct all resulting light curves for Milky Way extinction. To do so, we adopt $E(B-V)$ from the dust maps provided by \cite{dustmaps1} and \cite{dustmaps2}, and use a \cite{fitzpatrick_massa_2007} extinction model with $R_V=3.1$. We neglect host galaxy extinction, which is difficult to calculate from light curve data alone.

\par ZTF alert data is calculated from difference imaging and point spread function (PSF) photometry to measure the time-variable flux of transients. However, poor template subtraction can lead to a complex background, causing light curves to asymptote well above zero flux and creating spurious detections after the transient events (and potentially before, if forced photometry is included). Our ZTF light curves do not include forced photometry, so we only see false detections at the tail end of the light curves. While we could add a constant term to our parametric model to account for a flux offset, we find that including such an offset leads to oversubtraction among SNe with long plateaus. Therefore, we instead filter these spurious data points before fitting. Following the procedure described in Appendix~\ref{sec:clipping}, we clip the tail ends of \numLCclipped\ (\fracLCclipped) light curves to some extent.

\par Because we do \textit{not} utilize redshift in the primary version of Superphot+, we do not convert light curves into rest-frame. When incorporating redshift information into our classifier in Section~\ref{sec:redshift}, we add redshifts and $k$-corrected peak absolute magnitudes as additional input features rather than altering the light curves or their derived fit parameters.

\subsection{Dataset Pruning}\label{sec:pruning}

\par To refine our spectroscopic dataset before model fitting and training, we exclude light curves that fail to satisfy certain criteria. First, we keep only light curves with at least five points of signal-to-noise ratio (SNR) greater than 3 in each of the $g$- and $r$-bands (\textit{after} clipping light curve tails). This number is selected so all light curves have either (1) somewhat constrained fit parameters across the entire light curve or (2) strongly constrained fits in at least one portion of the light curve, depending on the sampling cadence. This cut on the number of observations removes \numPrunedNObs\ (\fracPrunedNObs) light curves from the dataset, leaving \numRemainingNObs\ events. Additionally, we include only light curves whose brightness variability in both bands cannot solely be attributed to measurement error. Quantitatively, we remove transients whose maximum amplitude in either band is less than three times that band's mean flux uncertainty. We also remove transients in which the standard deviation of all fluxes in a single band is less than that band's mean flux uncertainty. These two cuts eliminate \numPrunedVar\ (\fracPrunedVar) light curves from the remaining spectroscopic set. This is a smaller fraction than is removed by the observational cut, consistent with ZTF registering an event as a transient only after sufficient brightening relative to the template flux.

\par A summary of our data quality cuts on all TNS classes in our dataset is shown in Table~\ref{table:data_cuts}. We are left with \numPrunedAllTypes\ transients from all spectroscopic classes. We note that longer-duration transients (such as SLSNe and TDEs) have much smaller fractions of their total sample pruned, as there is more time for ZTF to sufficiently sample each light curve before they fade in brightness. In contrast, very rapid transients such as M dwarf stellar flares or cooler transients such as SNe Ib/c are more heavily pruned.

\subsection{Class Selection for Training}\label{subsec:sel}

After pruning poor-quality light curves from our dataset, we filter the remaining sample to only include SNe spectroscopically classified as either SN Ia, SN Ib/c, SN II, SN IIn, or SLSN-I (following \citealt{superphot,superraenn}), including rarer subtypes as detailed below:

\begin{itemize}
    \item Type Ia SNe (SNe Ia): SNe Ia have distinct Si spectroscopic features (while lacking those of H/He) near peak. They often exhibit secondary peaks in the near-infrared (which can appear in the $r$-band; ~\citealt{Kasen_2006}). Their progenitors are usually white dwarfs that experience thermonuclear runaway as they exceed the Chandrasekhar limit, although diversity in progenitor scenarios exists~\citep{blondin_2012}. Due to their high intrinsic rates and bright peak magnitudes ($M_\mathrm{B}\simeq-19.3$), which make them observable at higher redshifts for a fixed magnitude limit, Type Ia SNe dominate our dataset. In addition to Branch normal SNe Ia \citep{branchnormal}, we have included SNe Ia-91T-like, SNe Ia-CSM, and SNe Ia-91bg-like in this category.
    \item Type Ib/c SNe (SNe Ib/c): Type Ib/c SNe are core-collapse SNe without H spectroscopic features. SNe Ic additionally lack He lines. SNe Ib/c tend to be optically redder at peak compared to SNe Ia. The progenitor stars of SNe Ib/c have likely been stripped of their H/He envelopes, potentially by a binary companion \citep{Filippenko_2005}. The optical light curves are predominantly powered by the radioactive decay of $^{56}$Ni and $^{56}$Co. We include broad-lined SNe Ic (SNe Ic-BL) and calcium-rich SNe Ib (SNe Ib-Ca-rich) in this category. 
    \item Type II SNe (SNe II): SNe II are core-collapse SNe with H spectroscopic features. They are primarily powered by H recombination following collapse of a red supergiant (or potentially a blue supergiant), creating a post-peak plateau in their light curves. We include both SNe IIL and SNe IIP subtypes in this category, as there is debate whether these subclasses are truly distinct \citep{Sanders_2015, Rubin_2016}.
    \item Type IIn SNe (SNe IIn): SNe IIn are primarily characterized by narrow H emission lines during the photospheric phase \citep{Smith_2014}. Their light curves are extremely heterogeneous, in overall brightness, duration, and shape (see \citealt{nyholm2020type} for a recent sample). They are primarily powered by shocks arising from the interaction of the SN ejecta and pre-existing circumstellar material (CSM), which was likely deposited by luminous blue variable (LBV) progenitors. We include Type II superluminous SNe (SLSNe-II) as a subset of SNe IIn, as their hydrogen emission lines can strongly resemble those of SNe IIn. However, it is uncertain whether they are part of the IIn continuum or a distinct class (e.g., see \citealt{Gal_Yam_2012} and \citealt{pessi2023}). Merging SLSNe-II with SNe IIn leads to improved classifier performance compared to grouping SLSNe-II with SLSNe-I or leaving SLSN-II as a distinct label.
    \item Type I superluminous SNe (SLSNe-I): SLSNe-I are exceptionally bright SNe ($M_\mathrm{B}\lesssim-20$) that lack signatures of H/He/Si in their near-peak spectra. Their light curves tend to be bluer and longer in duration compared to SNe Ia. Their exact progenitor channel is uncertain. Some evidence ~\citep{Nicholl_2017} suggests that they are powered by the rapid spindown of a newly born magnetar \citep{Metzger_2013}. However, potential signs of CSM interaction have also been noted \citep{Hosseinzadeh_2022}.
\end{itemize}

We remove \numSNother\ objects that do not belong to these five spectroscopic classes, \numSNpeculiar\ of which arguably belong to the above classes but are marked as peculiar (``pec"). This selection leaves a training sample of \numSpecBeforeChisq\ SNe. From Table~\ref{table:data_cuts}, \numSNrare\ pruned events not used in training are LBVs, SNe IIb, SNe Ibn, SNe Iax, or TDEs. We apply Superphot+ to these events in Section~\ref{sec:misctype}, as they are the most prevalent possible contaminants of SN-like datasets without spectroscopic labels. We exclude AGNs and cataclysmic variables (CVs) from this subsequent analysis, as we assume they will be separated from our SN-like dataset by a more general classifier (e.g, the ALeRCE light curve classifier).

\subsection{Photometric Dataset}\label{subsec:phot}

\par In addition to a ``spectroscopic'' training dataset, we also collate a ``photometric'' dataset. This consists of ZTF light curves that are high-quality and SN-like in behavior, but do not have a spectroscopic classification. Our photometric set will serve as our test set and be classified by Superphot+.

\par To collate this dataset, we use ALeRCE's \citep{alerce} two ``top-level" classifiers. One of these classifiers uses two-band light curves to distinguish between SNe, stochastic phenomena (e.g., AGNs, CVs) and periodic variables (the ``light curve'' classifier; \citealt{alerce}). \cite{alerce} find that ALeRCE's light curve classifier is highly successful, with an F$_1$-score of 0.97 and SN completeness of 100\% (these metrics are defined in Section~\ref{sec:metrics}). The other classifier uses image cutouts (the ``stamp'' classifier; \citealt{alerce_stamp}) to potentially label objects as asteroids or bogus in addition to SNe, stochastic variables, or periodic variables. \cite{alerce_stamp} report 87\% SN completeness for the stamp classifier at the time of training, with a $\simeq5$\% false positive rate.

\par First, we gather \numLCAlerceTL\ light curves that are photometrically classified as a SN-like transient by ALeRCE's light curve classifier with 50\% or greater confidence but do not have associated spectroscopic labels (as of October 2023). We then clip spurious light curve tails as described in Appendix \ref{sec:clipping}, followed by the same observational and variability cuts that we applied to the spectroscopic dataset. These cuts prune \numPrunedPhotNObs\ and \numPrunedPhotVar\ light curves, respectively, leaving \numPhotBeforeStamp\ high-quality light curves without spectroscopic labels.

\par Through visual inspection, we find that these cuts do not sufficiently eliminate non-SNe transients from our photometric set. Non-SNe sources include bogus detections, AGN-like variables, and very noisy or low-amplitude variable stars. Therefore, we also remove any light curve not marked as a SN by the ALeRCE stamp classifier. This leaves \numPhotBeforeChisq\ events, which means that less than 50\% of the events labeled ``SN-like" by ALeRCE's light curve classifier are also labeled as SN-like by the stamp classifier. This result is surprising and suggests either a much higher false positive rate than reported for ALeRCE's light curve classifier, or a much lower SN completeness than reported for the stamp classifier. Investigation into the performance of top-level classifiers is left to other work.

\subsection{Properties of the Reduced Datasets}\label{subsec:reduced_datasets}

\par After pruning our datasets, we are left with \numSpecBeforeChisq\ light curves in the spectroscopic training set, and \numPhotBeforeChisq\ light curves in the photometric test set. The breakdown of the spectroscopic set is as follows:

\begin{itemize}
    \item SN Ia: \numIa\ (\fracIa\%)
    \item SN II: \numII\ (\fracII\%)
    \item SN Ib/c: \numIbc\ (\fracIbc\%)
    \item SN IIn: \numIIn\ (\fracIIn\%)
    \item SLSN-I: \numSLSN\ (\fracSLSN\%)
\end{itemize}

\par We find that \fracInBTS\ of these light curves are also in the ZTF Bright Transient Survey (ZTF BTS; \citealt{ztf_bts}) catalog, which aims to spectroscopically classify all light curves brighter than 19 magnitude at peak that pass certain quality cuts\footnote{ZTF BTS requires light curves (1) have two constraining measurements within 7.5 days of the brightness peak, and (2) do not set within a month after maximum light.}. Of our light curves not in ZTF BTS's catalog, \fracBrightNotBTS\ are brighter than 19 mag but do not pass ZTF BTS's quality cuts. Our class fractions approximately match those from the entire ZTF BTS dataset~\citep{ztf_bts2}, with the exception of SLSNe-I. Our cuts yield a SLSN-I fraction (\fracSLSN\%) almost double that of ZTF BTS (\fracSLSNBTS\%). This is not unexpected, as (1) programs targeting SLSNe-I (e.g., FLEET, \citealt{fleet2}) increase their prevalence in TNS, and (2) SLSNe-I have longer timescales that allow for more observations at a fixed cadence; they therefore pass our first quality cut more frequently than the other SN classes.

\par The distribution of number of observations (in either band) per light curve above various SNRs is shown in Figure~\ref{fig:snr_hist}. Because we require at least five points of SNR~$\geq 3$ per band, there are no events in our pruned dataset with fewer than ten combined observations above this SNR. Most light curves in our pruned dataset (\fracLCtwentyPoints\%) have $\geq 20$ datapoints with $\mathrm{SNR} \geq 3$, and most points with SNR $\geq 3$ (\fracPointsSNRFive\%) also have $\mathrm{SNR} \geq 5$. This demonstrates that the quality of the average light curve in our training set is significantly higher than our minimum quality cuts, and our cuts only remove events that are of significantly lower quality than the rest of the dataset.

\par The final peak $r$-band apparent magnitude distribution of both the spectroscopic and photometric dataset is shown in Figure~\ref{fig:appm_hist_unc}. The distribution of spectroscopically classified SNe matches our expectations, as it exhibits a single peak at approximately 18.5 mag, where ZTF BTS attempts to enforce high spectroscopic completeness. The photometric dataset includes a tail of events much brighter than any spectroscopically classified light curve; these events are likely remnant bogus objects, non-SNe, and detections in very high extinction regions. After fitting our pruned datasets, we only classify light curves with sufficiently good quality fits (determined by a reduced chi-squared metric). This cut is further detailed in Section~\ref{subsec:chisq}, and removes most of the very bright outliers within the photometric dataset. The remaining photometric events share a similar distribution to the spectroscopic dataset, except that the distribution peaks around 19 mag. The number of events in both datasets dimmer than 19.75 mag falls off rapidly, consistent with the ZTF limiting magnitude of $\sim$20.5 and our signal-to-noise and amplitude constraints.

\par We note that no constraints are imposed on the temporal coverage of the light curve, so our final spectroscopic sample includes \numOnlyRise\ light curves (\fracOnlyRise) with only pre-peak observations, and \numOnlyFall\ light curves (\fracOnlyFall) with only post-peak observations. These partial light curves, while less informative, will be common in realtime classification and are thus crucial to keep in our training set. Similarly, our photometric dataset includes \numOnlyRisePhot\ (\fracOnlyRisePhot) pre-peak and \numOnlyFallPhot\ (\fracOnlyFallPhot) post-peak light curves. We explore partial light curves more thoroughly in Section~\ref{sec:realtime}.

%% file: sec3_fitting.tex
\section{Parametric Model and Fitting Procedure}\label{sec:fitting}

\par After each light curve is pre-processed, we fit the SN flux in each band to a piecewise parametric model introduced by \cite{Villar_2019} and \cite{superphot}:

\begin{align}\label{eq:emp_model}
\begin{split}
    F(t) = &\dfrac{A}{1 + \exp\Big[\frac{-(t-t_0)}{\tau_\mathrm{rise}}\Big]} \times  \\
    &\begin{cases}
        1 - \beta(t-t_0), & \mathrm{ if\,\,} t - t_0 < \gamma \\\\
        (1 - \beta\gamma)\exp\Big[{\frac{\gamma - (t - t_0)}{\tau_\mathrm{fall}}}\Big], & \mathrm{otherwise}
    \end{cases}
\end{split}
\end{align}

\par This model has seven fit parameters and describes a rise in brightness followed by an approximately linear plateau, which then switches to an exponential decline $\gamma$ days after $t_0$. This general form captures the main characteristics of both core-collapse (CC) SNe and SNe~Ia. The effect of each parameter on the model is illustrated in Figure 2 of ~\cite{Villar_2019}. $A$ is the amplitude of the model, $t_0$ is roughly the phase of peak brightness, and $\tau_\mathrm{rise}$ and $\tau_\mathrm{fall}$ are the exponential timescales for the rise and decline of the light curve, respectively. $\beta$ and $\gamma$ represent the slope (relative to the amplitude) and duration of the plateau following peak, respectively. There are two versions of each parameter, as shown Table~\ref{table:priors}, corresponding to the fits separately derived from the set of $g$-band observations and the set of $r$-band observations. Finally, each band has an associated $\sigma_{\mathrm{extra}}$ parameter which serves as an extra uncertainty added in quadrature to each of the flux uncertainties. This extra uncertainty accounts for the limitations in the empirical model itself.

\par We expect the $g$- and $r$-band light curves for a given event to be correlated. As a result, when using Bayesian fitting techniques, we choose to express the $g$-band priors relative to the $r$-band priors for each parameter. We designate the $r$-band as the ``reference'' band since most light curves have better coverage in the $r$-band. Sampled $g$-band parameters, denoted by the ``$g$" subscript, are assumed to be the log of the ratio between the actual $g$- and $r$-band fit parameters. We choose to sample all of our parameter ratios in log space because an equal shift in either direction of the log parameter corresponds to an equal but inverse multiplicative scaling of the $g$- over $r$-band parameter ratio. The only exception is $t_{0,g}$, which is expressed as the difference (time delay) between the $g$- and $r$-band $t_0$ parameters. All sampled $g$-band ratios are then combined with the sampled $r$-band parameters before being used in our parametric model. This formulation correlates fit parameters across bands and constrains multi-band fits in regions sampled in only one band. It also leads to very narrow and informative priors, as we find that $g$- to $r$-band parameter ratios are quite similar across SN-like light curves.  Our choice of parameterization could bias fits for parameters where inter-band correlations are less physically justified. Examples include the amplitude ratio of SNe affected by extreme host reddening, or parameters derived from partial light curves (which would just reflect our fitting priors). We note that the latter issue would only arise among the \numPartial\ partial light curves (\fracPartial) in our spectroscopic sample.

\par The above choice of priors necessitates that we fit $g$- and $r$-bands simultaneously in a 14-dimensional parameter space. This differs from the method used in \cite{Villar_2019} and \cite{superphot}, where each band is fit independently twice, and the second iteration's prior distribution is the average of each band's posterior distribution from the first iteration. While both strategies encourage similar fits across bands of the same light curve, ours allows us to define the expected variation between bands \textit{a priori}. Furthermore, the final fits are less likely to be skewed by bands with fewer datapoints. However, fitting fourteen parameters simultaneously is more difficult and computationally expensive than fitting seven parameters.

\par To efficiently explore this combined high-dimensional space, we turn to a variety of modern sampling techniques as discussed in Appendix~\ref{sec:sampler}. We ultimately use nested sampling to fit the archival light curves used in this work, and stochastic variational inference for realtime light curve fitting. The efficiency of both of these samplers at early iterations relies heavily on the size of the prior volume; therefore, we spend significant effort refining the fit priors to best mirror the expected best-fit parameters of ZTF SNe. To achieve this, we start with broad, uniform priors, and then iteratively alternate between fitting our dataset and replacing our priors with the marginal posterior distributions combined from all our light curves\footnote{Note that, before combining posteriors, we oversample our fits to balance class prevalence, as described in Section~\ref{subsec:balance}.}. This process continues until the priors and population-level posteriors are sufficiently similar. All final priors are truncated Gaussians or truncated log-Gaussians, as detailed in Table~\ref{table:priors}. The priors for $A_r$ and $\sigma_{\mathrm{extra}, r}$ are expressed relative to the maximum $r$-band flux value of each light curve. We can see that our final priors mirror the dataset's combined marginal posterior distributions in Figure~\ref{fig:1d_distributions}.

\par For each light curve, our nested sampler returns a set of several hundred posterior samples (the exact number varying per light curve). The resulting fits for six representative SNe are shown in Figure~\ref{fig:ex_lc}. Note that our fits are tightly constrained for very well-sampled light curves, while the fits from poorly sampled light curves (e.g., those with only rise or decline information) are more likely to be prior-dominated. We will use our posteriors to estimate uncertainties in our final classifications.

\par We note that the same parametric model (that of \citealt{Villar_2019}) is used as part of a larger feature set in \cite{alerce}, although \cite{alerce} fit each band independently and do not use the $\sigma_\mathrm{extra}$ parameters. \cite{alerce} find that the resulting best-fit parameters are more effective than other extracted light curve features in differentiating between SN classes. However, ALeRCE uses the Levenberg-Marquardt fitting algorithm, implemented in the Python package \texttt{scipy} \citep{scipy} as \verb+curve_fit+. As we explore in Appendix~\ref{sec:sampler}, gradient-descent based minimization algorithms like Levenberg-Marquardt can often lead to poor optimal model fits.

\subsection{Fit Quality Metrics}\label{subsec:chisq} 

\par To evaluate the quality of our model fits, we calculate a modified reduced chi-squared value for each fit per light curve:
\begin{equation}
    \chi^2_\mathrm{red} = \frac{1}{N}\sum\frac{|\vec{f}_\mathrm{obs} - \vec{f}_\mathrm{model}|^2}{\vec{\sigma}_\mathrm{eff}^2}
\end{equation}
where $N$ is the number of datapoints and $\vec{\sigma}_{\mathrm{eff}}^2 = \vec{\sigma}^2 + \sigma_{\mathrm{extra}}^2$. This differs slightly from the traditional reduced chi-squared definition in that we do not augment the denominator to reflect the number of fit parameters (degrees of freedom). We opt for our modified metric as opposed to the traditional reduced chi-squared value because the latter breaks down for light curves with fewer than eight datapoints per band; the influence of priors would prevent the best fit from perfectly passing through the datapoints even for these sparser light curves, yielding infinite reduced chi-squared values.

\par We calculate the median reduced chi-squared values across all fits per light curve, and we find that \numSNSpec\ ($\sim$99\%) light curves in our spectroscopic dataset are fit with a median reduced chi-squared value less than 1.2. We then apply a chi-squared cut to remove the remaining \numRemovedChisqCut\ light curves. This set includes \IaCutChisq\ SNe Ia, \IICutChisq\ SNe II, \IInCutChisq\ SNe IIn, and \IbcCutChisq\ SNe Ib/c. From these events, we see that \numBadTemplates\ cannot be fit adequately by our empirical model as a result of either poor template subtraction, extreme outlier points, or extreme secondary peaks either before or after the primary peak\footnote{The samples with poor template subtraction pass the data pruning process due to the estimated flux baseline dramatically changing over time, causing the rise of the light curve to end up dimmer than the decline region in the subtracted light curve. This mode of incorrect subtraction mainly appears among pre-2019 light curves.}. An additional \numSubparFits\ light curves could have been fit better to lower the chi-squared value, leaving about \numGoodRemovedChisq\ well-fit light curves that are unnecessarily removed. However, not applying a chi-squared cut or using a more lenient threshold lets significantly more bogus fits through, degrading the quality of our training set.

\par We also apply a $\chi^2_\mathrm{red} \leq 1.2$ cut to our photometric dataset, which removes \numPhotRemovedChisqCut\ light curves. From both Figure~\ref{fig:appm_hist_unc} and visual inspection, we see that this cut successfully removes both abnormally bright light curves that are likely bogus, and light curves that are clearly not from SNe. We are left with a final photometric dataset of \numSNPhot\ (\fracRemainingPhotChisq) light curves.

We see the effect of our fit quality cut in Figure~\ref{fig:chisq_cutoff}, where we also overlay classification accuracy as a function of fit $\chi^2_\mathrm{red}$. This accuracy is calculated from the final training results in Section~\ref{sec:results}. We see in this plot that a cutoff value of 1.2 preserves most of our spectroscopic dataset and prevents accuracy from dropping at high $\chi^2_\mathrm{red}$ values due to poorly estimated fit parameters. Among the photometric dataset, the peak of the $\chi^2_\mathrm{red}$ distribution is also well below the cutoff value.

%% file: sec4_classifier.tex
\section{Classifier Details}\label{sec:classifier}

\subsection{Balancing the Training Set}\label{subsec:balance}
\par As detailed in Section~\ref{subsec:reduced_datasets}, our spectroscopic dataset is heavily imbalanced across classes, with there being over fifty times as many SN Ia (our majority class) as SLSN-I events (smallest class). Machine learning algorithms perform less efficiently on imbalanced training sets, preferring to over-classify the majority class. Therefore, we need to (1) make the most of every light curve in our minority classes and (2) re-balance the training set as part of our training process.

\par To accomplish the former, we use stratified $K$-fold cross-validation to calculate performance metrics from every sample in our spectroscopic set. In this work, we use $K=10$ folds, meaning the dataset is split into ten groups, with equal class fractions in each group, and each group is used as the test set for a separately trained classifier. We find that while fewer folds degrades classifier performance, more folds increases performance variation across folds. The remaining 90\% of each fold not used in the test set is further split 90-10 into a training and validation set. These two sets are oversampled independently for class balancing and then used to train the corresponding classifier.

\par We re-balance our training and validation sets by oversampling multiple model fits per minority-class light curve; the classifier treats each fit as a separate input. This procedure is discussed in detail, and compared to traditional oversampling techniques, in Appendix~\ref{sec:oversampling}. 

\subsection{Classification Metrics}\label{sec:metrics}

Before exploring classifier architectures, we define four metrics to evaluate classification performance. The per-class \textit{completeness} is the fraction of samples that belong to one class that are correctly classified as that class. The \textit{accuracy} is the micro-averaged completeness, or the fraction of light curves in the entire dataset that are classified correctly. The per-class \textit{purity} is the fraction of light curves classified as one class that are actually of that class. The \textit{F$_1$-score} is the harmonic mean of the completeness and purity. Purity, completeness, and F$_1$-score are calculated separately for each SN class. Each performance metric is calculated as follows:

\begin{equation}
    \text{Accuracy} = \frac{TP}{N}
\end{equation}
\begin{equation}
    \text{Completeness} = \frac{TP}{TP + FN}
\end{equation}
\begin{equation}
    \text{Purity} = \frac{TP}{TP + FP}
\end{equation}
\begin{equation}
    \text{F}_1 = \frac{2\times \text{Purity} \times \text{Completeness}}{\text{Purity} + \text{Completeness}}
\end{equation}
where $N$ is the total number of light curves in the dataset. $TP$ is the true positive count, which is the number of light curves within one class that are correctly predicted as that class. Likewise, $FP$ is the false positive count (the number of samples classified as one type that are not actually that type), and $FN$ is the true negative count (the number of samples of one class that are misclassified). All metrics are expressed in our results as 80\% confidence intervals, or the median value bounded by the second highest and second lowest value among our ten $K$-folds.

\par Because our dataset is highly imbalanced, quantifying classifier performance by accuracy will bias it toward correctly classifying SNe Ia at the expense of the rarer SN types. On the other hand, macro-averaged (i.e. class-averaged) statistics equally penalize low performance within any SN type. Because we value both per-class completeness and purity, as each has its respective science cases, we optimize our parameters by maximizing the macro-averaged F$_1$-score; this gives equal importance to all SN classes and balances completeness and purity values.

\subsection{Classifier Feature Selection and Architecture}\label{sec:features}

\par Photometric redshift estimates (``photo-zs'') are notably more common than spectroscopic estimates in current and future (e.g. LSST) SN datasets. Each photo-z is calculated from the broadband spectral energy distribution of the SN's host galaxy, and associating SNe with the correct host galaxy is not a trivial task. Sources highly offset from their closest galaxy, sources at high redshift, or sources near more than one galaxy can be attributed no or incorrect photometric redshift (see e.g., \citealt{Gagliano_2021}). Therefore, we first train our classifier on features that do not require redshift estimates. This is straightforward since we do not use redshifts during the pre-processing or fitting steps of Superphot+. We construct our redshift-independent classifier inputs from twelve out of the fourteen fit parameters, excluding both $A_r$ and $t_{0,r}$. Neither $A_r$ nor $t_{0,r}$ are intrinsic SN properties; the former is dependent on the phases each SN is observed (and associated apparent fluxes), while the latter is dependent on the absolute MJD of each SN.

\par ZTF has more complete and reliable redshift estimates compared to what we expect from the first few years of Rubin (as detailed in Section~\ref{sec:intro}). Therefore, we present a second version of our classifier that does use redshift information and is only trained on light curves with associated host galaxy redshifts. From our spectroscopic dataset, 16 light curves are missing redshift estimates on TNS. We exclude these light curves when training the redshift-inclusive classifier.

\par We also re-add the $r$-band amplitude ($A_r$) as an input feature for the redshift-inclusive classifier. Additionally, we include the redshift $z$ and the (cosmological $k$-corrected) $r$-band absolute magnitude $M_r$, calculated from the brightest measured $r$-band flux. The cumulative distributions of these two new inputs are plotted in Figure~\ref{fig:abs_mags}. We see that most events in our training set are at $z < 0.2$, with the exception of some SLSNe-I that have brighter absolute magnitudes. Because $A_r$ is the amplitude of the modeled peak relative to the brightest measured flux (which can be significantly offset for partial light curves), combining it with $M_r$ quantifies the absolute magnitude of the \textit{modeled} light curve at peak.

\par The shape parameters $\beta$, $\gamma$, $\tau_\mathrm{rise}$, and $\tau_\mathrm{fall}$ all have a first-order $(1+z)$ redshift dependence resulting from time dilation. When using redshifts as input features, we assume our classifier can learn to correct for this dilation. However, for our redshift-independent classifier, not correcting for this effect could skew fit parameters for farther SNe (such as many SLSNe-I) and in turn affect classifier performance. For our ZTF SN dataset, the majority of redshifts are less than 0.2, and less than 0.6 among SLSNe-I, which would not shift fit parameters enough to mimic other SN classes (as supported by Figure~\ref{fig:1d_distributions} and our use of log-normal priors). Therefore, we do not consider this a significant source of classification error within our redshift-independent classifier. However, for surveys that can observe SNe at farther redshifts (e.g. Rubin), time dilation can stretch light curves by much larger factors, potentially impacting classification. For these surveys, one can alternatively replace the four aforementioned input features with $\theta_1 = \beta \tau_\mathrm{rise}$, $\theta_2 = \tau_\mathrm{rise} / \tau_\mathrm{fall}$, and $\theta_3 = \tau_\mathrm{rise} / \gamma$, which would reduce our input vector's length by one and cancel out the $(1+z)$ multiplicative factors. We find that this alternate feature set reduces the F$_1$-score by $\sim$10\% for our ZTF dataset, but may improve performance for deeper surveys.

\par The output of each classifier is a vector of five values representing the ``pseudo-probability" of the input event belonging to each SN class. These are not true probabilities because while the vector elements sum to unity, they are not calibrated (i.e. confidence values do not match the fraction of true samples within events assigned that confidence), as shown in Figure~\ref{fig:calibration}. For our multi-class problem, the assigned label for each light curve is the class with the highest output pseudo-probability. The classification ``confidence'' is this highest probability. For single-class variants, we instead assign positive labels to an event if the positive pseudo-probability exceeds a pre-specified confidence threshold. We provide a more detailed discussion on calibration of these pseudo-probabilities and confidence thresholds in Section~\ref{sec:binary}.

\par We explore two different classifier architectures, neural networks and gradient-boosted machines (GBMs), in Appendix~\ref{sec:arch_compare}. We find that GBMs, constructed with the \texttt{LightGBM} package, yield optimal classifier performance across $K$-folds. We train separate GBMs to classify events with and without redshift information.

%% file: sec5_results.tex
\section{Classification Results}\label{sec:results}
In this Section, we test multiple variants of our classifier and quantify the efficacy of our pipeline. These variations include multi-class classification, single-class classification, and training with and without redshift information.

\subsection{Multi-Class Classification without Redshift Information}

\par First, we study the results of our five-way gradient-boosted machine. This classifier does not use any redshift information and is the ``default'' classification mode in Superphot+. Our trained model classifies the spectroscopic dataset with a F$_1$-score of \FNoRedshift\ and an accuracy of \accNoRedshift. The associated confusion matrices are shown in Figure~\ref{fig:no_redshift_cm}. The class-averaged completeness is \avgCompleteness. SNe Ia unsurprisingly have the highest completeness at \IaCompleteness\ due to their high prevalence in the spectroscopic dataset. On the other hand, SNe IIn are prone to the highest fraction of misclassifications, with a completeness of \IInCompleteness. This may be because SNe IIn are a highly diverse class of SNe in terms of observational properties (see e.g.,~\citealt{Nyholm_2020}, for a recent sample from the Palomar Transient Factory).

\par The class-averaged purity is \avgPurity, with SNe II and SNe Ia having the highest purities at \IIPurity\ and \IaPurity, respectively. In contrast, SLSNe-I and SNe Ib/c have notably lower purities (both \approxSLSNPurity). This is likely reflective of the imbalances in our training set; even a small fraction of true SNe Ia or SNe II can heavily contaminate the small samples of predicted SLSNe-I or SNe Ib/c. For example, only \fracIaPredIbc\ of SNe Ia are misclassified as SNe Ib/c, but this small fraction still translates to \numIaPredIbc\ SNe Ia. These contaminants account for over half of the \numPredIbc\ total predicted SNe Ib/c. We note that this difficulty is also observed in \cite{superphot}, again due to a significant class imbalance (albeit less extreme than our dataset's imbalance).

\par Feature importance analysis shows that Superphot+ relies most on fall timescales, followed by plateau durations, peak band ratios (a proxy for color), and rise timescales. It is then no surprise that Superphot+ most commonly misclassifies light curves of similar timescales. SNe Ia are most likely to be misclassified as SNe Ib/c (and vice versa), and SLSNe-I are most likely to be misclassified as SNe IIn (and vice versa). SNe II are misclassified about equally across other classes. We also find that very long-lived light curves, such as those of SNe IIn and SLSNe-I, are sometimes wrongly attributed long plateaus (causing SN II misclassifications), or SNe II without post-plateau sampling are fitted with no plateaus and slow fall timescales (causing SLSN-I misclassifications). We can clearly see overlaps in fitting parameters between classes in Figure~\ref{fig:1d_distributions}, each potentially impacting classifier performance.

Examples of misclassified light curves, along with their classification probabilities, are shown in Figure~\ref{fig:misclassified_lc}. By manual inspection, the main confounding factors among misclassified light curves of each type are:
\begin{itemize}
    \item SNe Ia: Most SNe Ia misclassified as SNe II, SNe IIn, or SLSNe-I have partial light curves and thus some unconstrained fit parameters. Many SNe Ia that are wrongly labeled as SNe Ib/c have secondary $r$-band peaks affecting fits, or $g-r$ colors similar to those of SNe Ib/c (i.e., redder than typical SNe Ia). We note that, since we only correct for Milky Way extinction, SNe Ia belonging to host galaxies with exceptionally high extinction will appear redder and more like SNe Ib/c.
    \item SNe II: SNe II with missing observations that prevent plateau constraints are more likely to be misclassified. This is especially problematic for long-lived events, which are often misclassified as SNe IIn.
    \item SNe IIn: The SN IIn class, in general, is particularly heterogeneous. Misclassified events are most often classified as SLSNe-I (similar timescales) or SNe II (fit with a plateau).
    \item SLSNe-I: Misclassified SLSNe-I often have slow declines, which are best fit as plateaus in our empirical model, meaning that most misclassified SLSNe-I are assigned high SN IIn and SN II probabilities.
    \item SNe Ib/c: Misclassified SN Ib/c light curves are often sparse or noisy. Some events are particularly blue around peak and thus misclassified as SNe Ia.
\end{itemize}

\par We next investigate light curve properties associated with especially poor classifier performance. First, we explore the impact of number of observations and light curve SNR on classification accuracy. We calculate the latter using the top 90th percentile of all the datapoints' SNRs within the light curve. Both metrics show weak positive correlations with classification accuracy, as evident in Figure~\ref{fig:snr_vs_accuracy}. More data points correlates more strongly with SN II and SN IIn accuracy. This makes sense, as light curves with many high-SNR observations are more likely to have well-sampled plateaus. Interestingly, SLSN-I and SN Ib/c classification worsens beyond $\sim$30 datapoints, as does SLSN-I and SN IIn classification at a 90th percentile SNR above $\sim$20. Most of the incorrectly classified light curves with over thirty observations are exceptionally long-lived and/or have secondary behavior beyond what our empirical model can capture, such as additional peaks, high variance within the fall region, or pre-explosion variations that are unable to be removed through template subtraction. The best way the model can capture this anomalous behavior is with extended rise times (resembling SLSNe-I and SNe IIn) or longer plateaus (resembling SNe II and SNe IIn). A small fraction ($<10$ objects) of SNe Ib/c with $>50$ observations show dramatically inconsistent template subtraction across the light curve, preventing clean light curve tail clipping, but this is not the main culprit of SNe Ib/c misclassification. Among SLSNe-I of SNR above 20, we find either (1) shorter, Ia-like evolution timescales, or (2) exceptionally long-lived light curves with II-like plateaus. Most misclassified, SNR $> 20$ SNe IIn have incomplete light curves and are classified with lower confidence, with the remainder exhibiting either shorter (Ia-like) or longer (SLSN-like) timescales than expected.

\par We additionally consider our classifier's performance within its most confident predictions. We regenerate our confusion matrices in Figure~\ref{fig:no_redshift_p07} including only the \numHCNoRedshift\ (\fracHCNoRedshift) events classified with confidence greater than 0.7 (a cutoff chosen to match \citealt{superphot} and \citealt{superraenn}). Our performance metrics improve substantially, with a new F$_1$-score, class-averaged purity, and class-averaged completeness of \FNoRedshiftHC, \avgPurityHC, and \avgCompletenessHC, respectively. Completeness increases most substantially for SLSNe-I (from \SLSNcompleteness\ to \SLSNcompletenessHC) and SNe IIn (from \IInCompleteness\ to \IInCompletenessHC). These classes also significantly improve in purity, with SLSN-I's increasing from \SLSNPurity\ to \SLSNPurityHC\ and SN IIn's increasing from \IInPurity\ to \IInPurityHC. However, \approxSLSNFracCutHC\ of both classes are removed by the high confidence cut (compared to \IaFracCutHC\ of SNe Ia), leaving only a handful of very confidently labeled events. We note that SNe Ib/c are, again, labeled with the lowest purity at \IbcPurityHC\ (with many labeled SNe Ib/c being true SNe Ia).

\subsection{Classification of Partial Light Curves}\label{sec:realtime}

\par Next, we explore the efficacy of Superphot+ as a realtime classifier by analyzing its performance on partial light curves. This is important to explore, as one major benefit of Superphot+ is computationally efficient fitting that can keep up with both ZTF and expected LSST alert streams, enabling realtime classification.

\par First, we consider that not all fit parameters will be informative for early-phase light curves. If the light curve's final observation is before the end of a plateau or peak, then the $t_0 > \gamma$ function of our piecewise model will be completely unconstrained, and the best-fit values for $\gamma$ and $\tau_\mathrm{fall}$ (for both bands) will reflect their priors. These features may skew our classifier towards incorrect labels for these partial light curves. Therefore, we train an alternate version of our GBM classifier without $\gamma$, $\tau_\mathrm{fall}$, or the corresponding $g$-/$r$-band ratios, calling this our ``early-phase'' classifier. We compare this classifier's performance to our ``full-phase'' classifier described in the previous section.

\par To compare the realtime performance of our early-phase and full-phase classifiers (both without redshift information), we randomly select up to \numLCsRealtime\ SNe per $K$-fold per class, and then truncate these selected light curves at a series of increasing phases, where phase is defined as the time after peak $r$-band brightness. Each truncated light curve is fit and classified, and the completeness, purity and F$_1$-score are calculated at each phase for each SN type for each fold. These metrics for both the early-phase (dashed) and full-phase (solid) classifiers are shown in Figure~\ref{fig:phase_vs_acc}, with uncertainty margins representing the full-phase $1\sigma$ uncertainties across 10 $K$-folds. The macro-averaged metrics are shown in black.

\par We find that the early-phase classifier outperforms the full-phase classifier for cutoff phases before $\sim$20 days, after which the full-phase classifier performs better. The early-phase classifier especially excels for light curves truncated near peak, with much higher phase $=0$ values for SN IIn purity (\IInEarlyPurityZero\ versus \IInFullPurityZero), SN Ia completeness (\IaEarlyCompletenessZero\ versus \IaFullCompletenessZero), and SN II completeness (\IIEarlyCompletenessZero\ versus \IIFullCompletenessZero) compared to the full-phase classifier. In contrast, the full-phase classifier yields significantly higher SN II completeness (\IIEarlyCompletenessLate\ versus \IICompleteness) at late phases. 

\par SLSN-I classification for both variants is stable from very early phases, as is SN IIn classification without post-peak features. This is expected, as both classes can be set apart by their slower rise evolution, which is constrained weeks before peak brightness. Both variants assign SN Ia and SN Ib/c labels with better completeness and purity near peak, where the peak color is fit more precisely; from Figure~\ref{fig:1d_distributions}, we infer SNe Ib/c are most distinguishable by redder colors. Classification accuracy for SNe II (and SNe IIn with post-peak features) only increases weeks after peak, as classification relies on constraint of their characteristic plateaus.

\par Our full-phase classifier is currently integrated as an ANTARES \citep{antares} filter, labeling events from the ZTF Alert Stream in real time\footnote{https://antares.noirlab.edu/}. Application of the early-phase classifier to sufficiently truncated light curves from the ANTARES alert stream is left for future work.

\subsection{Single-Class Performance and Calibration}\label{sec:binary}

\par High purity samples of a singular SN class are often required for population-specific studies (e.g., increasing the sample of spectroscopically classified SLSNe), at the cost of completeness. Here, we consider the performance of Superphot+ when optimized for binary (single-class) classification problems. We can reuse our trained multi-class GBM by selecting a target class and compressing all probabilities outside of the target class into a single ``negative'' probability. The problem simplifies from assigning an object one of five class labels to assigning one of two: positive and negative. In the multi-class problem, the assigned label is determined by the highest probability, which differs for every event. Setting a minimum confidence threshold would result in some objects receiving no assigned label, which we do not allow in our multi-class framework. For binary classification, we can adjust the confidence threshold required for a positive label; this choice inversely impacts purity and completeness of the target class. For example, requiring a very high classification confidence before assigning a positive label will lead to a smaller predicted dataset of that class but also fewer contaminants from other spectroscopic classes.

\par Before considering different confidence thresholds, we first consider whether the pseudo-probabilities from our classifier are well-calibrated. The calibration curve, shown in Figure~\ref{fig:calibration}, examines whether the pseudo-probabilities assigned by the classifier for a specific class is an over- or underestimate of the true probability. Pseudo-probabilities are ``well-calibrated" if the reported classifier probability matches the fraction of events correctly classified at that confidence. An ideal calibration curve perfectly follows a $y=x$ line for all classes. For Superphot+, the classifier assigns overconfident SN Ib/c pseudo-probabilities, but underconfident SN Ia values. This likely reflects the classifier's balance between SN Ib/c purity and SN Ia completeness. The other three class probabilities do not show strong biases.

\par Keeping in mind that our classifier is uncalibrated, we can now explore the effect of single-class confidence thresholds on the performance metrics detailed in Section~\ref{sec:metrics}. While receiver operator characteristic (ROC) curves \citep{roc} are commonly generated to summarize a classifier's performance, they tend to be overly optimistic for highly imbalanced datasets, such as our SN dataset. Therefore, we instead rely on purity-completeness curves (i.e. precision-recall curves in machine learning literature) to explore binary classifier performance \citep{roc_pr_curves, roc_pr_curves2}. The purity-completeness curve for each class in our dataset is shown in Figure~\ref{fig:pr_curve}, with $1\sigma$ uncertainties calculated across $K$-folds. A perfect classifier for a SN type follows the top-right corner, where both the purity and completeness are 1.0. A completely random classifier follows a horizontal line aligned with the target class's prevalence in the dataset. The diamonds correspond to a confidence cutoff of 0.5, where the assigned label corresponds to the highest (binary) pseudo-probability.

\par The area under the purity-completeness curve (AUPC) quantifies binary classifier performance. It benefits from not relying on choice of confidence threshold, unlike the binary F$_1$-score. We use Figure~\ref{fig:pr_curve} to directly compare Superphot+'s classification of SLSNe-I with that of FLEET~\citep{fleet, fleet2}, a binary classifier designed to isolate a high-purity SLSN-I (or TDE) dataset. FLEET's AUPC value is \fleetAUPC, with the uncertainty resulting from different random seed initializations. Superphot+'s SLSN-I AUPC value is \AUPC, where the larger uncertainty propagates from variance across $K$-folds. These overlapping AUPC values are promising, as it shows that Superphot+, which must balance the performance of multiple classes, boasts comparable binary performance to pipelines optimized for binary classification.

\par We can then choose a confidence threshold to optimize F$_1$-score for each target class. As an example, we retrain our GBM to output either SN Ia or CC SN labels, the latter of which includes our other four classes. Generating high-purity (Branch normal) SN Ia datasets is crucial for cosmological studies~\citep{Jones_2017}, though we note our Type Ia sample is only \fracBranchNormal\ Branch normal; 91bg-like SNe Ia (\numIaBG), 91T-like SNe Ia (\numIaT), and SNe Ia with CSM interaction (\numIaCSM) are also grouped into this classification. We see in Figure~\ref{fig:f1_vs_thresh} that our GBM's optimal confidence threshold is at $p=\bestIaThresh$, with a maximal F$_1$-score of \FBinary\ and accuracy of \accBinary. This matches our knowledge that our models return underconfident SN Ia pseudo-probabilities. We also show the corresponding purity matrix; SN Ia purity is high at \binaryPurityIa, at the cost of a lower CC SN purity. CC SN contamination is approximately double that in the curated SN Ia dataset from \cite{Jones_2017}, but a confidence cut can be applied to increase our classifier's SN Ia purity. Additionally, only \fracIaLabeledCC\ of SNe Ia are classified as core-collapse SNe.

\begin{deluxetable*}{cccccc}    
\tablecaption{Miscellaneous Transient Classifications} 
    \tablehead{\colhead{Object Type} & \colhead{SN Ia} & \colhead{SN II} & \colhead{SN IIn} & \colhead{SLSN-I} & \colhead{SN Ib/c}}
    \startdata
    SN Iax (12) & 0.250 (3) & 0 (0) & 0 (0) & 0 (0) & \textbf{0.750 (9)} \\
    SN Ibn (18) & \textbf{0.444 (8)} & 0.278 (5) & 0.056 (1) & 0.222 (4) & 0 (0) \\
    SN IIb (57) & 0.123 (7) & 0.228 (13) & 0.035 (2) & 0.018 (1) & \textbf{0.596 (34)} \\
    TDE (51) & 0.137 (7) & 0.098 (5) & 0.216 (11) &\textbf{0.549 (28)} & 0 (0) \\
    LBV (6) & 0.167 (1) & \textbf{0.500 (3)} & 0.167 (1) & 0 (0) & 0.167 (1) \\
    \hline
    Total (144) & 26 & 26 & 15 & 33 & 44 \\
    Phot. Frac. & 0.006 & 0.031 & 0.052 & 0.128 & 0.058
    \enddata
    \tablecomments{Summary of how miscellaneous transients are classified by our five-class, redshift-independent classifier. The absolute number of events is shown in parentheses. In general, SNe Iax and SNe IIb are labeled as SNe Ib/c. Most SNe Ibn are classified as SNe Ia. TDEs tend to be grouped with SLSNe-I. LBVs are mostly labeled as SNe II. We also show the fraction of contamination in Superphot+'s predicted datasets from these rarer classes, with the heaviest contamination at 12.8\% for predicted SLSNe-I. \label{table:misc_probs}}
\end{deluxetable*}

\subsection{Classification of Excluded Transient Types}\label{sec:misctype}

\par Superphot+'s output classes exclude rarer transient types that have SN-like light curves but lack the prevalence to constitute additional output classes. These classes include SNe Iax, Ibn, IIb, TDEs, and LBVs/other massive star outbursts. We expect many light curves from these classes to pass our data quality and fitting chi-squared cuts and thus contaminate our predicted datasets. We first fit light curves of these classes from our pruned spectroscopic set, and we remove the \numRareCutChisq\ events with median reduced chi-squareds above 1.2. We then classify the remaining \numSNrareChisq\ objects with our five-output GBM model to determine which labels they would likely be assigned, and summarize the results in Table~\ref{table:misc_probs}.

\par Most SNe Iax are not labeled as SNe Ia, but rather as SNe Ib/c. This is perhaps not surprising as Type Iax SNe tend to be redder than SNe Ia~\citep{snIax}. SNe IIb are also primarily labeled as SNe Ib/c because they tend to be redder and faster-evolving than SNe II without clear plateaus~\citep{snIIb}. In contrast, most SNe Ibn are classified as SNe Ia or SNe II, since their light curves evolve over faster time scales but are too blue to be mistaken for SNe Ib/c. TDEs are most commonly classified as SLSNe-I or SNe IIn since their light curves are bluer and decay over very long timescales (though many partial light curves are labeled as SNe Ia or SNe II). LBV eruptions are predominantly labeled as SNe II; while their timescales are shorter than those of SLSNe or SNe IIn, many light curves exhibit post-peak variability that are incorrectly fit as plateaus.

\par We can examine Superphot+'s photometric predictions for its five output classes and determine the level of contamination from the classes listed here. From Table~\ref{table:misc_probs}, we see that the most common label is SNe Ib/c, with \numRareLabeledIbc\ (out of \numSNrareChisq) rare-type transients predicted to be SNe Ib/c. While this is significant compared to the \numIbc\ true SNe Ib/c in our spectroscopic dataset, it is only \fracPredIbcActuallyRare\ of the \numPredIbc\ events labeled as SNe Ib/c by Superphot+, the majority being SN Ia misclassifications. Therefore, we find that rare-type transients are a minor contaminant for predicted SNe Ib/c. Across all classes, misclassifications from within our five main spectroscopic labes have the greatest impact on class purity. This supports our decision not to include the rarer classes listed here as additional output labels. However, we expect orders of magnitude more events from these classes to be detected by Rubin; a Rubin-tailored Superphot+ variant may include these classes as additional outputs.

\subsection{Inclusion of Redshift Information}\label{sec:redshift}

\par Here, we train a variant GBM that does use redshift information, as described in Section~\ref{sec:features}; this GBM includes redshifts and brightest absolute magnitudes as additional features. The resulting confusion matrices are shown in Figure~\ref{fig:redshift_cm}. By including redshift information, the accuracy increases from \accNoRedshift\ to \accRedshift, and the F$_1$-score increases from \FNoRedshift\ to \FRedshift. The most significant improvement is the SLSN-I purity from \SLSNPurity\ to \SLSNPurityRedshift, where contamination from true SNe Ia and SNe II is dramatically reduced. Without redshift information, partial SLSN-I light curves can be mistaken for SNe Ia or SNe II depending on how much of the rise and fall regions are missing; the disparity in peak absolute magnitudes fixes most of these misclassifications. We additionally see improved SN Ib/c purity from \IbcPurity\ to \IbcPurityRedshift, as brighter SNe Ia are less frequently mislabeled as SNe Ib/c.

\par We also consider the subsample classified with confidence above 0.7 in Figure~\ref{fig:redshift_p07}. Of the \numWRedshifts\ events with associated redshifts, \numWRedshiftsHC\ (\fracRedshiftHC) are classified with confidence above 0.7, with a F$_1$-score of \FRedshiftHC. This F$_1$-score is actually lower than that of the GBM trained without redshift information when the same confidence cut is applied. However, the GBM with redshift information is generally more confident; a much lower fraction of light curves (25\% versus 47\% for the redshift-independent classifier) is removed by the $p > 0.7$ cut. We conclude that including redshift information makes our classifier more confident when labeling events, but \textit{less} accurate among highly confident predictions. To determine why, we generate performance metrics among the most confident 3,200 events (chosen to match the number of events remaining after the redshift-independent high confidence cut). In this case, the performance metrics are nearly identical, but we see higher SLSN-I and SN IIn completeness when not using redshift information. This, combined with the number of high-confident SLSNe-I approximately doubling after including redshift information, leads us to conclude that redshift information is biasing our classifier to label high-redshift events as SLSNe-I by default. 

\par Finally, we train a SN Ia versus CC SN model that uses redshift information, and once again optimize the confidence threshold. With an optimal threshold of $p=\bestIaThreshRedshift$, the model returns F$_1=$ \FBinaryRedshift\ and a SN Ia purity of \binaryPurityIaRedshift, both which slightly exceed the corresponding binary metrics when not using redshift information.

%% file: sec7_alerce.tex
\section{Comparison with Other Classifiers}\label{sec:compare}

\par We compare Superphot+'s performance (both including and excluding redshift information) with those of three state-of-the-art pipelines from previous works: Superphot~\citep{superphot}, SuperRAENN~\citep{superraenn}, and ParSNIP~\citep{boone_2021}. Unlike Superphot+, all of these previous pipelines require redshift information, and all were originally trained on four-band Pan-STARRS Medium Deep Survey (PS1-MDS) light curves. Both the original Superphot and SuperRAENN papers use random forests for classification, whereas ParSNIP uses a gradient-boosted machine trained with LightGBM. We regenerate light curve encodings from each pipeline with our ZTF dataset for a fair comparison, and use those features to train identical GBMs. This isolates variations in performance as resulting from better light curve encapsulation by the selection of parametric (for Superphot+ and Superphot) or non-parametric (for SuperRAENN and ParSNIP) features. We train both multi-class and single-class (SN Ia) GBMs for each pipeline's feature set. We also train GBMs using only peak color ($A_r$, $A_g$) and redshift information ($z$, $M_{r}$) as a baseline, to isolate performance improvements from light curve shape information from each classifier.

\par The resulting accuracies and F$_1$-scores are shown in Table~\ref{table:binary_comparison}, with example light curves modeled by each pipeline in Figure~\ref{fig:classifier_compare_lc}. Training with only color and redshift information yields an accuracy of \accAbsMag\ and F$_1$-score of \FabsMag. Superphot, the precursor to our current pipeline, yields F$_1 =$ \FSuperphot\ and an accuracy of \accSuperphot. This is only marginally better than only using redshift and color features, indicating that fits suffer tremendously from broad, uniform priors. SuperRAENN yields F$_1=$ \FSuperraenn\ and an accuracy of \accSuperraenn, which is higher than metrics from the original paper. We find that SuperRAENN decodings from the same encoded light curve vary depending on the requested number of decoded time stamps. ParSNIP, using a variational auto-encoder, performs similarly with a F$_1 =$ \FParsnip\ and accuracy of \accParsnip. It excels at differentiating between SNe Ib/c and  SNe I/a since it can nonparametrically encode secondary $r$-band ``bumps" present in some SN Ia light curves; neither Superphot's nor Superphot+'s empirical model can capture these additional peaks.

\subsection{Comparisons to the ALeRCE Light Curve Classifier}\label{sec:alerce}

\par We next direct our attention to ALeRCE's light curve classifier, which is the only other publicly-running, multi-class SN classifier that does not use redshift information. Unlike Superphot+, which solely classifies within SNe, ALeRCE uses a hierarchical random forest to first distinguish between transients, stochastic sources, and periodic variables (the ``top-level classifier''), and then to categorize within these broad categories~\citep{alerce}. We use the top-level classifier to select a photometric dataset, as described in Section~\ref{subsec:phot}; here we explore ALeRCE's SN-specific light curve classifier (``ALeRCE-SN''). AleRCE-SN also uses the model fit parameters described in Section~\ref{sec:fitting} as part of a larger feature set, but it uses a gradient-descent based algorithm (through Python's \verb+scipy.curve_fit+ function) rather than our Bayesian sampling techniques. While faster, this fitting algorithm can yield poorer optimal fits and in turn more misclassifications. We will compare the performance of our classifier without redshift information to that of ALeRCE-SN to demonstrate the benefit of robust fitting techniques and careful class re-balancing.

\par One difficulty in directly comparing Superphot+ with ALeRCE-SN is that the latter only outputs pseudo-probabilities for four labels: SN Ia, SN Ib/c, SN II, and SLSN. It is unclear how we should treat our Type IIn true and predicted labels when calculating agreement between the two classifiers. During comparison, we only consider the events in our dataset also labeled by ALeRCE-SN, and we exclude all spectroscopic SNe IIn. After these cuts, we are left with \numSNAlerce\ events (\fracSNAlerce) from our dataset. This includes events that are not spectroscopically SNe IIn but are photometrically labeled as SNe IIn by Superphot+; for these events we instead use the label with second highest pseudo-probability. There are \numAlercePredIIn\ such light curves, with \numAlercePredIInRepredII, \numAlercePredIInRepredSLSN, \numAlercePredIInRepredIa, and \numAlercePredIInRepredIa\ events relabeled as predicted SNe II, SLSNe-I, SNe Ia, and SNe Ib/c, respectively. After these alterations, we can condense our five-class confusion matrices into four-class confusion matrices for direct comparison with ALeRCE-SN. Derived metrics will be underestimates compared to if we had trained a model without SN IIn labels, as our five-class model sacrifices some performance in all other classes to balance SN IIn performance.

\par The four-class confusion matrices for ALeRCE-SN and Superphot+ are shown in Figure~\ref{fig:four_type}. Using the shared dataset, Superphot+ has a F$_1$-score of \Ffourclass, which is better than ALeRCE-SN's F$_1=$ \Falerce. ALeRCE-SN tends to classify light curves with unconstrained fits as SLSNe-I, which drops its class-averaged purity. Superphot+ instead defaults to SN II or SN Ia labels when unsure.

\par A potential contributor to reduced ALeRCE-SN performance could be its inclusion of an absolute $t_0$ fit parameter, which is defined relative to a light curve's first observation. The first detection of incomplete or noisy light curves would be significantly offset from time of SN explosion, making the corresponding $t_0$ values uninterpretable and therefore biasing ALeRCE-SN's classifications. Superphot+, on the other hand, does not use $t_{0,r}$ at all (only a time delay between bands), so the classifier's behavior when applied to incomplete light curves is more predictable. This effect would be magnified when classifying the photometric dataset, which has more partial light curves.

%% file: sec8_newlabels.tex
\section{New ZTF Photometric Classifications}\label{sec:newlabels}

\subsection{Photometric Labels from Superphot+}
\par In this section, we apply Superphot+, trained without redshift information, to \numSNPhot\ SN-like light curves that have not been spectroscopically classified. Light curves in this ``photometric'' dataset, as collated in Section~\ref{subsec:phot}, pass the quality cuts described in Section~\ref{sec:pruning} and the reduced chi-squared fit cut described in Section~\ref{subsec:chisq}. The first few classification probabilities are shown in Table~\ref{table:new_classes}; a full version of this table is available as a machine-readable table in the online version and on Zenodo \citep{spp_zenodo}.

\par Superphot+ classifies \predFracIa\% of the photometric dataset as SNe Ia, \predFracII\% as SNe II, \predFracIIn\% as SNe IIn, \predFracSLSNI\% as SLSNe-I, and \predFracIbc\% as SNe Ib/c. The predicted SN Ia fraction is lower and the SLSN-I/SN IIn/SN Ib/c fractions are higher than those of the spectroscopic dataset. For example, we only expect \fracSLSN\% predicted SLSNe-I and \fracIbc\% predicted SNe Ibc in order to match the spectroscopic class fractions. If we assume the photometric dataset's true class breakdown matches the spectroscopic's true class breakdown (which may not be the case), we conclude that Superphot+ classifies many true SNe Ia as other classes, lowering CC SN purities and SN Ia completeness. Including redshift information as described in Section~\ref{sec:redshift} yields a negligible change in photometric class fractions, so we do not also analyze that variant here.

\par In Figure~\ref{fig:class_fractions}, we compare Superphot+'s spectroscopic and photometric class fractions with those from other SN datasets: the Pan-STARRS Medium-Deep Survey (PS-MDS) subset used in ~\cite{superphot}, and the Young Supernova Experiment Data Release 1 (YSE-DR1,~\citealt{yse_dr1}), which both use measurements from the Pan-STARRS telescopes. These Pan-STARRS fractions are quite similar, with the main difference being YSE-DR1's increased SN II (\fracII\% vs 22.8\%) and decreased SLSN-I (\fracSLSN\% vs 0.4\%) fractions. Because our spectroscopic sample is dominated by ZTF BTS SNe, its class breakdown is quite similar to that of ZTF BTS, as detailed in Section~\ref{subsec:reduced_datasets}. Therefore, we do not include a separate column for ZTF BTS in the figure.

 \subsection{Comparison with ALeRCE-SN's Predictions}

\par We also apply ALeRCE-SN to our photometric dataset for comparison, adding the resulting fractions to Figure~\ref{fig:class_fractions}. Because the photometric dataset is collated from ALeRCE's top-level predictions, every event in this dataset has been labeled by ALeRCE-SN. ALeRCE-SN classifies \alercePredSLSN\% of the photometric set as SLSNe, which is a higher fraction than Superphot+ but about equal to Superphot+'s combined SLSN-I and SN IIn predicted fraction. This is not surprising given ALeRCE-SN's low SLSN purity within the spectroscopic dataset, and ALeRCE-SN potentially labeling SN IIn contenders as SLSNe. Both photometric class compositions have fewer SNe Ia than expected in the photometric dataset. ALeRCE also underclassifies objects as SNe II (\alercePredII\%) compared to the spectroscopic \fracII\%.
 
\par We can now compare Superphot+'s and ALeRCE-SN's agreement when labeling the spectroscopic and photometric datasets. First, we generate the expected agreement matrix from each classifier's spectroscopic confusion matrix, as derived in Appendix B of ~\cite{superphot}. This predicts classification consistency assuming the two classifiers' latent spaces are completely independent:
\begin{equation}
    A = P^T_{\mathrm{Superphot+}}C_{\mathrm{ALeRCE-SN}}
\end{equation} 
Here, $P$ is the purity matrix and $C$ is the completeness matrix. The expected agreement score, which is simply the fraction of all samples expected to receive the same label from both classifiers, is \expectedAgreement.

\par Next, we generate true agreement matrices from our reduced spectroscopic and photometric datasets, which compare how both classifiers actually labeled events. The agreement matrices are shown in Figure~\ref{fig:true_agreement_phot}. The true agreement score for the spectroscopic dataset is \agreementSpec, which is much higher than the expected agreement score. All agreement scores are higher than expected, with the largest improvements among SNe Ia and SNe II. This improved agreement matches expectations, as the features used for classification are very much not independent; the same model is used to generate fit parameters for both pipelines.

\par To generate the agreement matrix for the photometric dataset, we first re-allocate Superphot+'s SN IIn predictions just as we did in Section~\ref{sec:alerce}. This reassigns \numAlercePhotPredIIn\ light curves to other predicted classes, mainly SNe II and SLSNe. The resulting photometric agreement matrix, with an overall agreement score of \agreementPhot, is shown in Figure~\ref{fig:true_agreement_phot} (bottom). This agreement matrix closely matches the expected agreement matrix, and somewhat mirrors the true agreement matrix for the spectroscopic set, with many ALeRCE-SN SLSNe-I classified as SNe II by Superphot+. However, the photometric agreement matrix magnifies the SLSN versus SN II disagreement we see in the spectroscopic agreement matrix. This could be interpreted as an ALeRCE-SN bias towards SLSN predictions or a Superphot+ bias towards SN II predictions, especially for uncertain light curves. Seeing how Superphot+ does not often assign SN II labels to partial light curves (see Section~\ref{sec:realtime}), an ALeRCE-SN SLSN bias is more likely. There is similar agreement within samples ALeRCE-SN labels as SNe Ib/c compared to the spectroscopic dataset.

\subsection{Correcting Class Fractions for Spectroscopic Bias}

\par Like any classifier, Superphot+ has its biases, and we can use those biases from classification of the spectroscopic dataset to ``correct'' our photometric dataset's class fractions. This will better inform us on potential systematic differences between the spectroscopic and photometric datasaets. We can correct our photometric class fractions by considering the fraction of each predicted class that belongs to other classes (i.e. the per-class purities). For example, the purity matrix from Figure \ref{fig:no_redshift_cm} demonstrates that of the spectroscopic SNe classified as Type IIn, \fracIInPredIIn\ are true SNe IIn, \fracIIPredIIn\ are true SNe II, and \fracIaPredIIn\ are true SNe Ia. We can use these values to adjust the corresponding class fractions in the photometric dataset (following \citealt{superphot,superraenn,yse_dr1}):
\begin{equation}
    \mathrm{frac}_{i, corr} = \sum_j \mathrm{frac}_j P_{i,j}
\end{equation}
Here, $P_{i,j}$ is the $i$th row and $j$th column of our purity matrix, or the fraction of SNe predicted to be class $j$ that are actually from class $i$. Computing this re-allocation for all five classes, we can convert our original class fractions to new, ``corrected'' class fractions. The new fractions better account for the biases that we know exist in our classifier. If the corrected photometric class fractions still differ from the spectroscopic class fractions, then there are potentially additional biases at play. Either the differing datasets are causing our classifier to behave differently than expected, or the spectroscopic and photometric class fractions intrinsically differ.

\par The results of these corrections are shown in Figure~\ref{fig:class_fractions}. Correcting Superphot+'s biases increases the SN Ia class fraction to \corrPredFracIa\%, closer to the spectroscopic SN Ia class fraction. We also compute corrections for ALeRCE-SN, which also brings its photometric fractions closer to the spectroscopic class fractions. For both classifiers, SNe II and SLSNe-I are still overrepresented among the corrected photometric fractions, while SNe Ia are underrepresented. The corrected ALeRCE-SN SLSNe-I fraction is still double that of the spectroscopic dataset. We note that while correcting for expected biases shifts both of our photometric fractions closer to the spectroscopic sample's, especially for ALeRCE-SN, we cannot use that information to modify individual predictions. Therefore, while ALeRCE-SN's corrected ratios are closer to the spectroscopic dataset's ratios (excluding SNe IIn), Superphot+ has less disparate uncorrected fractions among SLSNe-I and SNe II and therefore less biased individual classifications.

%% file: sec9_conclusions.tex
\section{Conclusions and Discussion}\label{sec:conclusion}
\par We have presented a novel photometric classifier, Superphot+, that does not require redshift information to assign one of five SN labels. We apply Superphot+ to ZTF light curves, but emphasize its easy adaptation to other current and future surveys. Superphot+ uses nested sampling and stochastic variational inference to rapidly fit light curves in all bands simultaneously, using correlated priors to optimize fits for inconsistently sampled light curves. These fit parameters are then oversampled to correct for class imbalances, and used as training features for a gradient-boosted machine. On the ZTF dataset of \numSNSpec\ spectroscopically classified events, Superphot+ achieves a class-averaged F$_1$-score of \FNoRedshift. We explore the addition of redshift information within Superphot+, finding better overall performance with F$_1 = $ \FRedshift, but worse performance when only considering high-confidence predictions. Superphot+'s performance exceeds or matches previous classifiers that require redshift information, and it outperforms ALeRCE's light curve classifier. This makes Superphot+, when trained on our ZTF dataset, the best-performing publicly available pipeline that does not require redshifts.

\par We assign new photometric labels to \numSNPhot\ SN-like ZTF light curves without spectroscopic labels, classifying \predFracIa\% as SNe Ia, \predFracII\% as SNe II, \predFracIIn\% as SNe IIn, \predFracSLSNI\% as SLSNe-I, and \predFracIbc\% as SNe Ib/c. Correcting for classifier bias updates these fractions to \corrPredFracIa\% SNe Ia, \corrPredFracII\% SNe II, \corrPredFracIIn\% SNe IIn, \corrPredFracSLSNI\% SLSNe-I, and \corrPredFracIbc\% SNe Ib/c. Before corrections, Superphot+ and ALeRCE-SN assign equal predicted SN Ib/c class fractions, and Superphot+ assigns SN IIn and SLSN-I labels approximately as frequently combined as ALeRCE-SN assigns SLSN labels. There is evidence of intrinsic differences within our spectroscopic and photometric datasets, supported by the lower than expected SN Ia prevalence from both classifiers.

\par Non-parametric light curve modeling is becoming more prevalent in the literature due to the computational bottleneck of Bayesian fitting to parametric models. However, with the use of stochastic variational inference and correlated priors, we have demonstrated that Superphot+ is computationally fast enough to keep up with both ZTF and expected LSST alert streams. In addition, parametric fitting benefits from enforcing light curve structure \textit{a priori}, improving modeling of very sparsely sampled light curves. Therefore, we argue that parametric classifiers are viable for the Rubin era, and we recommend Superphot+ for real-time ZTF and LSST photometric classification. 

\par Superphot+ has been integrated as a filter of the ANTARES alert broker \citep{antares}, though we have not yet incorporated the early-phase variant detailed in Section~\ref{sec:realtime}. In place of a top-level classifier to isolate SNe, we cross-check star catalogs and ignore light curves that were first observed over one year ago. This will undoubtedly lead to classification of non-SNe like AGNs and TDEs, but our existing cuts sufficiently reduce nightly SN counts such that we can visually inspect ``strange'' events. We save the fit parameters as light curve properties on ANTARES, enabling further downstream tasks like anomaly detection or realtime inference. We leave this exploration for future work.

\par In addition to the ZTF ANTARES filter detailed above, Superphot+ can readily be applied to LSST-like data when the survey begins in 2025. Indeed, the empirical fitting and subsequent classification described in this work has already been modified for six-band simulated LSST data to be used in the Extended LSST Astronomical Time-series Classification Challenge (ELASTICC). While a simple, ALeRCE-like top-level classifier was designed for this challenge, it has not been optimized to the same extent as Superphot+'s SN classification. The results of the ELASTICC challenge and Superphot+'s application to LSST-like datastreams is left to future work.

%% file: appendices.tex
\appendix

\section{Light Curve Clipping}\label{sec:clipping}

\par Many ZTF light curves suffer from improper template subtraction, leading to excessive constant flux datapoints long after the light curve has set. To remove these data points, we first calculate the ``maximum'' absolute slope of the light curve by drawing a line connecting the maximum flux measurement to the last observation of the light curve. If there is a chunk of datapoints at the end of the light curve with a sufficiently flat slope relative to this maximum slope, then we assume it to be an artifact of template subtraction, and remove it. To accomplish this, we start from the last observation and find the earliest observation such that the absolute slope between these points is less than 0.2 times the value of the ``maximum'' slope. The 0.2 cutoff was determined empirically by experimenting with various cutoff slopes. We then remove the entire section of data points that follows the observation. This process is done separately for the $g$- and $r$-bands, as the differences in band amplitudes correspond to differences in maximum slope from the last data point.

\par An example of this method applied to a SLSN-I is shown in Figure~\ref{fig:lc_clip_demo}. Due to improper template image subtraction, a constant non-zero flux is measured in both bands for an additional 1500 days after the SN's fall. As our fit model does not allow for a vertical offset, the incorrectly subtracted tail will lead to poorer fit parameters and artificially high chi-squared values. Our light curve clipping method successfully removes the tails (triangular datapoints) by determining that the flux change within the tail is substantially less than the flux change from peak to the last datapoint.

\par We do note that one downside of this method is that, in the case of transients with rapid declines, our light curve clipping strategy may remove a good fraction of post-peak points. For example, the radioactive tails of SN IIP light curves following plateau are sometimes excessively clipped. However, visual inspection shows that enough points are maintained to adequately constrain the fall timescale. Similarly, SNe Ia exhibit secondary ``bumps" in the $r$-band (e.g. see \citealt{Kasen_2006} for more on this well-known phenomena), which are sometimes clipped. However, this works in our favor, as secondary bumps not removed are often incorrectly fit with a plateau, which leads to misclassification of SNe Ia as SNe II. Finally, there is the potential for excessive clipping following spurious bright datapoints, but we do not see this in practice among the transients that pass our quality cuts.

\par Our procedure clips at least one datapoint from \numLCclipped\ out of \numBeforeProcessing\  light curves (45\%) in our unpruned spectroscopic dataset. The histogram of number of points clipped in each band among those clipped light curves is shown in Figure~\ref{fig:clip_hist}. While most light curves have less than ten points clipped per band, there is a long tail that extends beyond the figure up to a couple hundred. The light curves with an excessive number of points removed are those showing faulty template subtraction, with many points of constant flux extending out hundreds of days. The median number of points clipped is six in the $g$-band and five in the $r$-band. Data quality cuts are applied to the both datasets after light curves are clipped. Only analyzing spectroscopic light curves that pass our quality cuts yields a near identical number of clipped points distribution, meaning that we are not preferentially removing clipped light curves from the dataset. In contrast, only 24\% of the photometric dataset is clipped, with the median number of clipped points in either bands within this subset being four. This is expected, as light curves without spectroscopic followup tend to be more sparsely sampled and therefore have less excess points long after peak to be clipped.

\section{Sampler Selection}\label{sec:sampler}

While gradient-descent based optimizers (e.g., the Levenberg-Marquardt algorithm) can be used to rapidly obtain best fit parameters, the resulting fits are often suboptimal, and we cannot properly incorporate the $\sigma_\mathrm{extra}$ parameter during fitting. In contrast, traditional Metropolis-Hastings Markov chain Monte Carlo (MCMC) methods better explore the posterior space but take longer to converge and struggle computationally with the piecewise discontinuity at $t = t_0 + \gamma$. Therefore, to optimize the computational efficiency of our pipeline while still maintaining accuracy, we explore a variety of alternate sampling techniques to empirically fit our light curves. These include:
\begin{itemize}
    \item Importance nested sampling \citep{Feroz_2019}: Nested sampling uses shrinking ellipsoids to constrain regions of highest posterior density. Importance nested sampling uses information from previous as well as current sampled points to speed up convergence. In our code, we assume unimodality of the posterior space and use a single constraining ellipsoid \citep{single_ellipsoid}. We sample each light curve with 50 live points, 5,000 maximum iterations, and a stopping criterion of $\Delta \log Z \leq 0.5$. New live points are sampled with random walks from previous points \citep{random_walks}, which is more effective than uniform or slice sampling for our problem's dimensionality.
    \item No U-Turn Sampler (NUTS; \citealt{hoffman2011nouturn}): NUTS is a Hamiltonian Monte Carlo \citep{hmc} sampler, which creates an analog between finding posterior density maxima and minimizing the potential energy of a Hamiltonian system. This technique allows efficient and sometimes large steps across the parameter space. It makes no unimodality assumptions. The original Superphot~\citep{superphot} uses PyMC3's implementation of NUTS \citep{pymc3}.
    \item Stochastic variational inference (SVI; \citealt{hoffman2013stochastic}): SVI approximates the posterior by assuming an approximate posterior shape \textit{a priori}, which in our case is a multidimensional Gaussian. Our loss function is the negative of the evidence lower bound, which, when maximized, minimizes the disparity between our approximate and the true posterior distributions. Approximating the posteriors in this way significantly speeds up convergence compared to the other two methods, but SVI also tends to underestimate the variance of the marginal distributions.
    \item MIGRAD \citep{migrad}: MIGRAD is a very fast variable metric method, which iteratively approximates the negative log likelihood distribution as a multi-dimensional quadratic function. It performs very well near the true solution, but is prone to get stuck in local minima.
    \item CERES \citep{ceres}: CERES, like \verb+scipy.curve_fit+, uses the Levenberg-Marquardt algorithm (a combination of the Gauss-Newton method and gradient descent) to iteratively minimize the least-squares error of each fit. While the fastest of the algorithms, it is very prone to getting stuck in local minima or high-loss regions. This algorithm does not incorporate any $\sigma_\mathrm{extra}$ parameters, or prior distribution information (except for the parameter limits, and means for the initial guess). We include it here to demonstrate the shortcomings of gradient-descent based optimizers for our problem.
\end{itemize}

\par Our nested sampling script is implemented using the Python package \texttt{dynesty} \citep{speagle_2020}, whereas the NUTS and SVI algorithms are implemented using the \texttt{numpyro} package \citep{phan2019composable, bingham2019pyro}. The latter uses the JAX \citep{jax_paper} backend to speed up differentiation and fuse numerical functions, improving runtime drastically. MIGRAD is called through \texttt{iminuit}~\citep{iminuit}, which is a Python interface to the Minuit2 C++ library~\citep{minuit}. Minuit2 switches to a simplex algorithm~\citep{simplex} if MIGRAD cannot return a successful fit. The CERES fits are run through the \texttt{light-curve}~\citep{lc_python} package's \texttt{VillarFit} subroutine.

\par All techniques listed except MIGRAD and CERES are built on a Bayesian framework, in which the likelihood $p(\vec{f}| \vec{\theta})$, prior $\pi(\vec{\theta})$ and posterior $p(\vec{\theta} | \vec{f})$ probabilities are related following Bayes theorem:
\begin{equation}
    p(\vec{\theta} | \vec{f}) \propto p(\vec{f} | \vec{\theta}) \pi(\vec{\theta}),
\end{equation}
where $\vec{\theta}$ is the set of 14 model parameters described above, and $\vec{f}$ is the set of observed fluxes. We model the likelihood of each observation $p(f_i | \vec{\theta})$ as its probability if drawn from the multidimensional Gaussian with mean $F(t_i, \vec{\theta})$ and variance $\sigma_i^2 + \sigma_{extra}^2$. Here, $F(t_i, \vec{\theta})$ is the model flux calculated for parameters $\vec{\theta}$ at time $t_i$.

\par Bayesian techniques tend to be slower than gradient-based methods as they attempt to map out the entire posterior probability space rather than just find a nearby minimum within the space. However, they do tend to better handle problems with joint parameter constraints, which applies to our piecewise model. To ensure the piecewise transition happens after a (potentially infinitesimally small) plateau region, we enforce that the derivative of the first piecewise portion be $\leq 0$ at $t - t_0 = \gamma$. This works out to:
\begin{equation}
    e^{\frac{-\gamma}{\tau_{\mathrm{rise}}}} \times \Big(\frac{1}{\beta} - \tau_{\mathrm{rise}} - \gamma\Big) \leq \tau_{\mathrm{rise}}
\end{equation}
We also ensure that the derivative of the second piecewise portion is more negative than that of the first portion at $t - t_0 = \gamma$, to properly model a post-plateau drop off. This criterion simplifies to:
\begin{equation}
    \gamma \leq \frac{1 - \beta \tau_{\mathrm{fall}}}{\beta}
\end{equation}
This simultaneously ensures that the model flux is always positive. We enforce these two criteria for both bands independently, within every sampler except CERES.

\par To compare the efficiency and accuracy of each sampler's fits, we average both the mean runtime and median reduced chi-squared across twenty objects of each spectroscopic class. We use the median instead of mean reduced chi-squared value because some techniques treat each dimension independently, and sometimes a probable selection of parameters independently leads to a poor model fit jointly; using the median better handles these outlier $\chi_{\mathrm{red}}^2$ values. Additionally, if the median $\chi_{\mathrm{red}}^2$ is greater than 1.2, we consider it a ``poor'' fit, as we further justify in Section~\ref{subsec:chisq}. A comparison of the median reduced chi-squared values (among $\chi^2_{\mathrm{red}} < 1.2$ fits) and mean fitting time, as well as the fraction of poor fits, averaged across each true spectroscopic class is shown in Figure~\ref{fig:runtime_compare}. Additionally, we show how each sampler fits example light curves in Figure~\ref{fig:sampler_fit_compare}. We see that there is a clear tradeoff between runtime and fitting accuracy, with nested sampling and NUTS taking the longest yet yielding almost no poor fits. Nested sampling is about an order of magnitude faster than NUTS. Nested sampling, NUTS, and SVI have the lowest fraction of poor fits, which is expected from their Bayesian nature. Both SVI and MIGRAD run \textit{at least} an order of magnitude faster than nested sampling. SVI is limited by a one-time compilation so will be faster per fit after more consecutive fits. MIGRAD can potentially be sped up by an order of magnitude after compiling the likelihood function with numba. MIGRAD successfully fits about half the light curves, and struggles with handling the joint constraints between our model parameters. CERES seems to fail when fitting most of our light curves (returning the initial guess of the prior means), which could be due to it drawing directly from fit parameter space rather than log space, like our other samplers do.

\par Ultimately, we decide that because time is not a limiting factor for classifier training and adding new photometric labels to archival data, we use nested sampling for training and inference from archival datasets, such as the one analyzed in this paper. However, nested sampling proves too slow for consistent realtime fitting and classification, especially in anticipation of the LSST alert stream. Within the publicly available ANTARES filter, we use SVI as the default fitting method but switch to nested sampling on a per-light curve basis if the SVI fit is poor. We leave all five samplers as available options in the Superphot+ codebase.

\section{Bayesian Oversampling to Balance Classifier Training}\label{sec:oversampling}

\par To balance training across our five classes, we explore using only the median fit parameters in conjunction with (1) multiplying the loss contribution from each light curve by a class weight, or (2) using synthetic minority oversampling techniques (SMOTE, ~\cite{Chawla_2002}; similar to \citealt{superphot, superraenn}) to generate a balanced training set. Alternatively, we try a Bayesian oversampling approach where we draw multiple parameter sets from the posterior distribution of each minority-class light curve, and treat those draws as independent classifier inputs during training. The number of fits used per light curve depends on the relative training set abundance of that light curve classification, with the classifier using more parameter sets per light curve for less prevalent classes. This allows us to input an equal number of light curves from each class into our classifier, while also better accounting for the fit uncertainty within each light curve. The distributions of samples resulting from both SMOTE and our Bayesian oversampling technique are compared in Figure~\ref{fig:oversample_compare}, with the latter yielding improved classification accuracy and more robust performance metrics across classes and $K$-folds. We end up using 22,730 oversampled feature sets per class across our training and validation sets, corresponding to five sets of fit parameters per SN Ia light curve and \verb+round+$(22730/n)$ sets per light curve for each less prevalent class, with $n$ being the events per class in our training (or validation) set. We find that drawing more samples per SN class did not improve classifier performance, while drawing fewer samples hurts SN Ia classification. These fits are drawn from the light curve posterior space with replacement, so inputting the same set of model parameters into our classifier multiple times is possible among the less prevalent classes. Because of this, we oversample \textit{after} dividing the SNe between the training, validation, and test sets, so there is no possibility of repeat fits or fits from the same light curve appearing in both a training and test set.

\section{Classifier Architecture Selection}\label{sec:arch_compare}

\par Here, we explore two architecture options for SN classification. First, we train a multi-layer perceptron (MLP), which is a simple neural network with fully connected layers of ``neurons''; each neuron applies a nonlinear activation function to a linear combination of the input values. For simplicity, we only use dense layers in the MLP, with a constant number of neurons per hidden layer. Additionally, we include a 50\% dropout rate for each hidden layer, and a 20\% dropout rate on the input layer, as suggested by \cite{dropout}. Here, dropout refers to randomly masking out neuron values during training, which prevents a small subset of nodes from being the main determinant for discerning specific classes. As a result, nodes more evenly contribute to discerning important features within the latent space. The dropout applied to the input layer prevents the network from relying too heavily on specific fit parameters; instead, it more evenly weighs information from multiple regions of the light curve. This is important for accurately classifying partial light curves. To optimize the network architecture (number of hidden layers and neurons per layer), we perform a grid search over one to five hidden layers and 4 to 256 neurons per layer in intervals of $2^n$, and calculate the validation set's class-averaged F$_1$-score for each combination. We find that three hidden layers with 128 neurons each maximize the combined validation F$_1$-score across all $K$-folds. Across 10 folds, the MLP's F$_1$-score spans \FMLP\ across an 80\% confidence interval. For context, this is slightly lower than Superphot's reported F$_1$-score, which uses redshift information.

\par Next, we train a gradient-boosted machine (GBM) using LightGBM~\citep{lightgbm}. We attempt to optimize LightGBM's hyperparameters using a grid search, and find negligible performance improvements across different hyperparameter configurations. The results in this paper are obtained using the DART boosting strategy, \texttt{goss} sampling strategy, a max tree depth of 5, max number of leaves of 20, a regularization of 5, and 250 estimators. The LightGBM trains an order of magnitude faster than the MLP, and yields an F$_1$-score 80\% confidence interval of \FNoRedshift. While the median F$_1$ score is higher than that of the MLP, the two confidence intervals overlap (i.e. the worst-performing $K$-folds for the GBM report a lower F$_1$-score than the best-performing $K$-folds for the MLP). Because the $K$-folds were selected independently for the MLP and GBM, we thus argue that the lower MLP performance can potentially be attributed to the way the training set was randomly divided among folds for each architecture. Even so, we finalize the LightGBM model as the default classifier, as it is two orders of magnitude faster to train compared to the MLP.

%% file: superphot_plus.bbl
\begin{thebibliography}{}
\expandafter\ifx\csname natexlab\endcsname\relax\def\natexlab#1{#1}\fi
\providecommand{\url}[1]{\href{#1}{#1}}
\providecommand{\dodoi}[1]{doi:~\href{http://doi.org/#1}{\nolinkurl{#1}}}
\providecommand{\doeprint}[1]{\href{http://ascl.net/#1}{\nolinkurl{http://ascl.net/#1}}}
\providecommand{\doarXiv}[1]{\href{https://arxiv.org/abs/#1}{\nolinkurl{https://arxiv.org/abs/#1}}}

\bibitem[{Agarwal {et~al.}(2023)Agarwal, Mierle, \& Team}]{ceres}
Agarwal, S., Mierle, K., \& Team, T. C.~S. 2023, {Ceres Solver}, 2.2.
\newblock \url{https://github.com/ceres-solver/ceres-solver}

\bibitem[{{Aleo} {et~al.}(2023){Aleo}, {Malanchev}, {Sharief}, {Jones},
  {Narayan}, {Foley}, {Villar}, {Angus}, {Baldassare}, {Bustamante-Rosell},
  {Chatterjee}, {Cold}, {Coulter}, {Davis}, {Dhawan}, {Drout}, {Engel},
  {French}, {Gagliano}, {Gall}, {Hjorth}, {Huber}, {Jacobson-Gal{\'a}n},
  {Kilpatrick}, {Langeroodi}, {Macias}, {Mandel}, {Margutti}, {Matasi{\'c}},
  {McGill}, {Pierel}, {Ramirez-Ruiz}, {Ransome}, {Rojas-Bravo}, {Siebert},
  {Smith}, {de Soto}, {Stroh}, {Tinyanont}, {Taggart}, {Ward}, {Wojtak},
  {Auchettl}, {Blanchard}, {de Boer}, {Boyd}, {Carroll}, {Chambers},
  {DeMarchi}, {Dimitriadis}, {Dodd}, {Earl}, {Farias}, {Gao}, {Gomez},
  {Grayling}, {Grillo}, {Hayes}, {Hung}, {Izzo}, {Khetan}, {Kolborg},
  {Law-Smith}, {LeBaron}, {Lin}, {Luo}, {Magnier}, {Matthews}, {Mockler},
  {O'Grady}, {Pan}, {Politsch}, {Raimundo}, {Rest}, {Ridden-Harper}, {Sarangi},
  {Schr{\o}der}, {Smartt}, {Terreran}, {Thorp}, {Vazquez}, {Wainscoat}, {Wang},
  {Wasserman}, {Yadavalli}, {Yarza}, {Zenati}, \& {Young Supernova
  Experiment}}]{yse_dr1}
{Aleo}, P.~D., {Malanchev}, K., {Sharief}, S., {et~al.} 2023, \apjs, 266, 9,
  \dodoi{10.3847/1538-4365/acbfba}

\bibitem[{ALeRCE(2022)}]{alerce_docs}
ALeRCE. 2022, ALeRCE client¶.
\newblock \url{https://alerce.readthedocs.io/en/latest/}

\bibitem[{ANTARES(2024)}]{antares_docs}
ANTARES. 2024, NSF noirlab / community science and data center / antares /
  antares · GITLAB.
\newblock \url{https://gitlab.com/nsf-noirlab/csdc/antares/antares}

\bibitem[{{Astropy Collaboration} {et~al.}(2013){Astropy Collaboration},
  {Robitaille}, {Tollerud}, {Greenfield}, {Droettboom}, {Bray}, {Aldcroft},
  {Davis}, {Ginsburg}, {Price-Whelan}, {Kerzendorf}, {Conley}, {Crighton},
  {Barbary}, {Muna}, {Ferguson}, {Grollier}, {Parikh}, {Nair}, {Unther},
  {Deil}, {Woillez}, {Conseil}, {Kramer}, {Turner}, {Singer}, {Fox}, {Weaver},
  {Zabalza}, {Edwards}, {Azalee Bostroem}, {Burke}, {Casey}, {Crawford},
  {Dencheva}, {Ely}, {Jenness}, {Labrie}, {Lim}, {Pierfederici}, {Pontzen},
  {Ptak}, {Refsdal}, {Servillat}, \& {Streicher}}]{astropy:2013}
{Astropy Collaboration}, {Robitaille}, T.~P., {Tollerud}, E.~J., {et~al.} 2013,
  \aap, 558, A33, \dodoi{10.1051/0004-6361/201322068}

\bibitem[{{Astropy Collaboration} {et~al.}(2018){Astropy Collaboration},
  {Price-Whelan}, {Sip{\H{o}}cz}, {G{\"u}nther}, {Lim}, {Crawford}, {Conseil},
  {Shupe}, {Craig}, {Dencheva}, {Ginsburg}, {Vand erPlas}, {Bradley},
  {P{\'e}rez-Su{\'a}rez}, {de Val-Borro}, {Aldcroft}, {Cruz}, {Robitaille},
  {Tollerud}, {Ardelean}, {Babej}, {Bach}, {Bachetti}, {Bakanov}, {Bamford},
  {Barentsen}, {Barmby}, {Baumbach}, {Berry}, {Biscani}, {Boquien}, {Bostroem},
  {Bouma}, {Brammer}, {Bray}, {Breytenbach}, {Buddelmeijer}, {Burke},
  {Calderone}, {Cano Rodr{\'\i}guez}, {Cara}, {Cardoso}, {Cheedella}, {Copin},
  {Corrales}, {Crichton}, {D'Avella}, {Deil}, {Depagne}, {Dietrich}, {Donath},
  {Droettboom}, {Earl}, {Erben}, {Fabbro}, {Ferreira}, {Finethy}, {Fox},
  {Garrison}, {Gibbons}, {Goldstein}, {Gommers}, {Greco}, {Greenfield},
  {Groener}, {Grollier}, {Hagen}, {Hirst}, {Homeier}, {Horton}, {Hosseinzadeh},
  {Hu}, {Hunkeler}, {Ivezi{\'c}}, {Jain}, {Jenness}, {Kanarek}, {Kendrew},
  {Kern}, {Kerzendorf}, {Khvalko}, {King}, {Kirkby}, {Kulkarni}, {Kumar},
  {Lee}, {Lenz}, {Littlefair}, {Ma}, {Macleod}, {Mastropietro}, {McCully},
  {Montagnac}, {Morris}, {Mueller}, {Mumford}, {Muna}, {Murphy}, {Nelson},
  {Nguyen}, {Ninan}, {N{\"o}the}, {Ogaz}, {Oh}, {Parejko}, {Parley}, {Pascual},
  {Patil}, {Patil}, {Plunkett}, {Prochaska}, {Rastogi}, {Reddy Janga},
  {Sabater}, {Sakurikar}, {Seifert}, {Sherbert}, {Sherwood-Taylor}, {Shih},
  {Sick}, {Silbiger}, {Singanamalla}, {Singer}, {Sladen}, {Sooley},
  {Sornarajah}, {Streicher}, {Teuben}, {Thomas}, {Tremblay}, {Turner},
  {Terr{\'o}n}, {van Kerkwijk}, {de la Vega}, {Watkins}, {Weaver}, {Whitmore},
  {Woillez}, {Zabalza}, \& {Astropy Contributors}}]{astropy:2018}
{Astropy Collaboration}, {Price-Whelan}, A.~M., {Sip{\H{o}}cz}, B.~M., {et~al.}
  2018, \aj, 156, 123, \dodoi{10.3847/1538-3881/aabc4f}

\bibitem[{{Astropy Collaboration} {et~al.}(2022){Astropy Collaboration},
  {Price-Whelan}, {Lim}, {Earl}, {Starkman}, {Bradley}, {Shupe}, {Patil},
  {Corrales}, {Brasseur}, {N{"o}the}, {Donath}, {Tollerud}, {Morris},
  {Ginsburg}, {Vaher}, {Weaver}, {Tocknell}, {Jamieson}, {van Kerkwijk},
  {Robitaille}, {Merry}, {Bachetti}, {G{"u}nther}, {Aldcroft},
  {Alvarado-Montes}, {Archibald}, {B{'o}di}, {Bapat}, {Barentsen}, {Baz{'a}n},
  {Biswas}, {Boquien}, {Burke}, {Cara}, {Cara}, {Conroy}, {Conseil}, {Craig},
  {Cross}, {Cruz}, {D'Eugenio}, {Dencheva}, {Devillepoix}, {Dietrich},
  {Eigenbrot}, {Erben}, {Ferreira}, {Foreman-Mackey}, {Fox}, {Freij}, {Garg},
  {Geda}, {Glattly}, {Gondhalekar}, {Gordon}, {Grant}, {Greenfield}, {Groener},
  {Guest}, {Gurovich}, {Handberg}, {Hart}, {Hatfield-Dodds}, {Homeier},
  {Hosseinzadeh}, {Jenness}, {Jones}, {Joseph}, {Kalmbach}, {Karamehmetoglu},
  {Ka{l}uszy{'n}ski}, {Kelley}, {Kern}, {Kerzendorf}, {Koch}, {Kulumani},
  {Lee}, {Ly}, {Ma}, {MacBride}, {Maljaars}, {Muna}, {Murphy}, {Norman},
  {O'Steen}, {Oman}, {Pacifici}, {Pascual}, {Pascual-Granado}, {Patil},
  {Perren}, {Pickering}, {Rastogi}, {Roulston}, {Ryan}, {Rykoff}, {Sabater},
  {Sakurikar}, {Salgado}, {Sanghi}, {Saunders}, {Savchenko}, {Schwardt},
  {Seifert-Eckert}, {Shih}, {Jain}, {Shukla}, {Sick}, {Simpson},
  {Singanamalla}, {Singer}, {Singhal}, {Sinha}, {Sip{H{o}}cz}, {Spitler},
  {Stansby}, {Streicher}, {{{S}}umak}, {Swinbank}, {Taranu}, {Tewary},
  {Tremblay}, {Val-Borro}, {Van Kooten}, {Vasovi{'c}}, {Verma}, {de Miranda
  Cardoso}, {Williams}, {Wilson}, {Winkel}, {Wood-Vasey}, {Xue}, {Yoachim},
  {Zhang}, {Zonca}, \& {Astropy Project Contributors}}]{astropy:2022}
{Astropy Collaboration}, {Price-Whelan}, A.~M., {Lim}, P.~L., {et~al.} 2022,
  \apj, 935, 167, \dodoi{10.3847/1538-4357/ac7c74}

\bibitem[{Barbary(2017)}]{extinction_docs}
Barbary, K. 2017, extinction v0.3.0,  Zenodo, \dodoi{10.5281/zenodo.804967}

\bibitem[{Bellm {et~al.}(2018)Bellm, Kulkarni, Graham, Dekany, Smith, Riddle,
  Masci, Helou, Prince, Adams, Barbarino, Barlow, Bauer, Beck, Belicki, Biswas,
  Blagorodnova, Bodewits, Bolin, Brinnel, Brooke, Bue, Bulla, Burruss, Cenko,
  Chang, Connolly, Coughlin, Cromer, Cunningham, De, Delacroix, Desai, Duev,
  Eadie, Farnham, Feeney, Feindt, Flynn, Franckowiak, Frederick, Fremling,
  Gal-Yam, Gezari, Giomi, Goldstein, Golkhou, Goobar, Groom, Hacopians, Hale,
  Henning, Ho, Hover, Howell, Hung, Huppenkothen, Imel, Ip, Ivezi{\'{c}},
  Jackson, Jones, Juric, Kasliwal, Kaspi, Kaye, Kelley, Kowalski, Kramer,
  Kupfer, Landry, Laher, Lee, Lin, Lin, Lunnan, Giomi, Mahabal, Mao, Miller,
  Monkewitz, Murphy, Ngeow, Nordin, Nugent, Ofek, Patterson, Penprase, Porter,
  Rauch, Rebbapragada, Reiley, Rigault, Rodriguez, van Roestel, Rusholme, van
  Santen, Schulze, Shupe, Singer, Soumagnac, Stein, Surace, Sollerman, Szkody,
  Taddia, Terek, Sistine, van Velzen, Vestrand, Walters, Ward, Ye, Yu, Yan, \&
  Zolkower}]{ztf}
Bellm, E.~C., Kulkarni, S.~R., Graham, M.~J., {et~al.} 2018, Publications of
  the Astronomical Society of the Pacific, 131, 018002,
  \dodoi{10.1088/1538-3873/aaecbe}

\bibitem[{Bingham {et~al.}(2019)Bingham, Chen, Jankowiak, Obermeyer, Pradhan,
  Karaletsos, Singh, Szerlip, Horsfall, \& Goodman}]{bingham2019pyro}
Bingham, E., Chen, J.~P., Jankowiak, M., {et~al.} 2019, J. Mach. Learn. Res.,
  20, 28:1.
\newblock \url{http://jmlr.org/papers/v20/18-403.html}

\bibitem[{Blondin {et~al.}(2012)Blondin, Matheson, Kirshner, Mandel, Berlind,
  Calkins, Challis, Garnavich, Jha, Modjaz, Riess, \& Schmidt}]{blondin_2012}
Blondin, S., Matheson, T., Kirshner, R.~P., {et~al.} 2012, The Astronomical
  Journal, 143, 126, \dodoi{10.1088/0004-6256/143/5/126}

\bibitem[{Boone(2021)}]{boone_2021}
Boone, K. 2021, The Astronomical Journal, 162, 275,
  \dodoi{10.3847/1538-3881/ac2a2d}

\bibitem[{Boone \& Malanchev(2022)}]{parsnip_docs}
Boone, K., \& Malanchev, K. 2022, kboone/parsnip: v1.3.1, v1.3.1,  Zenodo,
  \dodoi{10.5281/zenodo.6980374}

\bibitem[{Bradbury {et~al.}(2018)Bradbury, Frostig, Hawkins, Johnson, Leary,
  Maclaurin, Necula, Paszke, Vander{P}las, Wanderman-{M}ilne, \&
  Zhang}]{jax2018github}
Bradbury, J., Frostig, R., Hawkins, P., {et~al.} 2018, {JAX}: composable
  transformations of {P}ython+{N}um{P}y programs, 0.3.13.
\newblock \url{http://github.com/google/jax}

\bibitem[{Bradley(1997)}]{roc}
Bradley, A.~P. 1997, Pattern Recognition, 30, 1145,
  \dodoi{https://doi.org/10.1016/S0031-3203(96)00142-2}

\bibitem[{{Branch} {et~al.}(1993){Branch}, {Fisher}, \&
  {Nugent}}]{branchnormal}
{Branch}, D., {Fisher}, A., \& {Nugent}, P. 1993, \aj, 106, 2383,
  \dodoi{10.1086/116810}

\bibitem[{Brooks {et~al.}(2011)Brooks, Gelman, Jones, \& Meng}]{hmc}
Brooks, S., Gelman, A., Jones, G., \& Meng, X.-L., eds. 2011, Handbook of
  Markov Chain Monte Carlo (Chapman and Hall/{CRC}), \dodoi{10.1201/b10905}

\bibitem[{Carrasco-Davis {et~al.}(2021)Carrasco-Davis, Reyes, Valenzuela,
  Förster, Est{\'{e} }vez, Pignata, Bauer, Reyes, S{\'{a}}nchez-S{\'{a}}ez,
  Cabrera-Vives, Eyheramendy, Catelan, Arredondo, Castillo-Navarrete,
  Rodr{\'{\i}}guez-Mancini, Ruz-Mieres, Moya, Sabatini-Gacit{\'{u}}a,
  Sep{\'{u}}lveda-Cobo, Mahabal, Silva-Farf{\'{a}}n, Camacho-I{\~{n}}iguez, \&
  Galbany}]{alerce_stamp}
Carrasco-Davis, R., Reyes, E., Valenzuela, C., {et~al.} 2021, The Astronomical
  Journal, 162, 231, \dodoi{10.3847/1538-3881/ac0ef1}

\bibitem[{{Chambers} {et~al.}(2016){Chambers}, {Magnier}, {Metcalfe},
  {Flewelling}, {Huber}, {Waters}, {Denneau}, {Draper}, {Farrow}, {Finkbeiner},
  {Holmberg}, {Koppenhoefer}, {Price}, {Rest}, {Saglia}, {Schlafly}, {Smartt},
  {Sweeney}, {Wainscoat}, {Burgett}, {Chastel}, {Grav}, {Heasley}, {Hodapp},
  {Jedicke}, {Kaiser}, {Kudritzki}, {Luppino}, {Lupton}, {Monet}, {Morgan},
  {Onaka}, {Shiao}, {Stubbs}, {Tonry}, {White}, {Ba{\~n}ados}, {Bell},
  {Bender}, {Bernard}, {Boegner}, {Boffi}, {Botticella}, {Calamida},
  {Casertano}, {Chen}, {Chen}, {Cole}, {Deacon}, {Frenk}, {Fitzsimmons},
  {Gezari}, {Gibbs}, {Goessl}, {Goggia}, {Gourgue}, {Goldman}, {Grant},
  {Grebel}, {Hambly}, {Hasinger}, {Heavens}, {Heckman}, {Henderson}, {Henning},
  {Holman}, {Hopp}, {Ip}, {Isani}, {Jackson}, {Keyes}, {Koekemoer}, {Kotak},
  {Le}, {Liska}, {Long}, {Lucey}, {Liu}, {Martin}, {Masci}, {McLean}, {Mindel},
  {Misra}, {Morganson}, {Murphy}, {Obaika}, {Narayan}, {Nieto-Santisteban},
  {Norberg}, {Peacock}, {Pier}, {Postman}, {Primak}, {Rae}, {Rai}, {Riess},
  {Riffeser}, {Rix}, {R{\"o}ser}, {Russel}, {Rutz}, {Schilbach}, {Schultz},
  {Scolnic}, {Strolger}, {Szalay}, {Seitz}, {Small}, {Smith}, {Soderblom},
  {Taylor}, {Thomson}, {Taylor}, {Thakar}, {Thiel}, {Thilker}, {Unger},
  {Urata}, {Valenti}, {Wagner}, {Walder}, {Walter}, {Watters}, {Werner},
  {Wood-Vasey}, \& {Wyse}}]{panstarrs}
{Chambers}, K.~C., {Magnier}, E.~A., {Metcalfe}, N., {et~al.} 2016, arXiv
  e-prints, arXiv:1612.05560.
\newblock \doarXiv{1612.05560}

\bibitem[{Chawla {et~al.}(2002)Chawla, Bowyer, Hall, \&
  Kegelmeyer}]{Chawla_2002}
Chawla, N.~V., Bowyer, K.~W., Hall, L.~O., \& Kegelmeyer, W.~P. 2002, Journal
  of Artificial Intelligence Research, 16, 321, \dodoi{10.1613/jair.953}

\bibitem[{Claeys {et~al.}(2011)Claeys, de~Mink, Pols, Eldridge, \&
  Baes}]{snIIb}
Claeys, J. S.~W., de~Mink, S.~E., Pols, O.~R., Eldridge, J.~J., \& Baes, M.
  2011, A\&A, 528, A131, \dodoi{10.1051/0004-6361/201015410}

\bibitem[{Davis \& Goadrich(2006)}]{roc_pr_curves2}
Davis, J., \& Goadrich, M. 2006, in Proceedings of the 23rd International
  Conference on Machine Learning, ACM, Vol.~06, \dodoi{10.1145/1143844.1143874}

\bibitem[{De~Jong {et~al.}(2019)De~Jong, Agertz, Berbel, Aird, Alexander,
  Amarsi, Anders, Andrae, Ansarinejad, Ansorge, {et~al.}}]{de20194most}
De~Jong, R.~S., Agertz, O., Berbel, A.~A., {et~al.} 2019, arXiv preprint
  arXiv:1903.02464

\bibitem[{de~Soto {et~al.}(2024)de~Soto, Villar, Berger, Gomez, Hosseinzadeh,
  Branton, Campos, DeLucchi, Kubica, Lynn, Malanchev, \& Malz}]{spp_zenodo}
de~Soto, K., Villar, A., Berger, E., {et~al.} 2024, {Superphot+: Real-Time
  Fitting and Classification of Supernova Light Curves}, 0.0.7,  Zenodo,
  \dodoi{10.5281/zenodo.10798425}

\bibitem[{Dembinski \& et~al.(2020)}]{iminuit}
Dembinski, H., \& et~al., P.~O. 2020, \dodoi{10.5281/zenodo.3949207}

\bibitem[{Dembinski {et~al.}(2023)Dembinski, Ongmongkolkul, Deil, Schreiner,
  Feickert, Burr, Watson, Rost, Pearce, Geiger, Abdelmotteleb, Desai,
  Wiedemann, Gohlke, Sanders, Drotleff, Eschle, Neste, Gorelli, Baak,
  Eliachevitch, \& Zapata}]{iminuit_docs}
Dembinski, H., Ongmongkolkul, P., Deil, C., {et~al.} 2023, scikit-hep/iminuit,
  v2.24.0,  Zenodo, \dodoi{10.5281/zenodo.8249703}

\bibitem[{Dillon {et~al.}(2017)Dillon, Langmore, Tran, Brevdo, Vasudevan,
  Moore, Patton, Alemi, Hoffman, \& Saurous}]{tfp}
Dillon, J.~V., Langmore, I., Tran, D., {et~al.} 2017, TensorFlow Distributions.
\newblock \doarXiv{1711.10604}

\bibitem[{Feroz {et~al.}(2019)Feroz, Hobson, Cameron, \& Pettitt}]{Feroz_2019}
Feroz, F., Hobson, M.~P., Cameron, E., \& Pettitt, A.~N. 2019, The Open Journal
  of Astrophysics, 2, \dodoi{10.21105/astro.1306.2144}

\bibitem[{Filippenko(1997)}]{filippenko}
Filippenko, A.~V. 1997, Annual Review of Astronomy and Astrophysics, 35, 309,
  \dodoi{10.1146/annurev.astro.35.1.309}

\bibitem[{{Filippenko}(2005)}]{Filippenko_2005}
{Filippenko}, A.~V. 2005, in Astronomical Society of the Pacific Conference
  Series, Vol. 332, The Fate of the Most Massive Stars, ed. R.~{Humphreys} \&
  K.~{Stanek}, 34, \dodoi{10.48550/arXiv.astro-ph/0412029}

\bibitem[{Fitzpatrick \& Massa(2007)}]{fitzpatrick_massa_2007}
Fitzpatrick, E.~L., \& Massa, D. 2007, The Astrophysical Journal, 663,
  320–341, \dodoi{10.1086/518158}

\bibitem[{Fletcher(1970)}]{migrad}
Fletcher, R. 1970, The Computer Journal, 13, 317,
  \dodoi{10.1093/comjnl/13.3.317}

\bibitem[{Foley {et~al.}(2013)Foley, Challis, Chornock, Ganeshalingam, Li,
  Marion, Morrell, Pignata, Stritzinger, Silverman, \& et~al.}]{snIax}
Foley, R.~J., Challis, P.~J., Chornock, R., {et~al.} 2013, The Astrophysical
  Journal, 767, 57, \dodoi{10.1088/0004-637x/767/1/57}

\bibitem[{Fremling {et~al.}(2020)Fremling, Miller, Sharma, Dugas, Perley,
  Taggart, Sollerman, Goobar, Graham, Neill, \& et~al.}]{ztf_bts}
Fremling, C., Miller, A.~A., Sharma, Y., {et~al.} 2020, The Astrophysical
  Journal, 895, 32, \dodoi{10.3847/1538-4357/ab8943}

\bibitem[{Frostig {et~al.}(2018)Frostig, Johnson, \& Leary}]{jax_paper}
Frostig, R., Johnson, M., \& Leary, C. 2018.
\newblock \url{https://mlsys.org/Conferences/doc/2018/146.pdf}

\bibitem[{Förster {et~al.}(2021)Förster, Cabrera-Vives, Castillo-Navarrete,
  Estévez, Sánchez-Sáez, Arredondo, Bauer, Carrasco-Davis, Catelan,
  Elorrieta, Eyheramendy, Huijse, Pignata, Reyes, Reyes, Rodríguez-Mancini,
  Ruz-Mieres, Valenzuela, Álvarez Maldonado, Astorga, Borissova, Clocchiatti,
  De~Cicco, Donoso-Oliva, Hernández-García, Graham, Jordán, Kurtev, Mahabal,
  Maureira, Muñoz-Arancibia, Molina-Ferreiro, Moya, Palma, Pérez-Carrasco,
  Protopapas, Romero, Sabatini-Gacitua, Sánchez, Martín, Sepúlveda-Cobo,
  Vera, \& Vergara}]{Forster_2021}
Förster, F., Cabrera-Vives, G., Castillo-Navarrete, E., {et~al.} 2021, The
  Astronomical Journal, 161, 242, \dodoi{10.3847/1538-3881/abe9bc}

\bibitem[{Gagliano {et~al.}(2023)Gagliano, Contardo, Foreman-Mackey, Malz, \&
  Aleo}]{gagliano2023}
Gagliano, A., Contardo, G., Foreman-Mackey, D., Malz, A.~I., \& Aleo, P.~D.
  2023, First Impressions: Early-Time Classification of Supernovae using Host
  Galaxy Information and Shallow Learning.
\newblock \doarXiv{2305.08894}

\bibitem[{Gagliano {et~al.}(2021)Gagliano, Narayan, Engel, \&
  and}]{Gagliano_2021}
Gagliano, A., Narayan, G., Engel, A., \& and, M. C.~K. 2021, The Astrophysical
  Journal, 908, 170, \dodoi{10.3847/1538-4357/abd02b}

\bibitem[{Gal-Yam(2012)}]{Gal_Yam_2012}
Gal-Yam, A. 2012, Science, 337, 927, \dodoi{10.1126/science.1203601}

\bibitem[{{Gal-Yam}(2021)}]{tns}
{Gal-Yam}, A. 2021, in American Astronomical Society Meeting Abstracts,
  Vol.~53, American Astronomical Society Meeting Abstracts, 423.05

\bibitem[{Gomez {et~al.}(2023{\natexlab{a}})Gomez, Berger, Blanchard,
  Hosseinzadeh, Nicholl, Hiramatsu, Villar, \& Yin}]{fleet2}
Gomez, S., Berger, E., Blanchard, P.~K., {et~al.} 2023{\natexlab{a}}, The
  Astrophysical Journal, 949, 114, \dodoi{10.3847/1538-4357/acc536}

\bibitem[{Gomez {et~al.}(2020{\natexlab{a}})Gomez, Berger, Blanchard,
  Hosseinzadeh, Nicholl, Villar, \& Yin}]{fleet}
---. 2020{\natexlab{a}}, The Astrophysical Journal, 904, 74,
  \dodoi{10.3847/1538-4357/abbf49}

\bibitem[{Gomez {et~al.}(2020{\natexlab{b}})Gomez, Berger, Blanchard,
  Hosseinzadeh, Nicholl, Villar, \& Yin}]{fleet_docs}
---. 2020{\natexlab{b}}, {FLEET Finding Luminous and Exotic Extragalactic
  Transients}, 1.0.0,  Zenodo, \dodoi{10.5281/zenodo.4013965}

\bibitem[{Gomez {et~al.}(2023{\natexlab{b}})Gomez, Villar, Berger, Gezari, van
  Velzen, Nicholl, Blanchard, \& Alexander}]{fleet_tde}
Gomez, S., Villar, V.~A., Berger, E., {et~al.} 2023{\natexlab{b}}, The
  Astrophysical Journal, 949, 113, \dodoi{10.3847/1538-4357/acc535}

\bibitem[{Graham {et~al.}(2018)Graham, Connolly, Ivezić, Schmidt, Jones,
  Jurić, Daniel, \& Yoachim}]{graham_2018}
Graham, M.~L., Connolly, A.~J., Ivezić, Z., {et~al.} 2018, The Astronomical
  Journal, 155, 1, \dodoi{10.3847/1538-3881/aa99d4}

\bibitem[{{Green}(2018)}]{dustmaps_docs}
{Green}, G. 2018, The Journal of Open Source Software, 3, 695,
  \dodoi{10.21105/joss.00695}

\bibitem[{Green {et~al.}(2023)Green, Edenhofer, Krughoff, Smith, Lenz, \&
  Malanchev}]{dustmaps_docs2}
Green, G., Edenhofer, G., Krughoff, S., {et~al.} 2023, gregreen/dustmaps:
  v1.0.11, v1.0.11,  Zenodo, \dodoi{10.5281/zenodo.8210973}

\bibitem[{Grisel {et~al.}(2024)Grisel, Mueller, Lars, Gramfort, Louppe, Fan,
  Prettenhofer, Blondel, Niculae, Nothman, Joly, Lemaitre, Estève, Vanderplas,
  du~Boisberranger, manoj kumar, Qin, Hug, Varoquaux, Layton, Jalali, Metzen,
  (Venkat)~Raghav, Schönberger, Yurchak, Jerphanion, Liu, la~Tour, Li, \&
  Lorentzen}]{sklearn_docs}
Grisel, O., Mueller, A., Lars, {et~al.} 2024, scikit-learn/scikit-learn:
  Scikit-learn 1.4.0, 1.4.0-1,  Zenodo, \dodoi{10.5281/zenodo.10532824}

\bibitem[{Harris {et~al.}(2020)Harris, Millman, van~der Walt, Gommers,
  Virtanen, Cournapeau, Wieser, Taylor, Berg, Smith, Kern, Picus, Hoyer, van
  Kerkwijk, Brett, Haldane, del R{\'{i}}o, Wiebe, Peterson,
  G{\'{e}}rard-Marchant, Sheppard, Reddy, Weckesser, Abbasi, Gohlke, \&
  Oliphant}]{numpy}
Harris, C.~R., Millman, K.~J., van~der Walt, S.~J., {et~al.} 2020, Nature, 585,
  357, \dodoi{10.1038/s41586-020-2649-2}

\bibitem[{Hinton {et~al.}(2012)Hinton, Srivastava, Krizhevsky, Sutskever, \&
  Salakhutdinov}]{dropout}
Hinton, G.~E., Srivastava, N., Krizhevsky, A., Sutskever, I., \& Salakhutdinov,
  R.~R. 2012, Improving neural networks by preventing co-adaptation of feature
  detectors,  arXiv, \dodoi{10.48550/ARXIV.1207.0580}

\bibitem[{Hoffman {et~al.}(2013)Hoffman, Blei, Wang, \&
  Paisley}]{hoffman2013stochastic}
Hoffman, M., Blei, D.~M., Wang, C., \& Paisley, J. 2013, Stochastic Variational
  Inference.
\newblock \doarXiv{1206.7051}

\bibitem[{Hoffman \& Gelman(2011)}]{hoffman2011nouturn}
Hoffman, M.~D., \& Gelman, A. 2011, The No-U-Turn Sampler: Adaptively Setting
  Path Lengths in Hamiltonian Monte Carlo.
\newblock \doarXiv{1111.4246}

\bibitem[{Hosseinzadeh {et~al.}(2022)Hosseinzadeh, Berger, Metzger, Gomez,
  Nicholl, \& Blanchard}]{Hosseinzadeh_2022}
Hosseinzadeh, G., Berger, E., Metzger, B.~D., {et~al.} 2022, The Astrophysical
  Journal, 933, 14, \dodoi{10.3847/1538-4357/ac67dd}

\bibitem[{Hosseinzadeh \& Dauphin(2021)}]{superphot_docs}
Hosseinzadeh, G., \& Dauphin, F. 2021, Superphot, v1.2.0,  Zenodo,
  \dodoi{10.5281/zenodo.5520623}

\bibitem[{Hosseinzadeh {et~al.}(2020)Hosseinzadeh, Dauphin, Villar, Berger,
  Jones, Challis, Chornock, Drout, Foley, Kirshner, \& et~al.}]{superphot}
Hosseinzadeh, G., Dauphin, F., Villar, V.~A., {et~al.} 2020, The Astrophysical
  Journal, 905, 93, \dodoi{10.3847/1538-4357/abc42b}

\bibitem[{Hsu {et~al.}(2023)Hsu, Blanchard, Berger, \& Gomez}]{hsu2023slsn}
Hsu, B., Blanchard, P.~K., Berger, E., \& Gomez, S. 2023, An Extensive
  $\textit{Hubble Space Telescope}$ Study of the Offset and Host Light
  Distributions of Type I Superluminous Supernovae.
\newblock \doarXiv{2308.07271}

\bibitem[{Hunter(2007)}]{matplotlib}
Hunter, J.~D. 2007, Computing In Science \& Engineering, 9, 90

\bibitem[{{Imbalanced-learn}(2024)}]{imblearn_docs}
{Imbalanced-learn}. 2024, imbalanced-learn documentation.
\newblock \url{https://imbalanced-learn.org/stable/}

\bibitem[{Ivezić \& {LSST Science Collaboration}(2018)}]{lsst_srd}
Ivezić, Z., \& {LSST Science Collaboration}. 2018, The LSST System Science
  Requirements Document - Rubin Observatory.
\newblock \url{https://docushare.lsst.org/docushare/dsweb/Get/LPM-17}

\bibitem[{James \& Roos(1975)}]{minuit}
James, F., \& Roos, M. 1975, Comput. Phys. Commun., 10, 343,
  \dodoi{10.1016/0010-4655(75)90039-9}

\bibitem[{Jones {et~al.}(2017)Jones, Scolnic, Riess, Kessler, Rest, Kirshner,
  Berger, Ortega, Foley, Chornock, Challis, Burgett, Chambers, Draper,
  Flewelling, Huber, Kaiser, Kudritzki, Metcalfe, Wainscoat, \&
  Waters}]{Jones_2017}
Jones, D.~O., Scolnic, D.~M., Riess, A.~G., {et~al.} 2017, The Astrophysical
  Journal, 843, 6, \dodoi{10.3847/1538-4357/aa767b}

\bibitem[{Kasen(2006)}]{Kasen_2006}
Kasen, D. 2006, The Astrophysical Journal, 649, 939, \dodoi{10.1086/506588}

\bibitem[{Ke {et~al.}(2017)Ke, Meng, Finley, Wang, Chen, Ma, Ye, \&
  Liu}]{lightgbm}
Ke, G., Meng, Q., Finley, T., {et~al.} 2017, in Advances in Neural Information
  Processing Systems, ed. I.~Guyon, U.~V. Luxburg, S.~Bengio, H.~Wallach,
  R.~Fergus, S.~Vishwanathan, \& R.~Garnett, Vol.~30 (Curran Associates, Inc.).
\newblock
  \url{https://proceedings.neurips.cc/paper_files/paper/2017/file/6449f44a102fde848669bdd9eb6b76fa-Paper.pdf}

\bibitem[{{Kessler} {et~al.}(2019){Kessler}, {Narayan}, {Avelino}, {Bachelet},
  {Biswas}, {Brown}, {Chernoff}, {Connolly}, {Dai}, {Daniel}, {Di Stefano},
  {Drout}, {Galbany}, {Gonz{\'a}lez-Gait{\'a}n}, {Graham}, {Hlo{\v{z}}ek},
  {Ishida}, {Guillochon}, {Jha}, {Jones}, {Mandel}, {Muthukrishna}, {O'Grady},
  {Peters}, {Pierel}, {Ponder}, {Pr{\v{s}}a}, {Rodney}, {Villar}, {LSST Dark
  Energy Science Collaboration}, \& {Transient and Variable Stars Science
  Collaboration}}]{plasticc_redshift}
{Kessler}, R., {Narayan}, G., {Avelino}, A., {et~al.} 2019, \pasp, 131, 094501,
  \dodoi{10.1088/1538-3873/ab26f1}

\bibitem[{Kisley {et~al.}(2023)Kisley, Qin, Zabludoff, Barnard, \&
  Ko}]{Kisley_2023}
Kisley, M., Qin, Y.-J., Zabludoff, A., Barnard, K., \& Ko, C.-L. 2023, The
  Astrophysical Journal, 942, 29, \dodoi{10.3847/1538-4357/aca532}

\bibitem[{Koposov {et~al.}(2023)Koposov, Speagle, Barbary, Ashton, Bennett,
  Buchner, Scheffler, Cook, Talbot, Guillochon, Cubillos, Ramos, Johnson, Lang,
  Ilya, Dartiailh, Nitz, McCluskey, \& Archibald}]{dynesty_docs}
Koposov, S., Speagle, J., Barbary, K., {et~al.} 2023, joshspeagle/dynesty:
  v2.1.3, v2.1.3,  Zenodo, \dodoi{10.5281/zenodo.8408702}

\bibitem[{Lema{{\^i}}tre {et~al.}(2017)Lema{{\^i}}tre, Nogueira, \&
  Aridas}]{imblearn}
Lema{{\^i}}tre, G., Nogueira, F., \& Aridas, C.~K. 2017, Journal of Machine
  Learning Research, 18, 1.
\newblock \url{http://jmlr.org/papers/v18/16-365.html}

\bibitem[{Leoni {et~al.}(2022)Leoni, Ishida, Peloton, \& Möller}]{Leoni_2022}
Leoni, M., Ishida, E. E.~O., Peloton, J., \& Möller, A. 2022, Astronomy \&
  Astrophysics, 663, A13, \dodoi{10.1051/0004-6361/202142715}

\bibitem[{{LSST Science Collaboration} {et~al.}(2009){LSST Science
  Collaboration}, Abell, Allison, Anderson, Andrew, Angel, Armus, Arnett,
  Asztalos, Axelrod, Bailey, Ballantyne, Bankert, Barkhouse, Barr, Barrientos,
  Barth, Bartlett, Becker, Becla, Beers, Bernstein, Biswas, Blanton, Bloom,
  Bochanski, Boeshaar, Borne, Bradac, Brandt, Bridge, Brown, Brunner, Bullock,
  Burgasser, Burge, Burke, Cargile, Chandrasekharan, Chartas, Chesley, Chu,
  Cinabro, Claire, Claver, Clowe, Connolly, Cook, Cooke, Cooray, Covey,
  Culliton, de~Jong, de~Vries, Debattista, Delgado, Dell'Antonio, Dhital,
  Stefano, Dickinson, Dilday, Djorgovski, Dobler, Donalek, Dubois-Felsmann,
  Durech, Eliasdottir, Eracleous, Eyer, Falco, Fan, Fassnacht, Ferguson,
  Fernandez, Fields, Finkbeiner, Figueroa, Fox, Francke, Frank, Frieman,
  Fromenteau, Furqan, Galaz, Gal-Yam, Garnavich, Gawiser, Geary, Gee, Gibson,
  Gilmore, Grace, Green, Gressler, Grillmair, Habib, Haggerty, Hamuy, Harris,
  Hawley, Heavens, Hebb, Henry, Hileman, Hilton, Hoadley, Holberg, Holman,
  Howell, Infante, Ivezic, Jacoby, Jain, R, Jedicke, Jee, Jernigan, Jha,
  Johnston, Jones, Juric, Kaasalainen, Styliani, Kafka, Kahn, Kaib, Kalirai,
  Kantor, Kasliwal, Keeton, Kessler, Knezevic, Kowalski, Krabbendam, Krughoff,
  Kulkarni, Kuhlman, Lacy, Lepine, Liang, Lien, Lira, Long, Lorenz, Lotz,
  Lupton, Lutz, Macri, Mahabal, Mandelbaum, Marshall, May, McGehee, Meadows,
  Meert, Milani, Miller, Miller, Mills, Minniti, Monet, Mukadam, Nakar, Neill,
  Newman, Nikolaev, Nordby, O'Connor, Oguri, Oliver, Olivier, Olsen, Olsen,
  Olszewski, Oluseyi, Padilla, Parker, Pepper, Peterson, Petry, Pinto, Pizagno,
  Popescu, Prsa, Radcka, Raddick, Rasmussen, Rau, Rho, Rhoads, Richards,
  Ridgway, Robertson, Roskar, Saha, Sarajedini, Scannapieco, Schalk, Schindler,
  Schmidt, Schmidt, Schneider, Schumacher, Scranton, Sebag, Seppala, Shemmer,
  Simon, Sivertz, Smith, Smith, Smith, Spitz, Stanford, Stassun, Strader,
  Strauss, Stubbs, Sweeney, Szalay, Szkody, Takada, Thorman, Trilling, Trimble,
  Tyson, Berg, Berk, VanderPlas, Verde, Vrsnak, Walkowicz, Wandelt, Wang, Wang,
  Warner, Wechsler, West, Wiecha, Williams, Willman, Wittman, Wolff,
  Wood-Vasey, Wozniak, Young, Zentner, \& Zhan}]{lsst_sb_ch9}
{LSST Science Collaboration}, Abell, P.~A., Allison, J., {et~al.} 2009, LSST
  Science Book, Version 2.0, 309--344.
\newblock \doarXiv{0912.0201}

\bibitem[{Malanchev(2021)}]{lc_python_docs}
Malanchev, K. 2021, light-curve/light-curve-python: Light-Curve Feature
  Extraction Library for python.
\newblock \url{https://github.com/light-curve/light-curve-python}

\bibitem[{{Malanchev} {et~al.}(2021){Malanchev}, {Pruzhinskaya}, {Korolev},
  {Aleo}, {Kornilov}, {Ishida}, {Krushinsky}, {Mondon}, {Sreejith}, {Volnova},
  {Belinski}, {Dodin}, {Tatarnikov}, {Zheltoukhov}, \& {(The SNAD
  Team)}}]{lc_python}
{Malanchev}, K.~L., {Pruzhinskaya}, M.~V., {Korolev}, V.~S., {et~al.} 2021,
  \mnras, 502, 5147, \dodoi{10.1093/mnras/stab316}

\bibitem[{Masci {et~al.}(2020)Masci, Laher, Rusholme, Shupe, Groom, Surace, \&
  Jackson}]{ztf_supplement}
Masci, F.~J., Laher, R.~R., Rusholme, B., {et~al.} 2020.
\newblock
  \url{https://web.ipac.caltech.edu/staff/fmasci/ztf/ztf_pipelines_deliverables.pdf}

\bibitem[{Masci {et~al.}(2018)Masci, Laher, Rusholme, Shupe, Groom, Surace,
  Jackson, Monkewitz, Beck, Flynn, Terek, Landry, Hacopians, Desai, Howell,
  Brooke, Imel, Wachter, Ye, Lin, Cenko, Cunningham, Rebbapragada, Bue, Miller,
  Mahabal, Bellm, Patterson, Jurić, Golkhou, Ofek, Walters, Graham, Kasliwal,
  Dekany, Kupfer, Burdge, Cannella, Barlow, Sistine, Giomi, Fremling,
  Blagorodnova, Levitan, Riddle, Smith, Helou, Prince, \& Kulkarni}]{ztf_data}
---. 2018, Publications of the Astronomical Society of the Pacific, 131,
  018003, \dodoi{10.1088/1538-3873/aae8ac}

\bibitem[{{Matheson} {et~al.}(2021){Matheson}, {Stubens}, {Wolf}, {Lee},
  {Narayan}, {Saha}, {Scott}, {Soraisam}, {Bolton}, {Hauger}, {Silva},
  {Kececioglu}, {Scheidegger}, {Snodgrass}, {Aleo}, {Evans-Jacquez}, {Singh},
  {Wang}, {Yang}, \& {Zhao}}]{antares2021}
{Matheson}, T., {Stubens}, C., {Wolf}, N., {et~al.} 2021, \aj, 161, 107,
  \dodoi{10.3847/1538-3881/abd703}

\bibitem[{Matplotlib(2023)}]{matplotlib_software}
Matplotlib. 2023, Matplotlib: Visualization with Python, v3.8.2,  Zenodo,
  \dodoi{10.5281/zenodo.10150955}

\bibitem[{Metzger {et~al.}(2013)Metzger, Vurm, Hascoët, \&
  Beloborodov}]{Metzger_2013}
Metzger, B.~D., Vurm, I., Hascoët, R., \& Beloborodov, A.~M. 2013, Monthly
  Notices of the Royal Astronomical Society, 437, 703,
  \dodoi{10.1093/mnras/stt1922}

\bibitem[{Microsoft(2023)}]{lightgbm_docs}
Microsoft. 2023,  Microsoft Corporation.
\newblock \url{https://lightgbm.readthedocs.io/en/v4.1.0/}

\bibitem[{{Mukherjee} {et~al.}(2006){Mukherjee}, {Parkinson}, \&
  {Liddle}}]{single_ellipsoid}
{Mukherjee}, P., {Parkinson}, D., \& {Liddle}, A.~R. 2006, \apjl, 638, L51,
  \dodoi{10.1086/501068}

\bibitem[{Muthukrishna {et~al.}(2019)Muthukrishna, Narayan, Mandel, Biswas, \&
  Hlo{\v{z}}ek}]{rapid}
Muthukrishna, D., Narayan, G., Mandel, K.~S., Biswas, R., \& Hlo{\v{z}}ek, R.
  2019, Publications of the Astronomical Society of the Pacific, 131, 118002,
  \dodoi{10.1088/1538-3873/ab1609}

\bibitem[{{Narayan} {et~al.}(2018){Narayan}, {Zaidi}, {Soraisam}, {Wang},
  {Lochner}, {Matheson}, {Saha}, {Yang}, {Zhao}, {Kececioglu}, {Scheidegger},
  {Snodgrass}, {Axelrod}, {Jenness}, {Maier}, {Ridgway}, {Seaman}, {Evans},
  {Singh}, {Taylor}, {Toeniskoetter}, {Welch}, {Zhu}, \& {ANTARES
  Collaboration}}]{antares2018}
{Narayan}, G., {Zaidi}, T., {Soraisam}, M.~D., {et~al.} 2018, \apjs, 236, 9,
  \dodoi{10.3847/1538-4365/aab781}

\bibitem[{Nelder \& Mead(1965)}]{simplex}
Nelder, J.~A., \& Mead, R. 1965, The Computer Journal, 7, 308,
  \dodoi{10.1093/comjnl/7.4.308}

\bibitem[{Nicholl {et~al.}(2017)Nicholl, Guillochon, \& Berger}]{Nicholl_2017}
Nicholl, M., Guillochon, J., \& Berger, E. 2017, The Astrophysical Journal,
  850, 55, \dodoi{10.3847/1538-4357/aa9334}

\bibitem[{NumPy(2024)}]{numpy_docs}
NumPy. 2024, NumPy 1.26.3 release notes.
\newblock \url{https://numpy.org/devdocs/release/1.26.3-notes.html}

\bibitem[{NumPyro(2019)}]{numpyro_docs}
NumPyro. 2019, NumPyro 0.12.1 documentation,  Uber Technologies, Inc.
\newblock \url{https://num.pyro.ai/en/0.12.1/}

\bibitem[{Nyholm {et~al.}(2020{\natexlab{a}})Nyholm, Sollerman, Tartaglia,
  Taddia, Fremling, Blagorodnova, Filippenko, Gal-Yam, Howell, Karamehmetoglu,
  {et~al.}}]{nyholm2020type}
Nyholm, A., Sollerman, J., Tartaglia, L., {et~al.} 2020{\natexlab{a}},
  Astronomy \& Astrophysics, 637, A73

\bibitem[{Nyholm {et~al.}(2020{\natexlab{b}})Nyholm, Sollerman, Tartaglia,
  Taddia, Fremling, Blagorodnova, Filippenko, Gal-Yam, Howell, Karamehmetoglu,
  Kulkarni, Laher, Leloudas, Masci, Kasliwal, Mor{\aa}, Moriya, Ofek,
  Papadogiannakis, Quimby, Rebbapragada, \& Schulze}]{Nyholm_2020}
---. 2020{\natexlab{b}}, Astronomy \& Astrophysics, 637, A73,
  \dodoi{10.1051/0004-6361/201936097}

\bibitem[{Pandas(2020)}]{pandas_docs}
Pandas. 2020, pandas-dev/pandas: Pandas, 2.0.3,  Zenodo,
  \dodoi{10.5281/zenodo.8092754}

\bibitem[{Paszke {et~al.}(2019)Paszke, Gross, Massa, Lerer, Bradbury, Chanan,
  Killeen, Lin, Gimelshein, Antiga, Desmaison, Kopf, Yang, DeVito, Raison,
  Tejani, Chilamkurthy, Steiner, Fang, Bai, \& Chintala}]{pytorch}
Paszke, A., Gross, S., Massa, F., {et~al.} 2019, in Advances in Neural
  Information Processing Systems 32, ed. H.~Wallach, H.~Larochelle,
  A.~Beygelzimer, F.~d'Alché Buc, E.~Fox, \& R.~Garnett (Curran Associates,
  Inc.), 8024--8035

\bibitem[{Pedregosa {et~al.}(2011)Pedregosa, Varoquaux, Gramfort, Michel,
  Thirion, Grisel, Blondel, Prettenhofer, Weiss, Dubourg, Vanderplas, Passos,
  Cournapeau, Brucher, Perrot, \& Duchesnay}]{sklearn}
Pedregosa, F., Varoquaux, G., Gramfort, A., {et~al.} 2011, Journal of Machine
  Learning Research, 12, 2825

\bibitem[{{Perley} {et~al.}(2020){Perley}, {Fremling}, {Sollerman}, {Miller},
  {Dahiwale}, {Sharma}, {Bellm}, {Biswas}, {Brink}, {Bruch}, {De}, {Dekany},
  {Drake}, {Duev}, {Filippenko}, {Gal-Yam}, {Goobar}, {Graham}, {Graham}, {Ho},
  {Irani}, {Kasliwal}, {Kim}, {Kulkarni}, {Mahabal}, {Masci}, {Modak}, {Neill},
  {Nordin}, {Riddle}, {Soumagnac}, {Strotjohann}, {Schulze}, {Taggart},
  {Tzanidakis}, {Walters}, \& {Yan}}]{ztf_bts2}
{Perley}, D.~A., {Fremling}, C., {Sollerman}, J., {et~al.} 2020, \apj, 904, 35,
  \dodoi{10.3847/1538-4357/abbd98}

\bibitem[{{Pessi} {et~al.}(2023){Pessi}, {Anderson}, {Folatelli}, {Dessart},
  {Gonz{\'a}lez-Gait{\'a}n}, {M{\"o}ller}, {Guti{\'e}rrez}, {Mattila},
  {Reynolds}, {Charalampopoulos}, {Filippenko}, {Galbany}, {Gal-Yam},
  {Gromadzki}, {Hiramatsu}, {Howell}, {Inserra}, {Kankare}, {Lunnan},
  {Martinez}, {McCully}, {Meza}, {M{\"u}ller-Bravo}, {Nicholl}, {Pellegrino},
  {Pignata}, {Sollerman}, {Tucker}, {Wang}, \& {Young}}]{pessi2023}
{Pessi}, P.~J., {Anderson}, J.~P., {Folatelli}, G., {et~al.} 2023, \mnras, 523,
  5315, \dodoi{10.1093/mnras/stad1822}

\bibitem[{Phan {et~al.}(2019)Phan, Pradhan, \& Jankowiak}]{phan2019composable}
Phan, D., Pradhan, N., \& Jankowiak, M. 2019, arXiv preprint arXiv:1912.11554

\bibitem[{Robitaille {et~al.}(2023)Robitaille, Tollerud, Aldcroft, van
  Kerkwijk, Lim, Bray, Droettboom, Sipőcz, Price-Whelan, Conseil, Bradley,
  Starkman, Dencheva, Mumford, Ginsburg, Jamieson, Craig, Seifert, mdmueller,
  StuartLittlefair, D'Avella, lglattly, Homeier, Günther, Vaher, Donath,
  perrygreenfield, Vinícius, Deil, \& Cara}]{astropy_docs}
Robitaille, T., Tollerud, E., Aldcroft, T., {et~al.} 2023, astropy/astropy:
  v5.3.1, v5.3.1,  Zenodo, \dodoi{10.5281/zenodo.8136839}

\bibitem[{Rose {et~al.}(2021)Rose, Baltay, Hounsell, Macias, Rubin, Scolnic,
  Aldering, Bohlin, Dai, Deustua, Foley, Fruchter, Galbany, Jha, Jones, Joshi,
  Kelly, Kessler, Kirshner, Mandel, Perlmutter, Pierel, Qu, Rabinowitz, Rest,
  Riess, Rodney, Sako, Siebert, Strolger, Suzuki, Thorp, Dyk, Wang, Ward, \&
  Wood-Vasey}]{romanhighlat}
Rose, B.~M., Baltay, C., Hounsell, R., {et~al.} 2021, A Reference Survey for
  Supernova Cosmology with the Nancy Grace Roman Space Telescope.
\newblock \doarXiv{2111.03081}

\bibitem[{Rubin {et~al.}(2016)Rubin, Gal-Yam, Cia, Horesh, Khazov, Ofek,
  Kulkarni, Arcavi, Manulis, Yaron, Vreeswijk, Kasliwal, Ben-Ami, Perley, Cao,
  Cenko, Rebbapragada, Woźniak, Filippenko, Clubb, Nugent, Pan, Badenes,
  Howell, Valenti, Sand, Sollerman, Johansson, Leonard, Horst, Armen, Fedrow,
  Quimby, Mazzali, Pian, Sternberg, Matheson, Sullivan, Maguire, \&
  Lazarevic}]{Rubin_2016}
Rubin, A., Gal-Yam, A., Cia, A.~D., {et~al.} 2016, The Astrophysical Journal,
  820, 33, \dodoi{10.3847/0004-637X/820/1/33}

\bibitem[{Saha {et~al.}(2014)Saha, Matheson, Snodgrass, Kececioglu, Narayan,
  Seaman, Jenness, \& Axelrod}]{antares}
Saha, A., Matheson, T., Snodgrass, R., {et~al.} 2014, SPIE Proceedings,
  \dodoi{10.1117/12.2056988}

\bibitem[{Saito \& Rehmsmeier(2015)}]{roc_pr_curves}
Saito, T., \& Rehmsmeier, M. 2015, PLOS ONE, 10,
  \dodoi{10.1371/journal.pone.0118432}

\bibitem[{{Salvatier} {et~al.}(2016){Salvatier}, {Wiecki{\^a}}, \&
  {Fonnesbeck}}]{pymc3}
{Salvatier}, J., {Wiecki{\^a}}, T.~V., \& {Fonnesbeck}, C. 2016, {PyMC3: Python
  probabilistic programming framework}, Astrophysics Source Code Library,
  record ascl:1610.016

\bibitem[{{S{\'a}nchez-S{\'a}ez} {et~al.}(2021){S{\'a}nchez-S{\'a}ez}, {Reyes},
  {Valenzuela}, {F{\"o}rster}, {Eyheramendy}, {Elorrieta}, {Bauer},
  {Cabrera-Vives}, {Est{\'e}vez}, {Catelan}, {Pignata}, {Huijse}, {De Cicco},
  {Ar{\'e}valo}, {Carrasco-Davis}, {Abril}, {Kurtev}, {Borissova}, {Arredondo},
  {Castillo-Navarrete}, {Rodriguez}, {Ruz-Mieres}, {Moya},
  {Sabatini-Gacit{\'u}a}, {Sep{\'u}lveda-Cobo}, \&
  {Camacho-I{\~n}iguez}}]{alerce}
{S{\'a}nchez-S{\'a}ez}, P., {Reyes}, I., {Valenzuela}, C., {et~al.} 2021, \aj,
  161, 141, \dodoi{10.3847/1538-3881/abd5c1}

\bibitem[{Sanders {et~al.}(2015)Sanders, Soderberg, Gezari, Betancourt,
  Chornock, Berger, Foley, Challis, Drout, Kirshner, Lunnan, Marion, Margutti,
  McKinnon, Milisavljevic, Narayan, Rest, Kankare, Mattila, Smartt, Huber,
  Burgett, Draper, Hodapp, Kaiser, Kudritzki, Magnier, Metcalfe, Morgan, Price,
  Tonry, Wainscoat, \& Waters}]{Sanders_2015}
Sanders, N.~E., Soderberg, A.~M., Gezari, S., {et~al.} 2015, The Astrophysical
  Journal, 799, 208, \dodoi{10.1088/0004-637X/799/2/208}

\bibitem[{{Schlafly} \& {Finkbeiner}(2011)}]{dustmaps2}
{Schlafly}, E.~F., \& {Finkbeiner}, D.~P. 2011, \apj, 737, 103,
  \dodoi{10.1088/0004-637X/737/2/103}

\bibitem[{{Schlegel} {et~al.}(1998){Schlegel}, {Finkbeiner}, \&
  {Davis}}]{dustmaps1}
{Schlegel}, D.~J., {Finkbeiner}, D.~P., \& {Davis}, M. 1998, \apj, 500, 525,
  \dodoi{10.1086/305772}

\bibitem[{Skilling(2006)}]{random_walks}
Skilling, J. 2006, Bayesian Analysis, 1, 833 , \dodoi{10.1214/06-BA127}

\bibitem[{Smith(2014)}]{Smith_2014}
Smith, N. 2014, Annual Review of Astronomy and Astrophysics, 52, 487–528,
  \dodoi{10.1146/annurev-astro-081913-040025}

\bibitem[{Speagle(2020)}]{speagle_2020}
Speagle, J.~S. 2020, Monthly Notices of the Royal Astronomical Society, 493,
  3132–3158, \dodoi{10.1093/mnras/staa278}

\bibitem[{Tonry {et~al.}(2018)Tonry, Denneau, Heinze, Stalder, Smith, Smartt,
  Stubbs, Weiland, \& Rest}]{atlas}
Tonry, J.~L., Denneau, L., Heinze, A.~N., {et~al.} 2018, Publications of the
  Astronomical Society of the Pacific, 130, 064505,
  \dodoi{10.1088/1538-3873/aabadf}

\bibitem[{Tyson(2002)}]{tyson_2002}
Tyson, J. 2002, SPIE Proceedings, \dodoi{10.1117/12.456772}

\bibitem[{Villar {et~al.}(2019)Villar, Berger, Miller, Chornock, Rest, Jones,
  Drout, Foley, Kirshner, Lunnan, Magnier, Milisavljevic, Sanders, \&
  Scolnic}]{Villar_2019}
Villar, V.~A., Berger, E., Miller, G., {et~al.} 2019, The Astrophysical
  Journal, 884, 83, \dodoi{10.3847/1538-4357/ab418c}

\bibitem[{Villar {et~al.}(2020)Villar, Hosseinzadeh, Berger, Ntampaka, Jones,
  Challis, Chornock, Drout, Foley, Kirshner, \& et~al.}]{superraenn}
Villar, V.~A., Hosseinzadeh, G., Berger, E., {et~al.} 2020, The Astrophysical
  Journal, 905, 94, \dodoi{10.3847/1538-4357/abc6fd}

\bibitem[{villrv(2020)}]{superraenn_docs}
villrv. 2020, villrv/SuperRAENN: v1.0 Release, v1.0,  Zenodo,
  \dodoi{10.5281/zenodo.3968715}

\bibitem[{Virtanen {et~al.}(2020)Virtanen, Gommers, Oliphant, Haberland, Reddy,
  Cournapeau, Burovski, Peterson, Weckesser, Bright, {van der Walt}, Brett,
  Wilson, Millman, Mayorov, Nelson, Jones, Kern, Larson, Carey, Polat, Feng,
  Moore, {VanderPlas}, Laxalde, Perktold, Cimrman, Henriksen, Quintero, Harris,
  Archibald, Ribeiro, Pedregosa, {van Mulbregt}, \& {SciPy 1.0
  Contributors}}]{scipy}
Virtanen, P., Gommers, R., Oliphant, T.~E., {et~al.} 2020, Nature Methods, 17,
  261, \dodoi{10.1038/s41592-019-0686-2}

\bibitem[{{W}es {M}c{K}inney(2010)}]{pandas}
{W}es {M}c{K}inney. 2010, in {P}roceedings of the 9th {P}ython in {S}cience
  {C}onference, ed. {S}t\'efan van~der {W}alt \& {J}arrod {M}illman, 56 -- 61,
  \dodoi{10.25080/Majora-92bf1922-00a}

\end{thebibliography}
